%% file: main.tex
\documentclass[prd,aps,longbibliography,preprintnumbers,nofootinbib,floatfix,superscriptaddress]{revtex4-2}
\usepackage{graphicx}
\usepackage{epsfig}
\usepackage{rotating}
\usepackage{amssymb}
\usepackage{dsfont}
\usepackage{psfrag}
\usepackage{amsmath,euscript,array,mathrsfs}
\usepackage{bbold}
\usepackage{bm}
\usepackage{multirow}
\usepackage{dcolumn}
\usepackage{hyperref}
\usepackage{caption}
\usepackage{subcaption}
\usepackage[skip=10pt plus1pt, indent=20pt]{parskip}
\usepackage{hyperref}
\usepackage{float}
\floatstyle{plaintop}
\restylefloat{table}
\usepackage{bbding}
\usepackage{natbib}
\usepackage{xcolor}
\newcommand{\beq}{\begin{equation}}
\newcommand{\eeq}{\end{equation}}
\newcommand{\beqs}{\begin{eqnarray}}
\newcommand{\eeqs}{\end{eqnarray}}

\newcommand{\Tr}{{\rm Tr}}

\def\hbar{\hspace{0pt}\raisebox{1pt}{$-$} \hspace{-7pt} h}

\newcommand{\be}{\begin{equation}}
\newcommand{\ee}{\end{equation}}

\newcommand{\bea}{\begin{eqnarray}}
\newcommand{\eea}{\end{eqnarray}}
\newcommand{\nn}{\nonumber}

\newcolumntype{C}[1]{>{\centering\let\newline\\\arraybackslash\hspace{0pt}}m{#1}}

\def\lbldef#1#2{\expandafter\gdef\csname #1\endcsname {#2}}

\def\href#1#2{#2}

\newcommand{\ber}{\begin{eqnarray}}
\newcommand{\eer}{\end{eqnarray}}
\newcommand{\beqar}{\begin{eqnarray}}

\newcommand{\eeqar}{\end{eqnarray}}

\newcommand{\dsl}
  {\kern.06em\hbox{\raise.15ex\hbox{$/$}\kern-.56em\hbox{$\partial$}}}

\newcommand{\eeqarr}{\end{eqnarray}}
\newcommand{\ZZ}{{\rm \kern 0.275em Z \kern -0.92em Z}\;}

\def\CC{{\mathchoice
{\rm C\mkern-8mu\vrule height1.45ex depth-.05ex
width.05em\mkern9mu\kern-.05em}
{\rm C\mkern-8mu\vrule height1.45ex depth-.05ex
width.05em\mkern9mu\kern-.05em}
{\rm C\mkern-8mu\vrule height1ex depth-.07ex
width.035em\mkern9mu\kern-.035em}
{\rm C\mkern-8mu\vrule height.65ex depth-.1ex
width.025em\mkern8mu\kern-.025em}}}

\def\RR{{\rm I\kern-1.6pt {\rm R}}}

\def\ZZ{{\rm Z}\kern-3.8pt {\rm Z} \kern2pt}
\def\IB{\relax{\rm I\kern-.18em B}}
\def\ID{\relax{\rm I\kern-.18em D}}
\def\II{\relax{\rm I\kern-.18em I}}
\def\IP{\relax{\rm I\kern-.18em P}}

\newcommand{\bear}{\begin{eqnarray}}
\newcommand{\eear}{\end{eqnarray}}

\def\6{\partial}

\def\bea{\begin{eqnarray}}
\def\eea{\end{eqnarray}}

\def\beqx{\begin{displaymath}}
\def\eeqx{\end{displaymath}}

\newcommand{\bmat}{\left(\begin{array}}
\newcommand{\emat}{\end{array}\right)}

\newcolumntype{L}[1]{>{\raggedright\let\newline\\\arraybackslash\hspace{0pt}}m{#1}}

\def\bo{{\raise-.3ex\hbox{\large$\Box$}}}               
\def\face{{\raise.2ex\hbox{$\displaystyle \bigodot$}\mskip-2.2mu \llap {$\ddot
        \smile$}}}                                   
\def\>{\rangle}                                      
\def\<{\langle}                                      

\def\leftrightarrowfill{$\mathsurround=0pt \mathord\leftarrow \mkern-6mu
        \cleaders\hbox{$\mkern-2mu \mathord- \mkern-2mu$}\hfill
        \mkern-6mu \mathord\rightarrow$}        
\def\dvec#1{\vbox{\ialign{##\crcr
        \leftrightarrowfill\crcr\noalign{\kern-1pt\nointerlineskip}
        $\hfil\displaystyle{#1}\hfil$\crcr}}}           
\def\Tr{{\rm Tr \,}}                                    

\def\-{\hphantom{-}}

\begin{document}

\preprint{CTPU-PTC-23-49}

\title{Lattice investigations of the chimera baryon spectrum in the $Sp(4)$ gauge theory}

\vspace{6mm}

\begin{abstract}

We report the results of lattice numerical studies of the $Sp(4)$ gauge theory coupled to fermions (hyperquarks) transforming
in the fundamental and two-index antisymmetric representations of the gauge group.
This strongly-coupled theory is the minimal candidate for the ultraviolet completion of composite Higgs models that facilitate the mechanism of partial compositeness for generating the top-quark mass.
We measure the spectrum of the low-lying, half-integer spin, bound states composed of two fundamental and one antisymmetric hyperquarks, dubbed chimera baryons,  in the quenched approximation. 

In this first systematic, non-perturbative study, we focus on the three lightest parity-even chimera-baryon states, in analogy with QCD, denoted as  $\Lambda_{\rm CB}$, $\Sigma_{\rm CB}$ (both with spin $1/2$), and $\Sigma_{\rm CB}^\ast$(with spin $3/2$).
The spin-$1/2$ such states are candidates of the top partners.
The extrapolation of our results to the continuum and massless-hyperquark limit is performed using formulae inspired by QCD heavy-baryon Wilson chiral perturbation theory.
Within the range of hyperquark masses in our simulations, we find that $\Sigma_{\mathrm{CB}}$ is not heavier than $\Lambda_{\mathrm{CB}}$.

\end{abstract}

\author{Ed~Bennett}
\email{e.j.bennett@swansea.ac.uk}
\affiliation{Swansea Academy of Advanced Computing, Swansea University,
Fabian Way, SA1 8EN Swansea, Wales, UK}

\author{Deog~Ki~Hong}
\email{dkhong@pusan.ac.kr}
\affiliation{Department of Physics, Pusan National University, Busan 46241, Korea}
\affiliation{Extreme Physics Institute, Pusan National University, Busan 46241, Korea}

\author{Ho~Hsiao}
\email{thepaulxiao.sc09@nycu.edu.tw}
\affiliation{Institute of Physics, National Yang Ming Chiao Tung University, 1001 Ta-Hsueh 
Road, Hsinchu 30010, Taiwan}

\author{Jong-Wan~Lee}
\email{j.w.lee@ibs.re.kr}
\affiliation{Particle Theory and Cosmology Group, Center for Theoretical Physics of the Universe,
Institute for Basic Science (IBS), Daejeon, 34126, Korea}
\affiliation{Extreme Physics Institute, Pusan National University, Busan 46241, Korea}

\author{C.-J.~David~Lin}
\email{dlin@nycu.edu.tw}
\affiliation{Institute of Physics, National Yang Ming Chiao Tung University, 1001 Ta-Hsueh Road, Hsinchu 30010, Taiwan}
\affiliation{Center for High Energy Physics, Chung-Yuan Christian University, Chung-Li 32023, Taiwan}
\affiliation{Centre for Theoretical and Computational Physics, National Yang Ming Chiao Tung University, 1001 Ta-Hsueh Road, Hsinchu 30010, Taiwan}

\author{Biagio~Lucini}
\email{b.lucini@swansea.ac.uk}
\affiliation{Department of Mathematics, Faculty  of Science and Engineering,
Swansea University, Fabian Way, SA1 8EN Swansea, Wales, UK}
\affiliation{Swansea Academy of Advanced Computing, Swansea University,
Fabian Way, SA1 8EN Swansea, Wales, UK}

\author{Maurizio~Piai}
\email{m.piai@swansea.ac.uk}
\affiliation{Department of Physics, Faculty  of Science and Engineering,
Swansea University, Singleton Park, SA2 8PP, Swansea, Wales, UK}

\author{Davide~Vadacchino}
\email{davide.vadacchino@plymouth.ac.uk}
\affiliation{Centre for Mathematical Sciences, University of Plymouth, Plymouth, PL4 8AA, United Kingdom}

\date{\today}

\maketitle
\flushbottom
\newpage
\tableofcontents

%
%

\section{Introduction}
\label{Sec:intro}

The discovery of the Higgs boson~\cite{ATLAS:2012yve,CMS:2012qbp} has exacerbated the need for a deeper understanding of the origin of electroweak symmetry breaking (EWSB).
On the one hand, no firm experimental evidence has been found of violations of the standard model (SM) predictions.  On the other hand, a plethora of considerations, in particular the triviality of the scalar sector~\cite{Aizenman:1981zz, Frohlich:1982tw, Luscher:1987ay, Luscher:1987ek, Luscher:1988uq, Bernreuther:1987hv, Kenna:1992np, Gockeler:1992zj, Wolff:2009ke, Weisz:2010xx, Hogervorst:2011zw, Siefert:2014ela} (and most likely of the whole Higgs-Yukawa sector~\cite{Molgaard:2014mqa, Bulava:2012rb, Chu:2018ldw}) implies that the SM cannot be a fundamental theory, but it rather provides an effective field theory (EFT) description, valid up to some large, but finite, ultraviolet (UV) cut-off scale, beyond which the SM has to be completed.
The challenge is that any theory serving as the UV completion of the SM must contain a light scalar state that can be interpreted as the observed Higgs boson, while also reproducing the observed SM phenomenology, up to the TeV scale, and down to the current (high) level of precision.

Composite Higgs models (CHMs)~\cite{Kaplan:1983fs,Georgi:1984hp,Dugan:1984hq}, for example those in Refs.~\cite{Contino:2003ve,Agashe:2004rs,Ferretti:2013kya,Ferretti:2016upr, Cacciapaglia:2019bqz,Ferretti:2014qta,Barnard:2013zea,Ma:2015gra,Vecchi:2015fma,Appelquist:2020bqj}---see also the reviews in Refs.~\cite{Contino:2010rs,Panico:2015jxa,Witzel:2019jbe,Cacciapaglia:2019vcb,Cacciapaglia:2020kgq,Bennett:2023wjw}---have been attracting attention in recent years, because they can naturally accommodate a light Higgs boson.
In these models, a novel strongly coupled  sector is introduced, based upon an asymptotically-free gauge theory coupled to fermions (hyperquarks).
At variance with technicolor models, the SM Higgs boson emerges as one of the pseudo-Nambu-Goldstone bosons (PNGBs), associated with a global symmetry of the new strong interaction, to provide a UV completion for the standard model.
The global symmetry is broken both spontaneously (by the condensates forming dynamically) and explicitly, hence the PNGBs develop a potential due to (small) symmetry breaking effects. Such effects may arise either within the strong-coupling sector itself (e.g., hyperquark mass terms) or due to its coupling to external fields (e.g., couplings to  SM fields).
As the Higgs fields are identified with a subset those that describe the PNGBs in the low-energy EFT description of the theory, EWSB is triggered by the interplay among different symmetry-breaking effects, along the lines of vacuum alignment analysis~\cite{Peskin:1980gc} and radiative EWSB~\cite{Coleman:1973jx}—for recent studies in the context of CHMs, see for instance Refs.~\cite{Arkani-Hamed:2002ikv,Contino:2003ve,Agashe:2004rs,Contino:2010rs,Golterman:2015zwa,Golterman:2017vdj,Banerjee:2023ipb}.

Since CHMs involve strongly coupled dynamics requiring a non-perturbative treatment, it is natural to rely on lattice calculations for their investigation.
Our collaboration has been performing such calculations for a particular UV-completion that is built with the $Sp(4)$ gauge theory containing two flavors ($N_{f} = 2$) of Dirac fermions in the fundamental, $(f)$, representation~\cite{Barnard:2013zea,Ferretti:2013kya}.
We denote these fundamental hyperquarks by $Q^{i\,a}$, where $a=1,\ldots, 4$ is the hypercolor index and $i=1,2$ the flavor one.
Because of the pseudoreality of this representation of the gauge group, the approximate global symmetry acting on the $(f)$ hyperquarks is $SU(4)$, which is broken spontaneously to $Sp(4)$~\cite{Peskin:1980gc}.
This results in five PNGBs, four of which can be interpreted as the SM complex scalar doublet, provided
the $SU(2)\times U(1)$ SM gauge group  is chosen as an appropriate subgroup of the components of $SU(4)$.
The $SU(4)/Sp(4)$ coset leads to the minimal CHM amenable to lattice treatment, in the sense that it gives candidates for the SM Higgs doublet with only one additional Goldstone mode.
Previous publications~\cite{Bennett:2017kga, Bennett:2019jzz, Bennett:2019cxd} reported on the meson spectra of this theory obtained from both quenched and dynamical lattice simulations. An extended study of meson spectra computed in the quenched approximation, for
various $Sp(2N)$ groups, and matter fields transforming in several different representations of the group, is in preparation~\cite{Bennett:quenched}.

It is possible to extend CHMs to address the flavor problem, or at least
its most challenging aspect: to generate the large mass for the SM top quark,
without  spoiling the SM successful description of flavor-changing neutral current 
processes and precision electroweak observables.
To address this challenge, the idea of (top) partial compositeness was introduced in Ref.~\cite{Kaplan:1991dc} (see also the discussions in Refs.~\cite{Grossman:1999ra,Gherghetta:2000qt,Lodone:2008yy,Chacko:2012sy,Grojean:2013qca}).
If one couples the theory to hyperquarks transforming in two different representations of the gauge group, 
and embeds the SM gauge group as an appropriate  subgroup of  the global symmetry  of the new sector, some of the
bound states formed by hyperquarks in different representations can be arranged to 
carry the same quantum numbers as the top quark.
Such bound states can be identified as top partners.
The top quark then acquires its mass by mixing with the top partners.
In  the $Sp(4)$ gauge theory with $N_{f}=2$ $(f)$ hyperquarks, top partial compositeness can be achieved by adding to the theory $n_{f}=3$  Dirac fermions in the two-index antisymmetric, $(as)$, representation of the gauge group~\cite{Bennett:2022yfa, Barnard:2013zea}.
We  denoted the $(as)$ hyperquarks by $\Psi^{k\,ab}$, with $k=1,2,3$ the flavor index.
Because the $(as)$ representation is real, the global symmetry for three flavors is $SU(6)$, spontaneously broken to $SO(6)$~\cite{Peskin:1980gc}—see also the CHMs in Refs.~\cite{ Cacciapaglia:2019ixa,Cai:2020njb}.
One can gauge the $SU(3)$ subgroup of the unbroken $SO(6)$, and identify it with the  QCD gauge group~\cite{Barnard:2013zea,Ferretti:2013kya}.
We call chimera baryons the hypercolor singlet bound states formed by one $\Psi$ and two  $Q$ fields.
Spin-1/2 chimera baryons can then act as candidate top partners. 
See Refs.~\cite{Cossu:2019hse,DelDebbio:2022qgu,Bergner:2021ivi,Ayyar:2019exp,Ayyar:2018zuk,Ayyar:2017qdf} for recent work on candidate  top partners in other gauge theories.

For the purposes of this paper, we consider the strongly coupled theory in isolation, hence there are no SM fields nor interactions.
We present our measurements of the masses of chimera baryons sourced by the following operators:
\beqs
 {\mathcal{O}}_{\rho}^{ijk,5} &\equiv& Q^{i\,a}_{\alpha} (C\gamma^{5})_{\alpha\beta} Q^{j\,b}_{\beta} \Omega^{ad}\Omega^{bc} \Psi^{k\,cd}_{\rho} \, , \label{eq:chim_bar_src} \\ 
 {\mathcal{O}}_{\rho}^{ijk,\mu} &\equiv& Q^{i\,a}_{\alpha} (C\gamma^{\mu})_{\alpha\beta} Q^{j\,b}_{\,\beta} \Omega^{ad}\Omega^{bc} \Psi^{k\,cd}_{\rho} \, ,\label{eq:chim_bar_src_mu}
 \eeqs
where $a,b,c,d$ are hypercolor indices, $\alpha, \beta$, $\rho$ are spinor indices, $i,j,k$ are flavor indices,
$\gamma^5$ and $\gamma^{\mu}$ are $4\times 4$ Dirac matrices,
and $C$ is the charge conjugation matrix.
 The symplectic matrix, $\Omega$, is defined as
\begin{equation}
\label{eq:Omega_def}
\Omega \equiv \left ( 
\begin{array}{cccc}
 0 & 0 & 1 & 0\\
 0 & 0 & 0 & 1\\
 -1 & 0 & 0 & 0\\
 0 & -1 & 0 & 0\\
\end{array}
\right ) \, .
 \end{equation}

We restrict our attention to operators for which the $SU(4)$ index is off-diagonal, $i\not= j$.
For mesons this requirement ensures that there is no disconnected contraction in computing two-point correlation functions
\footnote{Pioneering lattice studies of the flavor-singlet meson sector of candidate completions for CHMs can be found for example in Refs.~\cite{Arthur:2016ozw,Drach:2021uhl,Bennett:2023rsl}.}.
For chimera baryons it removes from the calculations diagrams involving $(f)$-type contractions within the initial and final state.
The operator ${\mathcal{O}}^{5}$ annihilates spin-1/2 composite states. Following an analogy with the $\Lambda$ baryon in QCD, to which we return later in this section, we denote the lightest state of this type as $\Lambda_{\mathrm{CB}}$.
The operator in Eq.~(\ref{eq:chim_bar_src_mu}), ${\mathcal{O}}^{\mu}$,  overlaps with both spin-1/2 and 3/2 states, and we denote the lightest ones by $\Sigma_{\mathrm{CB}}$ and $\Sigma^{\ast}_{\mathrm{CB}}$, respectively.
Both $\Lambda_{\mathrm{CB}}$ and $\Sigma_{\mathrm{CB}}$ baryons can be candidate  top partners~\cite{Gripaios:2009pe,Banerjee:2022izw}.
We report the quantum numbers of the three chimera baryons in Tab.~\ref{tab:chimerabaryons}, together with 
some of the properties of the analogous particle in QCD.
\footnote{  Other exotic hypercolor-singlet bound states can exist in the $Sp(4)$ gauge theory. However, these states are expected to be unstable and heavier than the chimera baryons. Furthermore,  given our primary focus on spin-1/2 states as top partners, we only study those chimera baryons in this work.}
Our lattice calculations of the masses of $\Lambda_{\mathrm{CB}}$, $\Sigma_{\mathrm{CB}}$ and $\Sigma^{\ast}_{\mathrm{CB}}$ are performed in the quenched approximation.
The determination of these masses is of importance in constructing a viable UV-complete composite Higgs model with partial compositeness, because it affects both the mass of the top quark, and direct and indirect 
new physics searches for top partners.

\begin{table}[t]
\begin{center}
\caption{The chimera baryons of interest in this paper, the interpolating operators that source them, 
their quantum numbers---spin, $J$, and irreducible representation of the unbroken, global symmetry groups, $Sp(4)$ and $SO(6)$---and
the properties of the analogous QCD state---mass in MeV (rounded to unit), strangeness, $S$, isospin, $I$, spin, $J$~\cite{ParticleDataGroup:2022pth}.
See also Refs.~\cite{Bennett:2022yfa,Hsiao:2022kxf,Bennett:2023wjw}.
\label{tab:chimerabaryons}}
\begin{tabular}{|c|c|C{1.2cm}|C{1.2cm}|C{1.2cm}|C{6cm}|}
\hline\hline
{\rm Chimera Baryon} & {\rm ~Interpolating operator~} & $J$  & $Sp(4)$ & $SO(6)$  & {\rm QCD analogy} \cr
\hline
 $\Lambda_{{\rm CB}}$
 &
${\mathcal{O}}^{5}$\,, Eq.~(\ref{eq:chim_bar_src}) & ${1}/{2}$ & $5$ & $6$
& $\Lambda(1116)$, $S=-1$, $I=0$, $J=1/2$\cr
\hline
 $\Sigma_{{\rm CB}}$
 &
${\mathcal{O}}^{\mu}$\,, Eq.~(\ref{eq:chim_bar_src_mu})&  ${1}/{2}$ & $10$& $6$
& $\Sigma(1193)$, $S=-1$, $I=1$, $J=1/2$\cr
\hline
 $\Sigma^{\ast}_{{\rm CB}}$
 &
${\mathcal{O}}^{\mu}$\,, Eq.~(\ref{eq:chim_bar_src_mu})& ${3}/{2}$ & $10$& $6$
& $\Sigma^{\ast}(1379)$, $S=-1$, $I=1$, $J=3/2$\cr
\hline
\end{tabular}
\end{center}
\end{table}

The mechanism by which SM fermions (the top quark in particular) acquire a mass via their coupling to chimera baryon operators of the strongly coupled sector is rather different from that provided by the Yukawa couplings in the standard model, as well as from the coupling to the meson operators adopted in extended technicolor~\cite{Dimopoulos:1979es,Eichten:1979ah} and walking technicolor~\cite{Yamawaki:1985zg,Holdom:1984sk,Appelquist:1986an} theories.
In particular, the value of the dynamically generated scaling dimension of the chimera baryon operators enters non-trivially into the estimates of the resulting SM fermion masses---see for example the discussion in Sections.~IV.B and V.B of Ref.~\cite{Chacko:2012sy},
in the studies reported  in Refs.~\cite{Ayyar:2018glg,BuarqueFranzosi:2019eee,DeGrand:2015yna},
 in Section~2.4.2 of the review~\cite{Bennett:2023wjw}, and references therein.
Measuring these scaling dimensions is an ambitious task that requires dedicated methodology---see Ref.~\cite{Hasenfratz:2023sqa} for recent progress along these lines, but in a different theory---and that we leave for the future, as it goes far beyond the reach of the quenched approximation we adopt here.

Yang-Mills theories have a well-defined limit for a large number of colors~\cite{tHooft:1973alw}.
 $SU(N_c)$ theories coupled to a finite number of fermions give, in the large-$N_c$ limit, a good description of 
important  properties of strong interactions, such as Zweig's rule, or vector meson dominance~\cite{Witten:1979kh}.
Furthermore, baryons can be realized as solitons, in agreement with Skyrme's picture~\cite{Skyrme:1961vq,Witten:1983tx}.
If one naively expects baryons in $Sp(2N)$ gauge theories to be well-defined in the large-$N$ limit, yet baryons made of $2N$ fundamental hyperquarks are unstable, decaying into $N$ mesons, since the totally antisymmetric tensor can be decomposed into  products of symplectic structures,  schematically written as
\begin{equation}
\epsilon_{a_1a_2\cdots a_{2N}}=\Omega_{a_1a_2}\Omega_{a_3a_4}\cdots\Omega_{a_{2N-1}a_{2N}}\pm\cdots.
\end{equation}
One can still have a well-defined limit in two ways. Either one generalizes the rank-$2$ antisymmetric hyperquark to the antisymmetric 
rank-$N$ hyperquark, transforming as the Pfaffian of $Sp(2N)$, to form a color-singlet with $N$ fundamental hyperquarks. 
As an alternative, one can also consider the singlet state obtained with one (conjugate) antisymmetric fermion and two fundamental ones: this state exists for both $SU(N_c)$ and $Sp(N_c=2N)$ theories, the two large $N_c$ limits yielding a common, finite mass, and one can show that in $SU(3)$ this is an ordinary baryon. Because in the following the two species of fermions have different masses, we can make an analogy for the heavier, conjugate antisymmetric fermions with the strange quark and for the fundamental fermions with the up and down quarks, leading to the aforementioned association of the states of interest in this paper with the  
$\Lambda$, $\Sigma$, and $\Sigma^\ast$ states in QCD---see for instance Ref.~\cite{Corrigan:1979xf} for a discussion within 
 $SU(N_c)$ gauge theories.

This paper is organized in the following way.  In Section~\ref{Sec:lattice}, we describe lattice field theory basic definitions, such as  the simulation algorithm and the correlation functions that enter our  measurements of chimera baryon masses.  Section~\ref{Sec:results} describes our data analysis procedure in the extraction of the chimera-baryon masses.  It also details the strategy applied to the continuum and massless extrapolations of these masses.  We then summarize our findings  in Section~\ref{Sec:summary}. More technical details are relegated to the Appendices.

\section{Lattice numerical calculations}
\label{Sec:lattice}

Lattice field theory enables to perform   first-principle non-perturbative computations in quantum field theory.
Since little is known about chimera baryon spectra in $Sp(4)$ gauge theories~\cite{Bennett:2022yfa}, we adopt the quenched approximation, which significantly reduces the demands on computing resources, while allowing the exploration of parameter space, independent of the number of fields, $N_f$ and $n_f$.
Based upon experience gained from  quenched calculations of the spectrum of QCD, we envisage that this approximation gives reasonably accurate results in some of the regions of parameter space of interest, in which the number of fermions is not too large, 
or their mass is not too small.
Furthermore, performing this first study in the quenched approximation facilitates an extensive scan of the space of bare parameter, to yield benchmarking information for our future computations involving dynamical hyperquarks.

This section describes the lattice action and provides technical details necessary to reproduce our calculations.
More details, such as the specific features of our implementation of the heat bath algorithm for $Sp(4)$ gauge theory and the scale-setting procedure based on the gradient flow, can be found in Refs.~\cite{Bennett:2017kga,Bennett:2019cxd,Bennett:2021mbw,Bennett:2022ftz}.  We also define the interpolating operators and correlation functions relevant for this work, in Section~\ref{Sec:corrleation}. Some additional technical details can be found in the appendix, in particular  pertaining to our use of smearing. 

\subsection{Lattice action}
\label{Sec:action}
We discretize the four-dimensional $Sp(4)$ gauge theory on a spatially isotropic Euclidean lattice.
The  dynamics of the gauge degrees of freedom is described by the standard Wilson plaquette action, $S_{g}$,  given by
\beq
S_g\equiv\beta \sum_x \sum_{\mu <\nu} 
\left(
1-\frac{1}{4}\textrm{Re}\,{\Tr} \,\mathcal{P}_{\mu\nu}
\right)\,,
\label{eq:gauge_action}
\eeq
where $\beta \equiv 8/g^2$ is the bare lattice coupling. The plaquette,  $\mathcal{P}_{\mu\nu}$, is defined as
\beq
\mathcal{P}_{\mu\nu}(x)\equiv U_\mu (x)U_\nu(x+\hat{\mu})U^\dagger_\mu (x+\hat{\nu})U^\dagger_\nu(x)\, ,
\label{eq:plaquette}
\eeq
with the link variable, $U_{\mu}(x) \in Sp(4)$, transforming in the adjoint representation of the gauge group.  
The action $S_{g}$ is used in our Monte Carlo computations to generate gauge-field ensembles.

The hyperquarks, constituents of the chimera baryons,  are fermions whose dynamics is described by the Wilson-Dirac lattice action
\beq
S_f \equiv a^4 \sum_{i=1}^{N_f}\sum_x \overline{Q^{i}}(x) D^{(f)}_m Q^{i}(x)+
a^4 \sum_{j=1}^{n_f}\sum_x \overline{\Psi^{j}}(x) D^{(as)}_m \Psi^{j}(x),
\label{eq:fermion_action}
\eeq
where $a$ is the lattice spacing, while $i$ and $j$ are  flavor indices---hypercolor and spinor indexes are understood. 
Explicitly, we write the following, with $R=(f)$ for fermions transforming in the fundamental representation, and  $R=(as)$ 
in the case of the 2-index antisymmetric representation:
\beqs\label{eq:WDO}
\overline{\psi^{R\,j}}(x) D^{R}_{m} \psi^{R\,j}(x) &\equiv& 
\overline{\psi^{R\,j}}(x)(4/a+m^{R}_0) \psi^{R\,j}(x)  \\
& & -\frac{1}{2a}\sum_\mu \nonumber
\overline{\psi^{R\,j}}(x)\left\{(1-\gamma_\mu)U^{R}_\mu(x)\psi^{R\,j}(x+\hat{\mu})
+(1+\gamma_\mu)U^{R\,\dagger}_\mu(x-\hat{\mu})\psi^{R\,j}(x-\hat{\mu})\right\}\,, \label{eq:DiracF}
\eeqs
with $\psi^{(f)\,j} = Q^{j}$, $\psi^{(as)\,j} = \Psi^{j}$, and $U_{\mu}^{(f)} = U_{\mu}$.  The construction of the antisymmetric gauge link, $U^{(as)}_{\mu}$, is detailed in Ref.~\cite{Bennett:2022yfa}.
The symbol $m_0^{R}$ denotes the flavor diagonal (degenerate) bare mass of  hyperquarks, $\psi^{R}_j$, transforming in the corresponding representation, $R$, of the gauge group.

\subsection{Numerical strategy}
\label{Sec:implementation}

For this work we use  the open source HiRep code~\cite{DelDebbio:2008zf}, with the add-ons we developed
in the context of earlier publications in order to implement $Sp(4)$~\cite{Bennett:2017kga} ---see also the first lattice study of symplectic gauge group~\cite{Holland:2003kg} and the recent implementation of $Sp(2N)$ in the Grid environment~\cite{Bennett:2023gbe,Forzano:2023duk,Boyle:2015tjk,Boyle:2016lbp,Yamaguchi:2022feu}.
Gauge field ensembles are generated using one-plus-four combinations of heat bath plus over-relaxation update algorithms. Two successive configurations in the Markov chain are separated by twelve such updates of the whole lattice.
More details of the implementation of this procedure can be found in Ref.~\cite{Bennett:2017kga}.
Also, in every Markov chain, the initial $600$ configurations are treated as thermalization steps and discarded from the measurements of physical observables.
For each ensemble, we generate $200$ configurations.
We monitor the topological charge and its evolution, to ascertain that there is no evidence of topological freezing.
We denote the dimensionless lattice volume as $N_{t}\times N_{s}^{3}$, where $N_{t}$ and $N_{s}$ are the temporal and spatial lattice extents, respectively.
Periodic boundary conditions are imposed on gauge fields, in all directions.
For  hyperquark fields, periodic and anti-periodic boundary conditions are implemented in spatial and temporal directions, respectively.

We generate five ensembles with different values of the lattice bare coupling $\beta$.
We summarize in Tab.~\ref{tab:ENS} the defining properties of each ensemble. We set the scale of dimensionful physical observables by employing the gradient-flow 
method~\cite{Luscher:2010iy, Luscher:2011bx,Luscher:2013vga}.
The procedure outlined in Ref.~\cite{BMW:2012hcm} yields  the quantity $w_{0}/a$, where $w_{0}$ has dimension of an inverse mass.
This scale-setting exercise was already carried out and reported in detail in previous publications---see Table II of Ref.~\cite{Bennett:2019cxd}, as well as the extensive discussions in Ref.~\cite{Bennett:2022ftz}---hence we borrow results for $w_{0}/a$ from Ref.~\cite{Bennett:2019cxd}.
We notice that, in respect to Eq.~(2.3) of Ref.~\cite{BMW:2012hcm},  we use the different reference value ${\cal W}_0=0.35$, rather than $0.30$. The information presented in Tab.~\ref{tab:ENS} shows that the spread of our choices of the lattice bare coupling corresponds to a variation of the lattice spacing roughly by a factor of two, which 
allows us to perform a first extrapolation of our results towards the continuum limit.
In this work, when a dimensional quantity is expressed in units of $w_{0}$, the corresponding dimensionless quantity is denoted with the caret symbol.  For instance, $\hat{a} \equiv a/w_{0}$ and $\hat{m} \equiv w_{0} m$, where $m$ stands  for a generic mass. 
The lattice parameters being identical,  the relevant autocorrelation 
times can be found in Table III in Ref.~\cite{Bennett:2019cxd}.

\begin{table}
   \caption{
  Gauge ensembles generated for the $Sp(4)$ theory.
   We report the bare coupling $\beta$, the lattice size, $N_t\times N^3_s$, the average plaquette $\left < P \right >$, and the gradient-flow scale $w_0/a$. The gradient-flow scales are taken from Ref.~\cite{Bennett:2019cxd}.
   \label{tab:ENS}}
\begin{center}
    \input{tabs/table_2.tex}
   \end{center}
\end{table}

\subsection{Interpolating operators and correlation functions}
\label{Sec:corrleation}

Following the notation introduced in Ref.~\cite{Bennett:2022yfa}, we denote the generic structure of the chimera baryon interpolating operators, built out of two  $(f)$ and one $(as)$ hyperquarks, as 
\beq\label{eq:ocb}
\mathcal{O}_{\rm{CB},\rho} (x) \equiv
\left (
{Q^{i\,a}}_\alpha(x) \frac{}{} \Gamma^{1\,\alpha\beta} {Q^{j\,b}}_\beta(x)
\right)
\Omega_{ad}\Omega_{bc} \Gamma^{2\, \rho\sigma} {\Psi^{k\,cd}}_{\sigma}(x)\,,
\eeq
where $\Gamma^{1,2}$ are gamma matrices and $\Omega$ is the $4\times4$ symplectic matrix defined in Eq.~(\ref{eq:Omega_def}), with $a,\,b,\,c,\,d$ being $Sp(4)$-hypercolor, $i,\,j,\,k$  flavor, and $\alpha,\,\beta,\,\sigma,\,\rho$  spinor indices.
Operators given in Eqs.~(\ref{eq:chim_bar_src})~and~(\ref{eq:chim_bar_src_mu}) are special cases of this generic structure.  
The Dirac conjugate operator of $\mathcal{O}_{\rm CB,\rho} (x)$ is
\beq\label{eq:ocb_bar}
\overline{\mathcal{O}_{\rm{CB},\rho}} (x) \equiv \overline{\Psi^{k\,cd}}_\sigma (x) \Omega_{cb}\Omega_{da}  \Gamma^{2\, \sigma\rho}
\left (
{\overline{Q^{j\,b}}_\beta}(x)  \Gamma^{1\,\beta\alpha} \overline{Q^{i\,a}}_\alpha(x)
\right)\,.
\eeq

The zero momentum, two-point correlation functions of interest, restricted to consider only $i \neq j$, are written as
\beqs
C_{\rm{CB},\sigma\rho}(t) &\equiv& \sum_{\vec{x}} \langle \mathcal{O}_{\rm{CB},\sigma}(x) \overline{\mathcal{O}_{\rm{CB},\rho}}(0) \rangle \nn \\
&=& - \sum_{\vec{x}} \left ( \Gamma^2 {S_{\Psi}^{k\,cd }}_{c^\prime d^\prime} (x,0)  \overline{\Gamma^{2}} \right)_{\sigma\rho} \Omega_{cb}\Omega^{b^\prime c^\prime} \Omega_{ad}\Omega^{d^\prime a^\prime}
 \Tr \left [ \Gamma^{1}  S^{b}_{Q\,\,\,b^{\prime}}(x,0) \overline{\Gamma^{1}} S_{Q\,\,\,\,a^\prime}^{a}(x,0) \right ],
\eeqs
where $x\equiv(t,\vec{x})$, while $\overline{\Gamma} \equiv \gamma^0 \Gamma^\dagger \gamma^0$. The trace is taken over the spinor indices. The hyperquark propagators are
\beq
S^{\,i\,a}_{Q\,\,\,\,b\,\alpha\beta}(x,y) = 
\langle Q^{i\,a}_{\,\,\,\,\,\alpha}(x) \overline{Q^{i\,b}}_{\beta}(y) \rangle\,,
~{\textrm{and}}~S^{\,k\,ab}_{\Psi\,\,\,\,\,\,\,\,cd\,\alpha\beta}(x,y) =
 \langle \Psi^{k\,ab}_{\,\,\,\,\,\,\,\,\,\,\alpha}(x) 
\overline{\Psi^{k\,cd}}_{\beta}(y) \rangle\,.
\label{eq:fermion_prop}
\eeq

We are interested in operators with $(\Gamma^1,\,\Gamma^2)=(C\gamma^5,\,\mathbb{1})$ and $(C\gamma^\mu,\,\mathbb{1})$.  The former overlaps with the $\Lambda_{\rm CB}$ state, while the latter sources both $\Sigma_{\rm CB}$ and $\Sigma^\ast_{\rm CB}$ baryons. 
The chimera baryon interpolating operators in Eq.~(\ref{eq:ocb}) generally couple to  states with both even and odd parity.
In order to facilitate the investigation of the spectrum of $\Lambda_{\rm CB}$, $\Sigma_{\rm CB}$, and $\Sigma^\ast_{\rm CB}$ 
chimera baryons, which are all parity-even, we apply appropriate projection operators, as detailed in Section~\ref{Sec:results}.

Our main objective  is to study how the mass of the chimera baryons changes in response to the variation of the hyperquark masses, in particular because 
it would be interesting to explore the limits in which $m_{0}^{(f)}$ and $m_{0}^{(as)}$ approach zero.
The methodology we apply to the extraction of these hadronic masses is described in Section~\ref{Sec:results}.
We perform our calculations with several choices of $am_{0}^{(f)}$ and $am_{0}^{(as)}$, on each available ensemble,
and report our results  in Appendix~\ref{app:data}.

For sufficiently light hyperquarks, we expect the square of the  pseudoscalar meson mass to depend linearly on the hyperquark mass.
Information on the meson spectrum hence allows us to perform a combined extrapolation to continuum and massless-hyperquark limit.
As this is a quenched calculation, the results of the extrapolation towards the massless-hyperquark limit have to be taken with a grain of salt~\cite{Sharpe:1992ft, Bernard:1992mk}.
Yet, they provide useful input for future dynamical calculations—see Figs.~17-18 in Ref.~\cite{Bennett:2019jzz} for examples of the difference in mesons mass between quenched and dynamical fermions in the case of the fundamental representation.
We can also monitor the ratio between the masses of pseudoscalar and vector mesons, as an indicator of the relative size of explicitly breaking of the global symmetry in the theory.

The meson interpolating operators for $(f)$ and $(as)$ hyperquarks are 
\beq
\label{eq:om}
 {\mathcal{O}}^{(f)}_{{\mathrm{M}}}(x) = \overline{{Q^{i\,a}}}_\alpha(x) \Gamma^{\alpha\beta}_{{\mathrm{M}}} {Q^{j\,b}}_\beta(x) \,,
 ~{\textrm{and}}~~
 {\mathcal{O}}^{(as)}_{{\mathrm{M}}}(x) = \overline{{\Psi^{k\,ab}}}_{\alpha}(x) \Gamma_{{\mathrm{M}}}^{\alpha\beta} {\Psi^{m\,cd}}_{\beta}(x) \,, 
\eeq
respectively. We can set $\Gamma_{{\mathrm{M}}} = \gamma^{5}$ for the pseudoscalar, and  $\Gamma_{{\mathrm{M}}} = \gamma^{\mu}$ for the vector mesons. Imposing  the restriction $i \neq j$ and $k \neq m$,  no disconnected diagrams contribute to the two-point correlation function, which read
\beq
\label{eq:meson_2pt}
C_{\rm M}^{(f)}(t) \equiv \sum_{\vec{x}} \langle \mathcal{O}^{(f)}_{\rm M}(x) \mathcal{O}_{\rm M}^{{(f)}\dagger}(0) \rangle = - \sum_{\vec{x}} \Tr \left [\gamma^5 \Gamma_{\rm M} S^{\,i\,a}_{Q\,\,\,\,a^\prime} (x,0) \bar{\Gamma}_{\rm M} \gamma^5 S^{\,i\,b \dagger}_{Q\,\,\,\,b^\prime}(x,0)\right]\,, 
\eeq
for mesons made with $(f)$ hyperquarks, and
\beq
\label{eq:meson_2pt_as}
C_{\rm M}^{(as)}(t) \equiv \sum_{\vec{x}} \langle \mathcal{O}^{(as)}_{\rm M}(x) \mathcal{O}_{\rm M}^{{(as)}\dagger}(0) \rangle = - \sum_{\vec{x}} \Tr \left [\gamma^5 \Gamma_{\rm M} S^{\,k\,a b}_{\Psi\,\,\,\,\, \,a^\prime b^\prime} (x,0) \bar{\Gamma}_{\rm M} \gamma^5 S^{\,m\, c d \dagger}_{\Psi\,\,\,\,\,\,\,\, c^\prime d^\prime}(x,0)\right]\,,
\eeq
for $(as)$ hyperquarks.
The traces are taken over  spinor indices.
The propagators of $(f)$ and $(as)$ hyperquarks  are given in Eq.~(\ref{eq:fermion_prop}).

The masses of the mesons are extracted from the large-$t$ behavior of  correlation functions.
For convenience, we label the pseudoscalar meson  masses as $m_{\rm PS}$ and $m_{\rm ps}$ and  the masses of the vector meson  as $m_{\rm V}$ and $m_{\rm v}$, with upper case subscripts referring to $(f)$ hyperquarks and lower case one to $(as)$ hyperquarks.
It is well known that numerical results of lattice computations of quantities involving baryons are noisy, and in this work we resort to modifying the correlation functions and the propagators used for chimera baryons and mesons, by applying two smearing techniques: the Wuppertal smearing~\cite{Gusken:1989qx} for the hyperquark fields and the APE smearing~\cite{APE:1987ehd} for the gauge fields.
We describe in Appendix~\ref{app:smearing} 
our implementation of these smearing procedures.

\section{Data analysis and numerical results}
\label{Sec:results}

In this section, we discuss the strategy of our analysis  and report numerical results for the spectrum of the low-lying chimera baryons.
In Section~\ref{Sec:projection}, we describe how we extract ground-state masses with definite spin and parity quantum numbers, by applying appropriate spin and parity projections on the correlation functions.
In Section~\ref{Sec:quenched} we report our measurements of the masses of the pseudoscalar and vector mesons, for both $(f)$ and $(as)$ hyperquarks.  
We then apply relations inspired by Wilson chiral perturbation theory to analyze
the spectra for various hyperquark masses, and we extrapolate to the continuum and massless-hyperquark limit.
The process and 
our results are presented in Section~\ref{Sec:mexpt}.
We employ the Akaike information criterion (AIC)~\cite{Akaike:1998zah} to  optimize for the best analysis procedure over various fitting ansatze and different selections of the data points to be included in this investigation.   In addition,
we also manually check the results, to demonstrate the correctness of the automated analysis.

We anticipate here that throughout  this work, in the data analysis of correlation functions, 
estimates of the statistical errors are obtained via the bootstrap method. 
For each measurement we generate $800$ bootstrap samples.
Technical details on the intermediate steps are relegated to the appendix. In particular, fit 
results of the ground-state masses are presented in Appendix~\ref{app:data},
while the choices of smearing parameters are reported in Appendix~\ref{app:fitting}.

\subsection{Spin and Parity projection}
\label{Sec:projection}
Correlation functions involving the (chimera) baryon operators in Eqs.~(\ref{eq:ocb}) and~(\ref{eq:ocb_bar})  can be further decomposed into components  with different spin and parity quantum numbers~\cite{Fucito:1982ip, Bennett:2022yfa,Hsiao:2022kxf}. 
We denote by ${\mathcal{O}}^{\mu}_{{\rm CB},\rho}$ the operator  with  Dirac matrix structure 
$(\Gamma^1,\Gamma^2)=(C\gamma_\mu,\mathbb{1})$, with $\mu$ running from 1 to 3. It overlaps with both spin-$1/2$ and $3/2$ states.
The corresponding two-point function with vanishing momentum, $\vec{p} = \vec{0}$, can be written as
\beq
C_{{\rm CB},\sigma\rho}^{ \mu \nu} (t) \equiv \sum_{\vec{x}} \langle \mathcal{O}^{\mu}_{\textrm{CB}\,,\sigma}(x) \overline{\mathcal{O}^{\nu}_{\textrm{CB}\,,\rho}}(0) \rangle \, .
\label{eq:corr_mu}
\eeq 
The lightest baryons  dominate the large Euclidean-time behaviors of the spin-1/2 and 3/2 components of $C_{{\rm CB},\sigma\rho}^{\mu \nu }$, and we identify them with $\Sigma_{\mathrm{CB}}$ and $\Sigma^{\ast}_{\mathrm{CB}}$ (see Section~\ref{Sec:intro}), respectively.  We define the following two correlation functions:
\beq
C_{\Sigma_{\mathrm{CB}},\sigma\rho}(t) \equiv \left[P^{1/2}_{\mu\nu}C_{{\rm CB}}^{\mu \nu}(t)\right ]_{\sigma\rho}
~{\textrm{and}}~~
C_{\Sigma^{\ast}_{\mathrm{CB}},\sigma\rho}(t) \equiv \left[ P^{3/2}_{\mu\nu}C_{{\rm CB}}^{\mu \nu}(t) \right]_{\sigma\rho} \, ,
\label{eq:corr_spin_projected}
\eeq
where the spin projectors~\cite{UKQCD:1996ssj} are (for $\mu,\,\nu$ = 1,\,2,\,3)
\beq
P^{1/2}_{\mu\nu} \equiv \frac{1}{3} \gamma^\mu \gamma^\nu\,
~{\textrm{and}}~~
P^{3/2}_{\mu\nu} \equiv \delta^{\mu\nu} -\frac{1}{3}\gamma^\mu \gamma^\nu\,.
\eeq

We define as  ${\mathcal{O}}^{5}_{{\rm CB}, \rho}$ the operator obtained from Eq.~(\ref{eq:ocb}) by considering $(\Gamma^1,\Gamma^2)=(C\gamma^5,\mathbb{1})$.
This operator only overlaps with spin-$1/2$ states,  the ground state of which is
 the $\Lambda_{\mathrm{CB}}$ introduced in Section~\ref{Sec:intro}.  Therefore, we define
\beq
\label{eq:corr_lambda_cb}
 C_{\Lambda_{\rm CB},\sigma\rho} (t)\equiv \sum_{\vec{x}} \langle 
 {\mathcal{O}}^{5}_{{\rm CB}, \sigma}(x)
 \overline{{\mathcal{O}}^{5}_{{\rm CB}, \rho}}(0)
 \rangle \, .
\eeq
For notational simplicity, in the rest of this article we will not write explicitly the spinor indices, $\sigma$ and $\rho$, 
 in the correlation functions  in Eqs.~(\ref{eq:corr_spin_projected}) and (\ref{eq:corr_lambda_cb}), but leave them understood.  Furthermore, we use the symbol $C_{\mathrm{CB}}(t)$ to
denote generically $C_{\Lambda_{\mathrm{CB}}}(t)$, $C_{\Sigma_{\mathrm{CB}}}(t)$, or $C_{\Sigma^{\ast}_{\mathrm{CB}}}(t)$.

 The chimera baryon interpolating operators, ${\mathcal{O}}^{5}_{{\rm CB},\rho}$ and ${\mathcal{O}}^{\mu}_{{\rm CB}, \rho}$, couple to both  even- and odd-parity states.
At large Euclidean time, due to the use of anti-periodic boundary conditions in the temporal direction for hyperquark fields, the two-point correlation function of a chimera baryon, following the convention in Ref.~\cite{Montvay:1994cy},  behaves asymptotically as 
\beq\label{eq:c_CB_eo}
C_{\rm CB}(t) \xrightarrow{0 \ll t \ll T} P_{+} \left[ c_+e^{-m^+ t} - c_- e^{-m^-(T-t)} \right] + P_{-} \left[ c_- e^{-m^- t} - c_+ e^{-m^+ (T-t)} \right]\,,
\eeq
where the parity projectors are
\beq
\label{eq:parity_projectors}
P_\pm \equiv \frac{1 \pm \gamma^0}{2} \,,
\eeq
while
$m^\pm$ are the masses of the even- and odd-parity states, 
and $c_\pm$  the corresponding baryon-to-vacuum matrix elements. 
We define  even- and odd-parity correlation functions, 
$C^{+}_{\rm CB}(t)$ 
and $C^{-}_{\rm CB} (t)$, by  applying the $P_{\pm}$ projectors:
\beq
C^{\pm}_{\rm CB} (t)\equiv P_{\pm} \, C_{\rm CB}(t)\,.
\eeq 
For finite but large extent of the temporal lattice, $T$,
we therefore find that the projected correlation functions 
at large Euclidean time,
$0\ll t \ll T$, 
behave as
\beq
C^{\pm}_{\rm CB} (t) \longrightarrow c_{\pm} e^{-m^\pm t} -c_{\mp} e^{-m^\mp (T-t)}\, .
\label{eq:corr_eo}
\eeq
To improve statistics, in the analysis we employ the averaged correlator, 
\beq\label{eq:cCB}
\overline{C}_{\rm CB}^\pm (t) = \frac{C_{\rm CB}^\pm (t) -C_{\rm CB}^\mp (T-t)}{2} \, ,
\eeq
which exhibits the same asymptotic behavior as in Eq.~(\ref{eq:corr_eo}).

\begin{figure}
\centering
\begin{subfigure}{0.32\textwidth}
    \includegraphics[width=\textwidth]{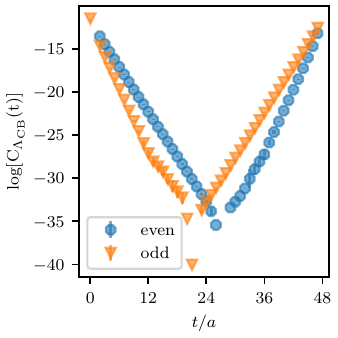}
    \caption{}
    \label{fig:P_lnc}
\end{subfigure}
\hfill
\begin{subfigure}{0.32\textwidth}
    \includegraphics[width=\textwidth]{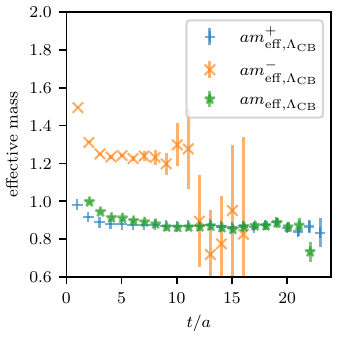}
    \caption{}
    \label{fig:P_m}
\end{subfigure}
\hfill
\begin{subfigure}{0.32\textwidth}
    \includegraphics[width=\textwidth]{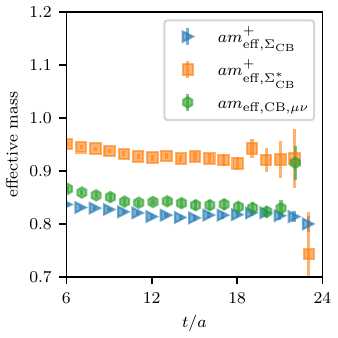}
    \caption{}
    \label{fig:spin_m}
\end{subfigure}
\caption{Illustrative examples of chimera baryon correlation functions, (a),  and effective mass plots, (b) and (c),
 obtained on the ensemble QB1 with hyperquark masses $am_0^{(f)}=-0.77$ and  $am_0^{(as)}=-1.1$.
(\subref{fig:P_lnc}): the parity-projected correlators, $C^{\pm}_{\Lambda_{\rm CB}}(t)$.
(\subref{fig:P_m}) effective masses, $a m_{\rm eff}$, extracted from correlation function obtained with, $C^{\pm}_{\Lambda_{\rm CB}}(t)$, and without parity projection, $C_{\Lambda_{\rm CB}}(t)$.
(\subref{fig:spin_m}): effective masses, $a m_{\rm eff}$, extracted from correlation functions upon which spin and parity projections are applied, $C^{+}_{\Sigma_{\rm CB}}(t)$ and $C^{+}_{\Sigma^{\ast}_{\rm CB}}(t)$, or without any projection, $C_{{\rm CB}, \mu\nu}(t)$.
}
\label{fig:projection}
\end{figure}

For both even- and odd-parity states, we define the effective masses as
\beq
am^{\pm}_{\rm eff, CB}(t) = \ln{\left [ \frac{\overline{C}^{\pm}_{\rm CB}(t)}{\overline{C}^{\pm}_{\rm CB}(t+1)} \right] }\, ,
\label{eq:meff_cb}
\eeq
and restrict our attention to  ranges of Euclidean time $0\ll t < T/2$.   
From Eqs.~(\ref{eq:corr_eo})
and~(\ref{eq:cCB}),  one expects  that 
$am^{\pm}_{\rm eff, CB}(t)$, when plotted against time, will asymptotically display a plateau  dominated by either the even-parity or odd-parity ground states, in $\overline{C}_{\rm CB}^+ (t)$ and $\overline{C}_{\rm CB}^- (t)$, respectively.  
By studying and comparing the resulting effective mass plots,  we determine  the parity of the lowest-lying chimera baryon state for each choice of  spin and global symmetry quantum numbers of interest, as listed in Tab.~\ref{tab:chimerabaryons}.  
As a cross-check of our results, we consider also the effective mass computed with  unprojected correlation functions, $C_{\rm CB}$.  In analogy with Eq.~(\ref{eq:meff_cb}), for $0\ll t < T/2$, we define it as 
\beq
am_{\rm eff, CB}(t) = \ln{\left [ \frac{C_{\rm CB}(t)}{C_{\rm CB}(t+1)} \right] }\, .
\label{eq:meff_cb_unproj}
\eeq
Given the asymptotic behavior expected in Eq.~(\ref{eq:c_CB_eo}),  the value of the plateau in $am_{\rm eff, CB}(t)$ should appear at a value compatible with the lightest  between $am^+_{\rm eff, CB}(t)$ and $am^-_{\rm eff, CB}(t)$.

In order to graphically illustrate how projectors affect the effective mass extraction, we present  in Fig.~\ref{fig:P_lnc} the parity-projected correlation functions, 
$C^{\pm}_{\Lambda_{\rm CB}}(t)$, obtained from the ensemble QB1 (see Tab.~\ref{tab:ENS}) with the bare hyperquark masses in the Wilson-Dirac action set to $am_{0}^{f}=-0.77$ and $am_{0}^{as}=-1.05$. 
Notice  the logarithm scale on the vertical axis. 
 The  lattice used to generate this ensemble has Euclidean time extent $T/a = 48$.
By comparing the slopes with the  behavior expected in Eq.~(\ref{eq:c_CB_eo}), one can infer that the parity-even state is lighter than its parity-odd partner, and hence that 
 the $\Lambda_{\rm CB}$ chimera baryon (a candidate top partner) has even parity.

Figure~\ref{fig:P_m} shows the  effective masses, $am^{\pm}_{{\rm eff}, \Lambda_{\rm CB}}$, extracted with and without applying parity projectors. For the $\Lambda_{\rm CB}$ state, the plot clearly demonstrates that $m^{+}_{{\rm eff}, \Lambda_{\rm CB}} < m^{-}_{{\rm eff}, \Lambda_{\rm CB}}$. Furthermore, examination of the  effective mass extracted from the unprojected correlator, $am_{{\rm eff}, \Lambda_{\rm CB}}$, confirms the hierarchy between the masses of the two parity eigenstates.  
It is worthy of notice that in Fig.~\ref{fig:P_m} we can clearly discern the emergence of a plateau for $am^{-}_{{\rm eff}, \Lambda_{\rm CB}}$ at smaller $t/a$.  This negative-parity ground state happens to be substantially heavier, but not parametrically so. It would be interesting to perform
a systematic study of the spectra of this and other heavy baryons, but doing so would go beyond the purposes of the present study, and requires the use of dedicated numerically strategies to optimize the signal. We postpone such a study to the future.

Following  the same procedure, applied to the correlation functions involving the operator ${\cal O}^{\mu}_{{\rm CB},\rho}$,
 we also demonstrate that $m^{+}_{{\rm eff}, \Sigma_{\rm CB}} < m^{-}_{{\rm eff}, \Sigma_{\rm CB}}$, 
 as well as that  $m^{+}_{{\rm eff}, \Sigma^{\ast}_{\rm CB}} < m^{-}_{{\rm eff}, \Sigma^{\ast}_{\rm CB}}$.  Therefore, it is established that $\Lambda_{\rm CB}$, $\Sigma_{\rm CB}$, and $\Sigma^\ast_{\rm CB}$ are all parity even, and we only discuss their masses (denoted as $m_{\Lambda_{\rm CB}}$, $m_{\Sigma_{\rm CB}}$ and $m_{\Sigma^{\ast}_{\rm CB}}$) in the rest of this paper.  
These baryon masses are extracted by performing single-exponential fits of the data for $\overline{C}^{+}_{\rm CB}$ to Eq.~(\ref{eq:corr_eo}) in the interval $0 \ll t \leq T/2$.  
The choice of fit range is guided by the range of the plateau of the effective mass, and can be optimized by tracking the value of $\chi^{2}/N_{\rm d.o.f.}$.

Besides parity, we perform also spin projections, as defined in Eq.~(\ref{eq:corr_spin_projected}), for the correlator $C_{{\rm CB}}^{\mu\nu}(t)$.
By doing so, we can discriminate between $\Sigma_{\rm CB}$ and $\Sigma^{\ast}_{\rm CB}$ states.
Figure~\ref{fig:spin_m} displays the effective masses computed from $C^{\mu\nu,+}_{{\rm CB}}(t)$ measured on ensemble QB1 with spin projections and same hyperquark masses as in Figs.~\ref{fig:P_lnc} and \ref{fig:P_m}.
This plot shows the expected hierarchy, $m_{\Sigma_{\rm CB}} < m_{\Sigma^{\ast}_{\rm CB}}$.  Furthermore, we also display the effective mass obtained from $C_{{\rm CB}}^{\mu\nu}(t)$ with neither spin nor parity projections.
As expected, the plateau value is compatible with that of the $\Sigma_{\rm CB}$ baryon, but contamination with the heavier states results in some deterioration of the signal quality.

\subsection{Mass hierarchy and hyperquark-mass dependence of chimera baryons}
\label{Sec:quenched}

\begin{figure}
\centering
\begin{subfigure}{0.48\textwidth}
    \includegraphics[width=\textwidth]{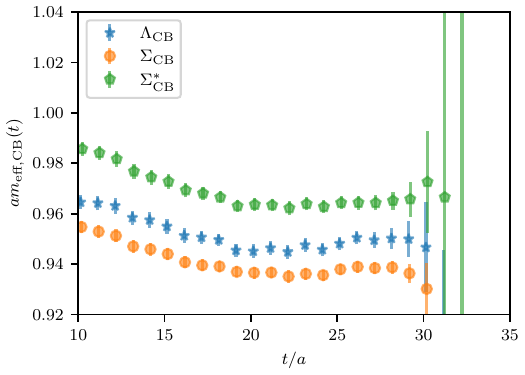}
    \caption{}
    \label{fig:mh_h}
\end{subfigure}
\hfill
\begin{subfigure}{0.48\textwidth}
    \includegraphics[width=\textwidth]{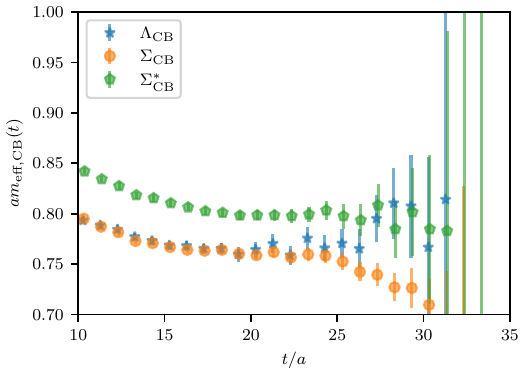}
    \caption{}
    \label{fig:mh_l}
\end{subfigure}
\caption{
Effective masses of the lightest, even-parity chimera baryons measured in the ensemble QB4, for two representative choices of  hyperquark masses, $am_0^{(as)}=-0.81$ and (\subref{fig:mh_h})~$am_0^{(f)}=-0.6$, or (\subref{fig:mh_l})~$am_0^{(f)}=-0.69$.
}\label{fig:mh}
\end{figure}

One interesting feature we observe is the hierarchy between the ground-state chimera baryons in the three channels of interest. 
Figure~\ref{fig:mh} shows $am_{{\rm eff}, \Lambda_{\rm CB}}(t)$, $am_{{\rm eff}, \Sigma_{\rm CB}}(t)$, and $am_{{\rm eff}, \Sigma^{\ast}_{\rm CB}}(t)$ for two representative choices of bare hyperquark masses, $(am_0^{(f)}, am_0^{(as)})=(-0.6,-0.81)$ and $(am_0^{(f)}, am_0^{(as)})=(-0.69,-0.81)$, as measured in the ensemble QB4 (see Tab.~\ref{tab:ENS}).
In the former case,  we find convincing evidence  that $\Sigma_{\rm CB}$ is the lightest among these states.
In the latter case, the $(f)$-type bare hyperquark mass is reduced to $am_0^{(f)}=-0.69$, and as shown in the right panel of Fig.~\ref{fig:mh},  $\Lambda_{\rm CB}$ and $\Sigma_{\rm CB}$ become almost degenerate, their masses cannot be discriminated with given present uncertainties.  For all choices we make of bare hyperquark masses, and  in all available ensembles in Tab.~\ref{tab:ENS}, 
$\Sigma^{\ast}_{\rm CB}$ is always the heaviest amongst the three lowest-lying  parity-even baryon states, and  $\Lambda_{\rm CB}$ is never lighter than $\Sigma_{\rm CB}$.
More detailed investigations of the hierarchy in the chimera-baryon masses, in particular its dependence on the hyperquark masses, will be discussed in this and the next subsections.

In Ref.~\cite{Bennett:2019cxd}, the mass spectrum of the lightest mesons composed of $(f)$ and $(as)$ hyperquark constituents has been reported, based upon measurements  using the same quenched ensembles as in Tab.~\ref{tab:ENS}, 
while varying the hyperquark bare masses. 
For this work, starting from the same choices of bare masses as in Ref.~\cite{Bennett:2019cxd}, we extend the parameter space into the regimes of lighter as well as heavier hyperquarks. 
The inclusion of data points with smaller hyperquark masses makes the massless extrapolation more reliable. 
It also enables access to a wide range of the value of the ratio between $(f)$ and $(as)$ hyperquark masses.
Our aim is to better understand the interplay between these hyperquark masses and the hierarchy amongst $m_{\Lambda_{\rm CB}}$, $m_{\Sigma_{\rm CB}}$ and $m_{\Sigma^{\ast}_{\rm CB}}$. 
To this purpose, we find it convenient to use the square of the mass of the pseudoscalar meson as a reference scale, as in Ref.~\cite{Bennett:2019cxd}, denoting the masses of the pseudoscalar mesons composed of $(f)$ and $(as)$ hyperquarks by $m_{\rm PS}$ and $m_{\rm ps}$, respectively.
For light masses we then expect $(m_{\rm PS,ps})^{2} \sim m^{(f),(as)}$.

At large Euclidean time, the meson two-point correlation functions in Eqs.~(\ref{eq:meson_2pt}) and (\ref{eq:meson_2pt_as}) are expected to behave as
\beq\label{eq:meson_corr}
C_{\rm M}^R(t) \xrightarrow{t /a  \gg1} A \left[ e^{-m_{\rm M}^R t} + e^{-m_{\rm M}^R (T-t)} \right]\,,
\eeq
where $m_{\rm M}^R$ is the mass of the ground-state meson, M, composed of hyperquarks transforming in the representation $R$, and $A$ is the relevant matrix element. 
Following Eq.~(\ref{eq:meson_corr}),
the meson effective mass can be computed through
\beq
am^{R}_{\rm eff, M}(t) \equiv \cosh^{-1}{\left[ \frac{C^R_{\rm M}(t+1)+C^R_{\rm M}(t-1)}{2C^R_{\rm M}(t)} \right ]}\, ,
\eeq
in the range of Euclidean time $0 < t < T-1$.  We then determine the fit interval for extracting the meson mass by performing a correlated fit of $C^R_{\rm M}(t)$ to Eq.~(\ref{eq:meson_corr}), by identifying a suitable plateau in $am^{R}_{\rm eff, M}(t)$.

\begin{figure}
\centering
\begin{subfigure}{\textwidth}
    \includegraphics[width=0.5\textwidth]{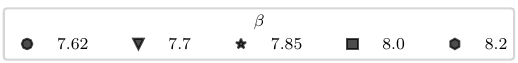}
\end{subfigure}
\begin{subfigure}{0.328\textwidth}
    \includegraphics[width=\textwidth]{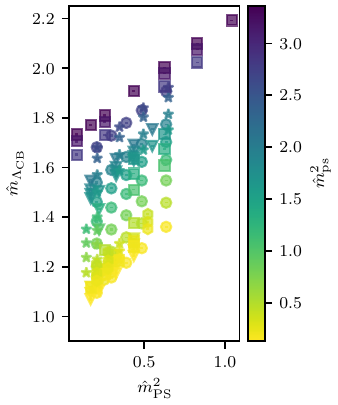}
    \caption{}
    \label{fig:lam_mf}
\end{subfigure}
\hfill
\begin{subfigure}{0.328\textwidth}
    \includegraphics[width=\textwidth]{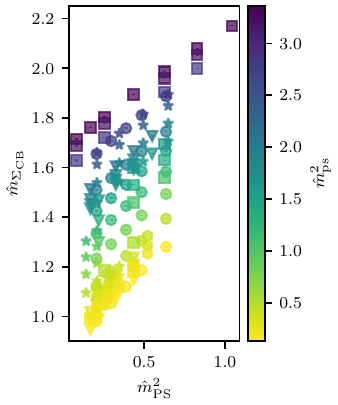}
    \caption{}
    \label{fig:sig_mf}
\end{subfigure}
\hfill
\begin{subfigure}{0.328\textwidth}
    \includegraphics[width=\textwidth]{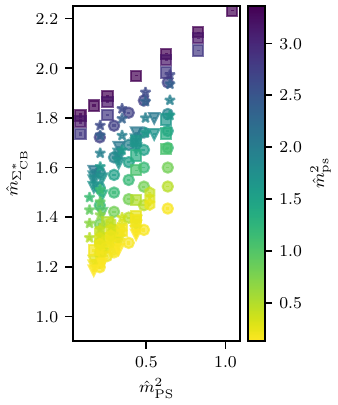}
    \caption{}
    \label{fig:sigs_mf}
\end{subfigure}
\vfill
\begin{subfigure}{0.328\textwidth}
    \includegraphics[width=\textwidth]{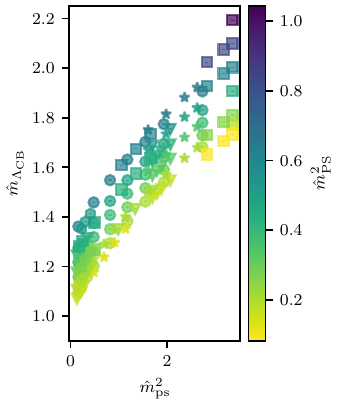}
    \caption{}
    \label{fig:lam_mas}
\end{subfigure}
\hfill
\begin{subfigure}{0.328\textwidth}
    \includegraphics[width=\textwidth]{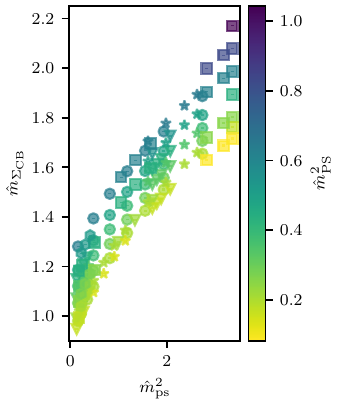}
    \caption{}
    \label{fig:sig_mas}
\end{subfigure}
\hfill
\begin{subfigure}{0.328\textwidth}
    \includegraphics[width=\textwidth]{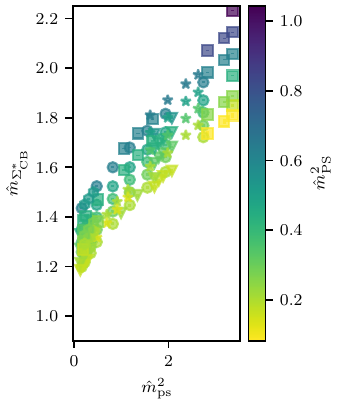}
    \caption{}
    \label{fig:sigs_mas}
\end{subfigure}
\vfill

\caption{
Masses of chimera baryons as functions of $\hat{m}^{2}_{\rm PS}$ and $\hat{m}^{2}_{\rm ps}$. In panels (\subref{fig:lam_mf}) to (\subref{fig:sigs_mf}), the values of $\hat{m}_{\Lambda_{\rm CB}}$, $\hat{m}_{\Sigma_{\rm CB}}$, and $\hat{m}_{\Sigma^{\ast}_{\rm CB}}$ are plotted against $\hat{m}^{2}_{\rm PS}$, while color-coding is used to denote  $\hat{m}^{2}_{\rm ps}$.  In panels (\subref{fig:lam_mas}) to (\subref{fig:sigs_mas}), the horizontal axis is $\hat{m}^{2}_{\rm ps}$, while color-coding denotes 
$\hat{m}^{2}_{\rm PS}$.
The plot markers represent different ensembles, characterized by
the $\beta$ values indicated in the legend. 
}
\label{fig:m_CB}
\end{figure}

\begin{figure}
\centering
\includegraphics[width=0.84\textwidth]{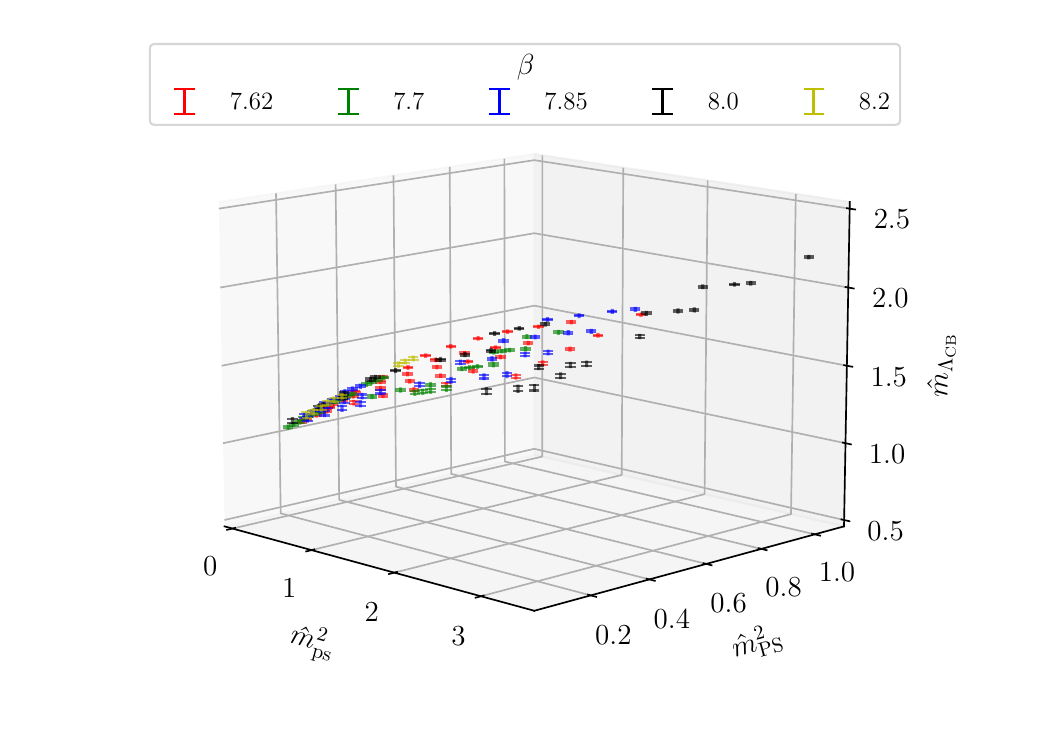}
\caption{
A 3-dimensional plot as an example that shows the chimera-baryon masses as functions of $\hat{m}^{2}_{\rm PS}$ and $\hat{m}^{2}_{\rm ps}$. Different colors refer to different $\beta$ values, as listed in the legend.
}\label{fig:m_CB_3d}

\end{figure}

In Figs.~\ref{fig:lam_mf}, \ref{fig:sig_mf}, and \ref{fig:sigs_mf}, we display $\hat{m}_{\Lambda_{\rm CB}}$, $\hat{m}_{\Sigma_{\rm CB}}$ and $\hat{m}_{\Sigma^{\ast}_{\rm CB}}$ as a function of $\hat{m}^{2}_{\rm PS}$.
For clarity of presentation, the value of $\hat{m}^{2}_{\rm ps}$ is color-coded. Conversely, in Figs.~\ref{fig:lam_mas}, \ref{fig:sig_mas}, and \ref{fig:sigs_mas}, 
the horizontal axis is $\hat{m}^{2}_{\rm ps}$, and the color coding corresponds to the value of  $\hat{m}^{2}_{\rm PS}$.  The data points shown in these six plots are obtained on the five available ensembles listed in Tab.~\ref{tab:ENS}, and are distinguished by the shape of the markers.  The meson masses take values in the range $\hat{m}_{\rm PS} \in [0.28,1.03]$ and $\hat{m}_{\rm ps} \in [0.35,1.84]$.  
The plots illustrate how  chimera-baryon masses decrease as 
either $\hat{m}^{2}_{\rm PS}$ or $\hat{m}^{2}_{\rm ps}$ is reduced, approaching a non-vanishing  limit
for $\hat{m}^{2}_{\rm PS}\rightarrow 0$ or $\hat{m}^{2}_{\rm ps}\rightarrow 0$.
To further demonstrate the dependence on both sources of explicit symmetry breaking (hyperquark masses and lattice spacing), we show all the data points together in the 3-dimensional plot in Fig.~\ref{fig:m_CB_3d} with $\hat{m}_{\Lambda_{\rm CB}}$ as an example. 
These baryon masses measured at different values of hyperquark masses lie on a surface for each value of $\beta$, and slightly decrease as we increase $\beta$.


To study how the mass hierarchy depends on the hyperquark masses, we conduct a thorough exploration across a wide range of $\hat{m}^{2}_{\rm PS}$ and $\hat{m}^{2}_{\rm ps}$.
The left panel of Fig.~\ref{fig:mh_mps2} shows that the ratio between $m_{\Lambda_{\rm CB}}$ and $m_{\Sigma_{\rm CB}}$ decreases at increasing $\hat{m}_{\rm ps}$, and tends to unity in the large-$\hat{m}^{2}_{\rm ps}$ regime.
The right panel of Fig.~\ref{fig:mh_mps2} exhibits a similar trend with respect to the variation of $m_{\rm PS}^{2}$.  Yet, $m_{\Lambda_{\rm CB}}/m_{\Sigma_{\rm CB}}$ is never consistent with $1$ in the region where the mass of the PS meson is large.
When the $(as)$ hyperquark is heavy (denoted by purple markers in the right panel of Fig.~\ref{fig:mh_mps2}), the ratio between $m_{\Lambda_{\rm CB}}$ and $m_{\Sigma_{\rm CB}}$ shows a mild dependence on $m_{\rm PS}^{2}$. In this regime, $m_{\Lambda_{\rm CB}}/m_{\Sigma_{\rm CB}}$ depends primarily on $\hat{m}^{2}_{\rm ps}$.
Within our whole range of hyperquark masses,  $\Lambda_{\rm CB}$ is never lighter than $\Sigma_{\rm CB}$.

\begin{figure}
\centering
\includegraphics[width=\textwidth]{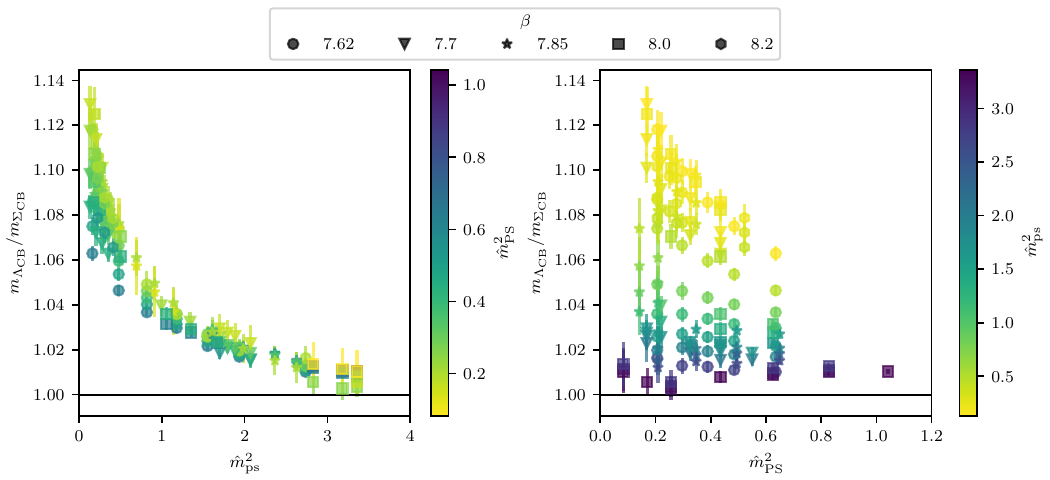}
\label{fig:mh_mps2}
\caption{Left:  ratio between the masses of $\Lambda_{\rm CB}$ and $\Sigma_{\rm CB}$,  plotted against $\hat{m}_{\rm ps}^2$, with the value of $\hat{m}_{\rm PS}^2$ color-coded.  Right: same ratio plotted against $\hat{m}_{\rm PS}^2$ while color-coding $\hat{m}_{\rm ps}^2$.  Different markers  denote different $\beta$ values as listed in the legend.}
\label{fig:mh_mps2}
\end{figure}

We conduct a similar, systematic investigation of  the ratio between $\hat{m}_{\Sigma_{\rm CB}}$ and $\hat{m}_{\Sigma^*_{\rm CB}}$, and the results are presented in Fig.~\ref{fig:ss_mps2}. 
We find that $\Sigma_{\rm CB}$ is always lighter than $\Sigma_{\rm CB}^*$,  their mass gap decreasing as $\hat{m}^{2}_{\rm PS}$ and $\hat{m}^{2}_{\rm ps}$ increase. 
This feature can be interpreted in terms of   heavy-hyperquark spin symmetry~\cite{Isgur:1989vq}. 
As the hyperquark masses increase, the effects of spin, which account for the mass difference between $\Sigma_{\rm CB}$ and $\Sigma^\ast_{\rm CB}$, are suppressed.

\begin{figure}
\centering
    \includegraphics[width=\textwidth]{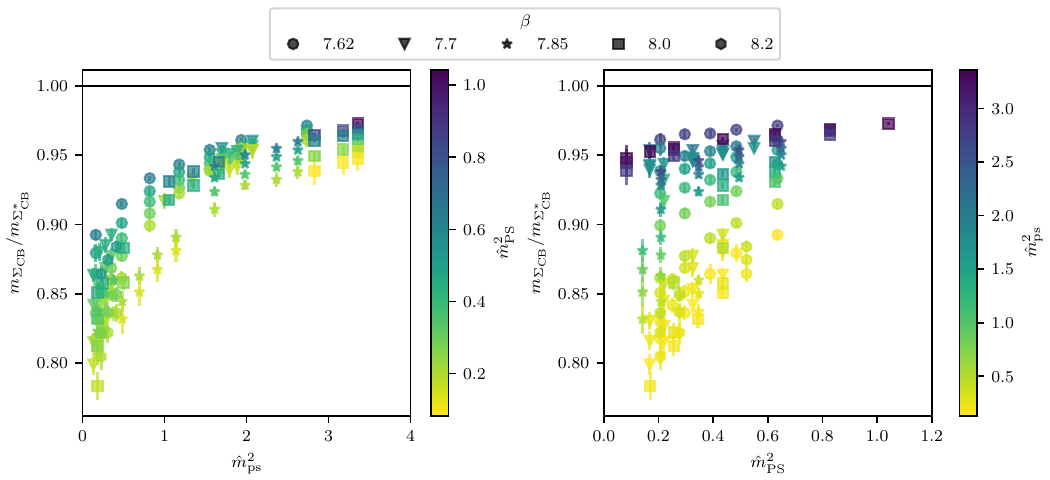}
    \label{fig:ss_mps2}
\caption{Left:  ratio between the masses of $\Sigma_{\rm CB}$ and $\Sigma^{\ast}_{\rm CB}$, plotted against $\hat{m}_{\rm ps}^2$, with  $\hat{m}_{\rm PS}^2$ color-coded.  Right: the same ratio plotted against $\hat{m}_{\rm PS}^2$, while color-coding
 $\hat{m}_{\rm ps}^2$.  Different markers denote different $\beta$ values as listed in the legend.}
\label{fig:ss_mps2}
\end{figure}
%

\subsection{Mass extrapolations and cross checks}
\label{Sec:mexpt}
We now discuss our extrapolation towards the continuum and the massless-hyperquark limit. 
Inspired by baryon chiral perturbation theory for QCD \cite{Jenkins:1990jv,Bernard:1995dp}, and for its lattice realization~\cite{Beane:2003xv},
we introduce the following ansatz and use it to carry out uncorrelated fits of our measurements of the chimera-baryon masses in terms of  polynomial functions of $\hat{m}_{\mathrm{PS}}$ and $\hat{m}_{\mathrm{ps}}$, as well as the lattice spacing, $\hat{a}$,
\beqs
\hat{m}_{\textrm{CB}} = \hat{m}_{\rm CB}^\chi &+& F_2 \hat{m}_{\rm PS}^2 + A_2 \hat{m}_{\rm ps}^2 + L_1 \hat{a}  \nn \\ 
&+& F_3 \hat{m}_{\rm PS}^3 + A_3 \hat{m}_{\rm ps}^3 + L_{2F} \hat{m}_{\rm PS}^2 \hat{a}+ L_{2A}  \hat{m}_{\rm ps}^2 \hat{a} \nn \\
&+& F_4 \hat{m}_{\rm PS}^4 + A_4  \hat{m}_{\rm ps}^4 + C_{4}  \hat{m}_{\rm PS}^2  \hat{m}_{\rm ps}^2\,,
\label{eq:fitting_func}
\eeqs
where ${\rm CB} = \Lambda_{\rm CB}$, $\Sigma_{\rm CB}$ or $\Sigma_{\rm CB}^\ast$, and all the hatted dimensionful quantities are expressed in units of the gradient flow scale, $w_0$. 
Here $\hat{m}_{\rm CB}^\chi$ represents the mass of the chimera baryon in the continuum and massless-hyperquark limit.

As anticipated,  the pseudoscalar meson mass squared plays the role of the hyperquark mass in each representation.
We call $F_j$ and $A_j$ the low energy constants (LECs)
associated with the corrections to $\hat{m}_{\rm CB}$ appearing at the $j$-th power in $\hat{m}_{\rm PS}$ and $\hat{m}_{\rm ps}$, respectively. 
As we are limited by the number of available lattice spacings and by the statistics, we retain terms up to the fourth power in the meson mass.
The coefficient $C_{4}$ controls the cross-term proportional to $\hat{m}^{2}_{\mathrm{PS}}\hat{m}^{2}_{\mathrm{ps}}$.
The $L_1$, $L_{2F}$, and $L_{2A}$  LECs are associated with the finite lattice spacing, $\hat{a}$,
for which we only consider the leading-order, linear in $\hat{a}$, as expected for Wilson-Dirac fermions. 
Note that the LECs, $\hat{m}^{\chi}_{\mathrm{CB}}$, $F_{j}$, $A_{j}$, $L_{j}$ and $C_{4}$, take different values for different chimera baryons.

 Chiral perturbation theory predicts the existence of terms logarithmic in the hyperquark masses, which we do not include in Eq.~(\ref{eq:fitting_func}).
 These additional terms have discernible effects only for light enough hyperquark masses (typically in the regime where the vector meson is more than twice heavier than the pseudoscalar meson), which is beyond the scope of this study.
 Furthermore, the quenched approximation results in diverging terms that in the limit where $\hat{m}_{\rm PS}$ or $\hat{m}_{\rm ps}$ approaches zero~\cite{Sharpe:1992ft, Bernard:1992mk}.  Therefore, we only investigate polynomial dependence of $\hat{m}_{\rm CB}$ on pseudoscalar meson masses in our analysis.  

Figure~\ref{fig:m_CB} shows clear evidence of a dependence of chimera-baryon masses on hyperquark masses and lattice spacing in our measurements. The result of a naive first attempt to fit our whole data set---tabulated in Appendix~\ref{app:data}---to Eq.~(\ref{eq:fitting_func}) is poor, as indicated by a large value of   $\chi^{2}/N_{\rm d.o.f.}$, and hence we do not report it here.
The truncated expansion in Eq.~(\ref{eq:fitting_func}) is expected to be valid only for light enough hyperquarks.
To test the possibility that a portion of our data points lie outside the range of validity of the expansion, we consider the effect of excluding data points collected at the largest available masses. To this purpose, we proceed systematically, according to

\begin{enumerate}
\item We start by  placing cuts, $\hat{m}_{\mathrm{PS, cut}} = \hat{m}_{\mathrm{ps, cut}} = 0.52$, and remove data points with $\hat{m}_{\rm PS} > \hat{m}_{\mathrm{PS, cut}}$ or $\hat{m}_{\rm ps} > \hat{m}_{\mathrm{ps, cut}}$, on all the five ensembles in Tab.~\ref{tab:ENS}.  The value, 0.52, is chosen such that there remain 13 data points in total. We verify by inspection that all these 13 measurements satisfy the condition $am_{\rm PS} < 1$ and $am_{\rm ps} < 1$. A fit to determine the 11 parameters in Eq.~(\ref{eq:fitting_func}) is then performed.  
\item After increasing $\hat{m}_{\mathrm{PS, cut}}$ and $\hat{m}_{\mathrm{ps, cut}}$, independently, in steps of  $0.05$, the above selection and fitting procedure is repeated. At each step, measurements for which $am_{\rm PS} > 1$ or $am_{\rm ps} > 1$ are also removed.
We stop when $\hat{m}_{\mathrm{PS, cut}}=1.07$ and $\hat{m}_{\mathrm{ps, cut}}=1.87$, at which point all the data points in Appendix~\ref{app:data} have been considered.
\end{enumerate}

\begin{table}
    \caption{List of the terms in Eq.~(\ref{eq:fitting_func}) that are included in the choices of fit ansatz used in our analysis. \label{tab:ansatz} }
\begin{center}
    \begin{tabular}{|L{1.8cm}|C{1.2cm}|C{1.2cm}|C{1.2cm}|C{1.2cm}|C{1.2cm}|C{1.2cm}|C{1.2cm}|C{1.2cm}|C{1.2cm}|C{1.2cm}|C{1.2cm}|}
    \hline
         Fit Ansatz & $\hat{m}_{\rm CB}^\chi$ & $\hat{m}_{\rm PS}^2$ & $\hat{m}_{\rm ps}^2$ & $\hat{m}_{\rm PS}^3$ & $\hat{m}_{\rm ps}^3$ & $\hat{m}_{\rm PS}^4$ & $\hat{m}_{\rm ps}^4$ & $\hat{m}_{\rm PS}^2 \hat{m}_{\rm ps}^2$ & $\hat{a}$ &  $\hat{m}_{\rm PS}^2 \hat{a}$ &  $\hat{m}_{\rm ps}^2 \hat{a}$ \\ \hline
         M2  & \checkmark & \checkmark & \checkmark & - & - & - & - & - & \checkmark & - & -\\ \hline
         M3  & \checkmark & \checkmark & \checkmark & \checkmark & \checkmark & - & - & - & \checkmark & \checkmark & \checkmark\\ \hline
         MF4 & \checkmark & \checkmark & \checkmark & \checkmark & \checkmark & \checkmark & - & - & \checkmark  & \checkmark & \checkmark \\ \hline
         MA4 & \checkmark & \checkmark & \checkmark & \checkmark & \checkmark & - & \checkmark & - & \checkmark  & \checkmark & \checkmark \\ \hline
         MC4 & \checkmark & \checkmark & \checkmark & \checkmark & \checkmark & - & - & \checkmark & \checkmark & \checkmark & \checkmark \\
         \hline
    \end{tabular} 
\end{center}
\end{table}

The above procedure results in $263$ distinct data sets, and $263$ fitting analyses,
each characterized by an unacceptably large $\chi^{2}/N_{\rm d.o.f.}$.
Furthermore, diffferent choices of initial values of the fit parameters lead to different results of the minimized $\chi^{2}/N_{\rm d.o.f.}$.
We interpret this result as evidence that the modeling of our data set encapsulated by Eq.~(\ref{eq:fitting_func}) is too general, so that the minimization of the $\chi^{2}/N_{\rm d.o.f.}$ with 11 fitting parameters is not well-converged.
Hence, some of the LECs cannot be determined by the available data.
In view of this, in this article, we do not report results obtained by fitting our data to Eq.~(\ref{eq:fitting_func}).  Instead,
 we explore a different numerical approach that allows for a variation of
  the set of free parameters included in the analysis, besides changing the number of incorporated measurements.

We summarize in Tab.~\ref{tab:ansatz} the five fit ansatze included in our analysis. They are all based upon
Eq.~(\ref{eq:fitting_func}), but are obtained by restricting the set of terms used in the fit, while setting the others to zero, to reduce the number of fitting parameters.
The first fit ansatz, dubbed M2, includes the polynomial terms in the first line of Eq.~(\ref{eq:fitting_func}), i.e.,  
$\hat{m}^{\chi}_{\mathrm{CB}}$ and corrections quadratic in pseudoscalar-meson masses or linear in lattice spacing.
In  M3, we also incorporate corrections up to the cubic terms in the pseudoscalar-meson masses, as well as the lattice-spacing corrections, $\hat{m}_{\textrm{PS}}^2 \hat{a}$ and $\hat{m}_{\textrm{ps}}^2 \hat{a}$. 
We further  include the three highest-order terms in Eq.~(\ref{eq:fitting_func}), one by one, in MF4, MA4, and MC4, corresponding to the addition of only $F_4 \hat{m}_{\textrm{PS}}^4$, $ A_4  \hat{m}_{\textrm{ps}}^4$, or $C_{4}  \hat{m}_{\textrm{PS}}^2  \hat{m}_{\textrm{ps}}^2$, respectively.

By combining the 5 fit ansatz with the $263$ data sets generated by imposing 
cuts on the data sets, we are left with $263 \times 5 = 1315$ different analysis procedures. 
Following the ideas in Ref.~\cite{Jay:2020jkz}, 
we select the best one by applying the Akaike information criterion (AIC).  
For each analysis procedure, one computes the quantity
\beq\label{eq:AIC}
{\rm AIC} \equiv \chi^2 + 2k + 2N_{\rm cut}\,,
\eeq
where $\chi^2$ is the standard chi-square, $k$ is the number of fit parameters, and $N_{\rm cut}$ is the number of data points removed by the introduction of the cuts $\hat{m}_{\mathrm{PS,cut}}$ and $\hat{m}_{\mathrm{ps,cut}}$.

\begin{figure}[t]
	\centering
	\includegraphics[width=1\textwidth]{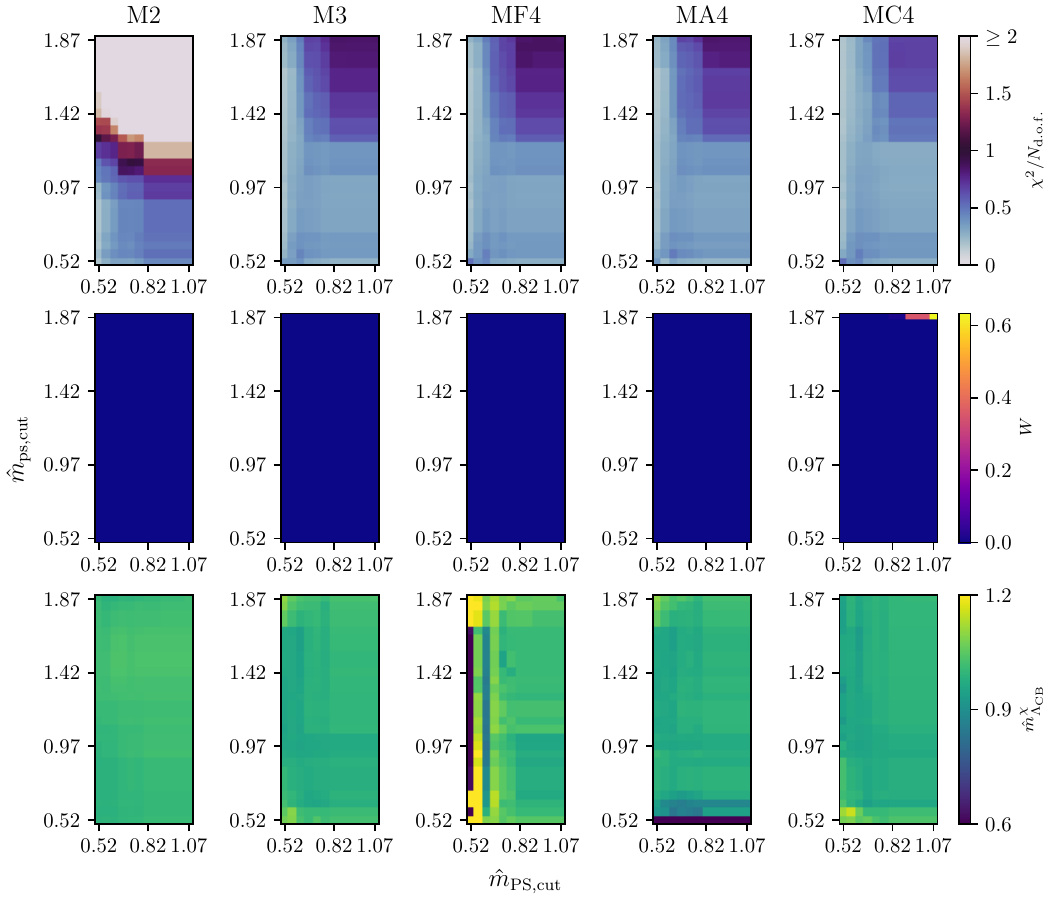}
	\caption{Heat-map plots of $\chi^2/N_{\rm d.o.f.}$, $W$ and $\hat{m}_{\rm CB}^\chi$ (top to bottom) for the analysis of $\hat{m}_{\Lambda_{\rm CB}}$ using different fit ansatze in Tab.~\ref{tab:ansatz}.  The horizontal and vertical axes are $\hat{m}_{\mathrm{PS, cut}}$ and $\hat{m}_{\mathrm{ps, cut}}$, respectively.}
	\label{fig:m_OC_AIC}
\end{figure}

\begin{table}[t]
    \centering
    \input{tabs/table_4.tex}
    \caption{The optimal choices of $\hat{m}_{\rm PS,cut}$ and $\hat{m}_{\rm ps,cut}$ for each fit ansatz in the continuum and massless-hyperquark extrapolation of $\hat{m}_{\Lambda_{\rm CB}}$.  Also shown are the corresponding value of $\chi^2 / N_{\rm d.o.f.}$, AIC and $W$.}\label{tab:cut_AIC_lam}
\end{table}

\begin{figure}[t]
	\centering
	\includegraphics[width=1\textwidth]{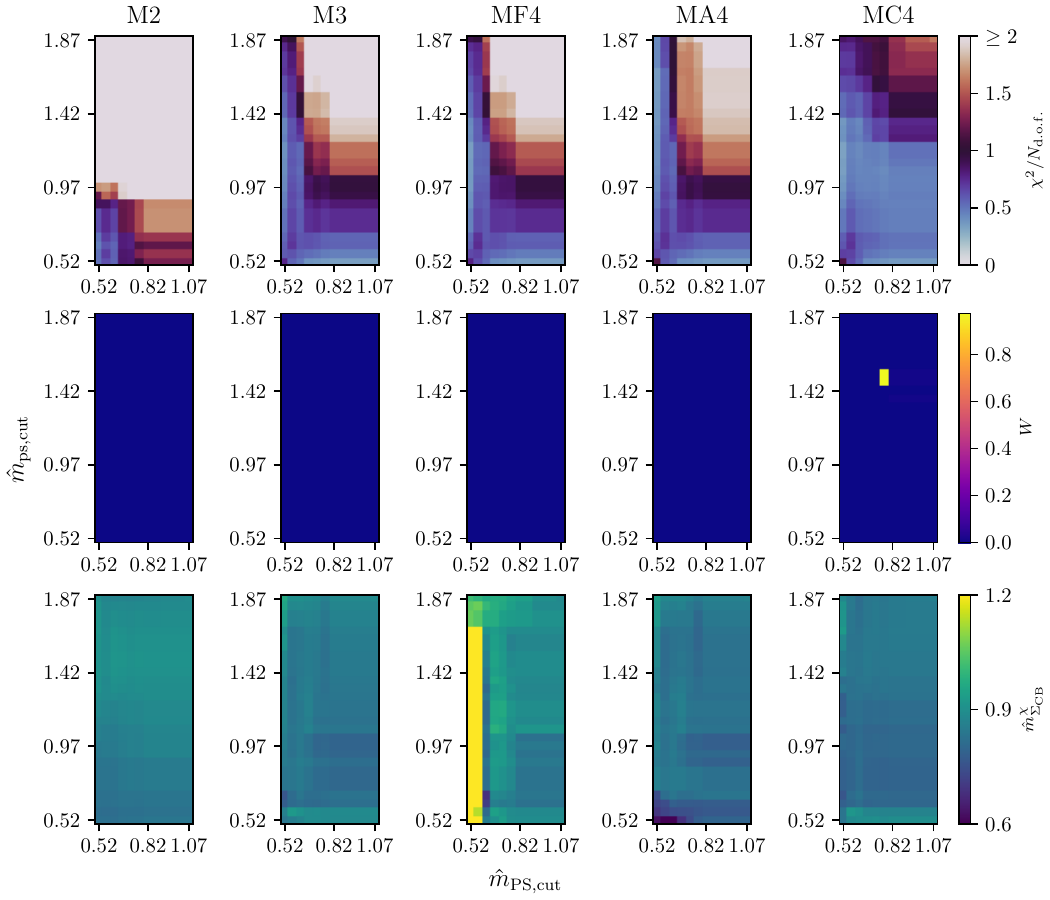}
	\caption{
	Heat-map plots of $\chi^2/N_{\rm d.o.f.}$, $W$ and $\hat{m}_{\rm CB}^\chi$ (top to bottom) for the analysis of $\hat{m}_{\Sigma_{\rm CB}}$ using different fit ansatze in Tab.~\ref{tab:ansatz}.  The horizontal and vertical axes are $\hat{m}_{\mathrm{PS, cut}}$ and $\hat{m}_{\mathrm{ps, cut}}$, respectively.}
	\label{fig:m_OV12_AIC}
\end{figure}

\begin{table}[h]
    \centering
    \input{tabs/table_5.tex}
    \caption{The optimal choices of $\hat{m}_{\rm PS,cut}$ and $\hat{m}_{\rm ps,cut}$ for each fit ansatz in the continuum and massless-hyperquark extrapolation of $\hat{m}_{\Sigma_{\rm CB}}$.  Also shown are the corresponding value of $\chi^2 / N_{\rm d.o.f.}$, AIC and $W$.}
    \label{tab:cut_AIC_sig}
\end{table}

\begin{figure}[t]
	\centering
	\includegraphics[width=1\textwidth]{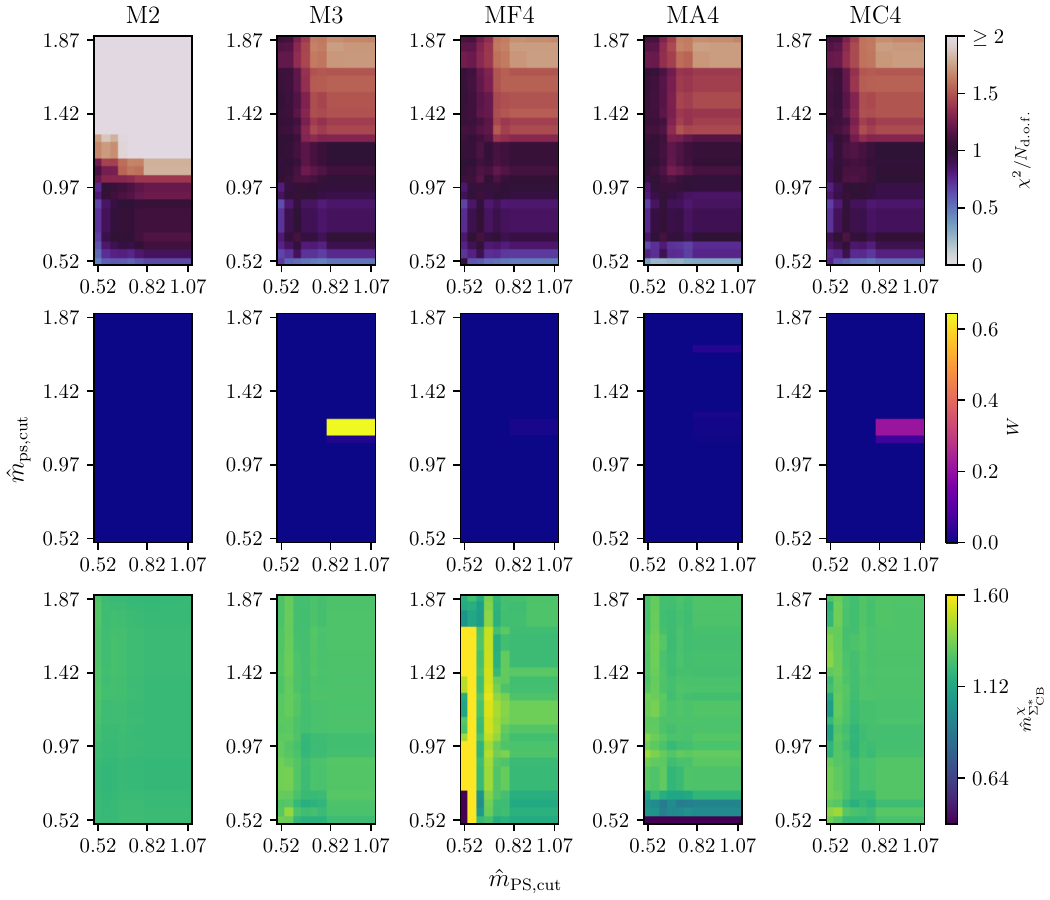}
	\caption{
	Heat-map plots of $\chi^2/N_{\rm d.o.f.}$, $W$ and $\hat{m}_{\rm CB}^\chi$ (top to bottom) for the analysis of $\hat{m}_{\Sigma^{\ast}_{\rm CB}}$ using different fit ansatze in Tab.~\ref{tab:ansatz}.  The horizontal and vertical axes are $\hat{m}_{\mathrm{PS, cut}}$ and $\hat{m}_{\mathrm{ps, cut}}$, respectively.
    }
	\label{fig:m_OV32_AIC}
\end{figure}
\begin{table}[h]
    \centering
    \input{tabs/table_6.tex}
    \caption{The optimal choices of $\hat{m}_{\rm PS,cut}$ and $\hat{m}_{\rm ps,cut}$ for each fit ansatz in the continuum and massless-hyperquark extrapolation of $\hat{m}_{\Sigma_{\rm CB}}$.  Also shown are the corresponding value of $\chi^2 / N_{\rm d.o.f.}$, AIC and $W$.}
    \label{tab:cut_AIC_sigs}
\end{table}

The corresponding probability weight is expected to be
\beq\label{eq:AIC_W}
W = \frac{1}{\mathcal{N}}\exp\left [ {-\frac{1}{2} {\rm AIC} } \right ]\,,
\eeq
where $\mathcal{N}$ is a normalization factor that ensures the sum of $W$ over all $1315$ analysis procedures equals to one. 
Maximizing the $W$ over ansatze and data sets is equivalent to minimizing the AIC.
We note that a smaller $\chi^2$ value can normally be obtained by considering more fit parameters, or by excluding data points that are not well described by the ansatz, e.g., points in the region of heavy hyperquark masses in our case. 
These correspond to the last two terms on the right-hand side of Eq.~(\ref{eq:AIC}).
They introduce a penalty by increasing the value of AIC, hence reducing $W$.
In Ref.~\cite{Jay:2020jkz},  the aim was to estimate a measured quantity by averaging over results from all analysis procedures with their probability weights. The $\chi^{2}$ therein was augmented to account for prior information. 
In this work, we focus on the standard $\chi^{2}$, with the aim of selecting the best analysis procedure.

Figures~\ref{fig:m_OC_AIC}, \ref{fig:m_OV12_AIC}, and \ref{fig:m_OV32_AIC} show, in  heat-map format, the $\hat{m}_{\mathrm{PS, cut}}$- and $\hat{m}_{\mathrm{ps, cut}}$-dependence of the $\chi^2/N_{\rm d.o.f.}$, the probability weight, $W$, in Eq.~(\ref{eq:AIC_W}), and the fitted $\hat{m}_{\rm CB}^\chi$, for measurements of the masses of $\Lambda_{\mathrm{CB}}$,  $\Sigma_{\mathrm{CB}}$, and $\Sigma^{\ast}_{\mathrm{CB}}$, respectively. In each row of a given figure, we display the results for the five distinct fitting strategies, M2, M3, MF4, MA4, and MC4, listed in Tab.~\ref{tab:ansatz}.
In all the plots, the horizontal and vertical axes correspond to $\hat{m}_{\mathrm{PS, cut}}$ and $\hat{m}_{\mathrm{ps, cut}}$, respectively.  
The center of each pixel in a heat map represents a set of cuts $(\hat{m}_{\mathrm{PS, cut}}, \hat{m}_{\mathrm{ps, cut}})$.
Notice that changing the values of $\hat{m}_{\mathrm{PS, cut}}$ and $\hat{m}_{\mathrm{ps, cut}}$ does not correspond to removing or including data points.  Therefore, in each heat map, the 336 pixels constituting the panel represent only $263$ distinct data sets.
This redundancy does not affect the normalization factor, ${\cal N}$, in Eq.~(\ref{eq:AIC_W}).

 In Tabs.~\ref{tab:cut_AIC_lam}, \ref{tab:cut_AIC_sig} and \ref{tab:cut_AIC_sigs}, we display the optimal choices of $\hat{m}_{\mathrm{PS, cut}}$ and $\hat{m}_{\mathrm{ps, cut}}$, as well as the corresponding value of $\chi^2/N_{\rm d.o.f}$, AIC and $W$, for all fitting methods in analyzing data of $\hat{m}_{\Lambda_{\rm CB}}$, $\hat{m}_{\Sigma_{\rm CB}}$, and $\hat{m}_{\Sigma^{\ast}_{\rm CB}}$, respectively.
 From these tables, as well as by inspection of Figs.~~\ref{fig:m_OC_AIC}, \ref{fig:m_OV12_AIC}, and \ref{fig:m_OV32_AIC}, we conclude that the best analysis procedure for the continuum and massless-hyperquark extrapolation of  $\hat{m}_{\Lambda_{\rm CB}}$ is the use of MC4 fit ansatz, with $\hat{m}_{\rm PS, cut} = 1.07$ and $\hat{m}_{\rm ps, cut} = 1.87$, while that of $\hat{m}_{\Sigma_{\rm CB}}$ is MC4 with $\hat{m}_{\rm PS, cut} = 0.77$ and $\hat{m}_{\rm ps, cut} = 1.47$.  Regarding $\hat{m}_{\Sigma^{\ast}_{\rm CB}}$, we find the optimal procedure to be the M3 ansatz with  $\hat{m}_{\rm PS, cut} = 0.82$ and $\hat{m}_{\rm ps, cut} = 1.17$.
It is noteworthy that the $\chi^2/N_{\rm d.o.f}$ values corresponding to the optimal fit, as determined through the minimization of the AIC, exhibit close proximity to unity in a posteriori examination.

Our results of the estimates of the LECs from the best analysis procedures are presented in Tab.~\ref{tab:best_fit}.
We notice for example that $F_{i}$ and $A_{i}$ are not compatible for at least $\Lambda_{\rm CB}$ and $\Sigma_{\rm CB}$ chimera baryons, confirming the necessity of using separate expansions in $\hat{m}_{\rm PS}$ 
and $\hat{m}_{\rm ps}$.  
Moreover, the $L_{1}$ coefficient for $\hat{m}_{\Sigma^{\ast}_{\rm CB}}$ is significantly larger than that for $\hat{m}_{\Lambda_{\rm CB}}$ and $\hat{m}_{\Sigma_{\rm CB}}$, indicating that lattice artifacts are expected to be more sizeable in this baryon mass, which is the heaviest of the three.
\begin{table}
\begin{center}
\input{tabs/table_7.tex}
\end{center}
   \caption{Low-energy constants in Eq.~(\ref{eq:fitting_func}) for each chimera baryon, as determined by the best analysis procedure selected from those presented in Tabs.~\ref{tab:cut_AIC_lam}, \ref{tab:cut_AIC_sig}, and \ref{tab:cut_AIC_sigs}. The missing coefficients are set to zero, as a result of the AIC-driven analysis.}
   \label{tab:best_fit}
\end{table}

To demonstrate the robustness of  the AIC-driven analysis, we perform cross checks by fitting the data 
obtained by fixing lattice spacing and mass of the pseudoscalar meson in one of the representations.
We first consider fixing the value of $\hat{m}_{\mathrm{ps}}$ and the lattice spacing.  In this case, the fit function, Eq.~(\ref{eq:fitting_func}), reduces to
\beq\label{eq:sense_AS}
\hat{m}_{\textrm{CB}} =
\tilde{m}_{\rm CB}^\chi + \tilde{F}_2 \hat{m}_{\textrm{PS}}^2 + \tilde{F}_3 \hat{m}_{\textrm{PS}}^3\,.
\eeq
We can then choose our data points at particular fixed $\hat{m}_{\mathrm{ps}}$, $\hat{a}$, and fit them to Eq.~(\ref{eq:sense_AS}) to determine $\tilde{m}^{\chi}_{\mathrm{CB}}$, $\tilde{F}_{2}$ and $\tilde{F}_{3}$.
Comparing with Eq.~(\ref{eq:fitting_func}), it is anticipated that these three parameters depend on the chosen values of $\hat{m}_{\mathrm{ps}}$ and $\hat{a}$.
Nevertheless, we expect that for small enough values of $\hat{m}_{\mathrm{ps}}$ and $\hat{a}$,  we should see that $\tilde{m}^{\chi}_{\mathrm{CB}}$ approaches $m^{\chi}_{\mathrm{CB}}$ determined from the global fit discussed above (MC4 for $\hat{m}_{\Lambda_{\rm CB}}$ and $\hat{m}_{\Sigma_{\rm CB}}$, and M3 for $\hat{m}_{\Sigma^{\ast}_{\rm CB}}$).  
Similar expectations apply to $\tilde{F}_{2}$ and $\tilde{F}_{3}$.
Also, the  $\hat{m}_{\mathrm{ps}}$-dependence in $\tilde{F}_{2}$ should primarily be accounted by the cross term, $C_{4}\hat{m}^{2}_{\mathrm{PS}}\hat{m}^{2}_{\mathrm{ps}}$.

Analogously, Eq.~(\ref{eq:fitting_func}), for fixed $\hat{m}_{\rm PS}$ and $\hat{a}$, reduces to
\beq\label{eq:sense_f}
\hat{m}_{\textrm{CB}} =
\tilde{m}_{\rm CB}^\chi + \tilde{A}_2 \hat{m}_{\textrm{ps}}^2 + \tilde{A}_3 \hat{m}_{\textrm{ps}}^3\,.
\eeq
Here $\tilde{m}^{\chi}_{\mathrm{CB}}$, $\tilde{A}_{2}$ and $\tilde{A}_{3}$ depend on the chosen values of $\hat{m}_{\mathrm{PS}}$ and $\hat{a}$.  They are expected to approach the appropriate LECs from the global fit when $\hat{m}_{\mathrm{PS}}$ and $\hat{a}$ are small.
These cross checks serve as examinations of our analyses using the fitting strategies listed in Tab.~\ref{tab:ansatz}.

\begin{figure}[t]
	\centering
	\includegraphics[width=1\textwidth]{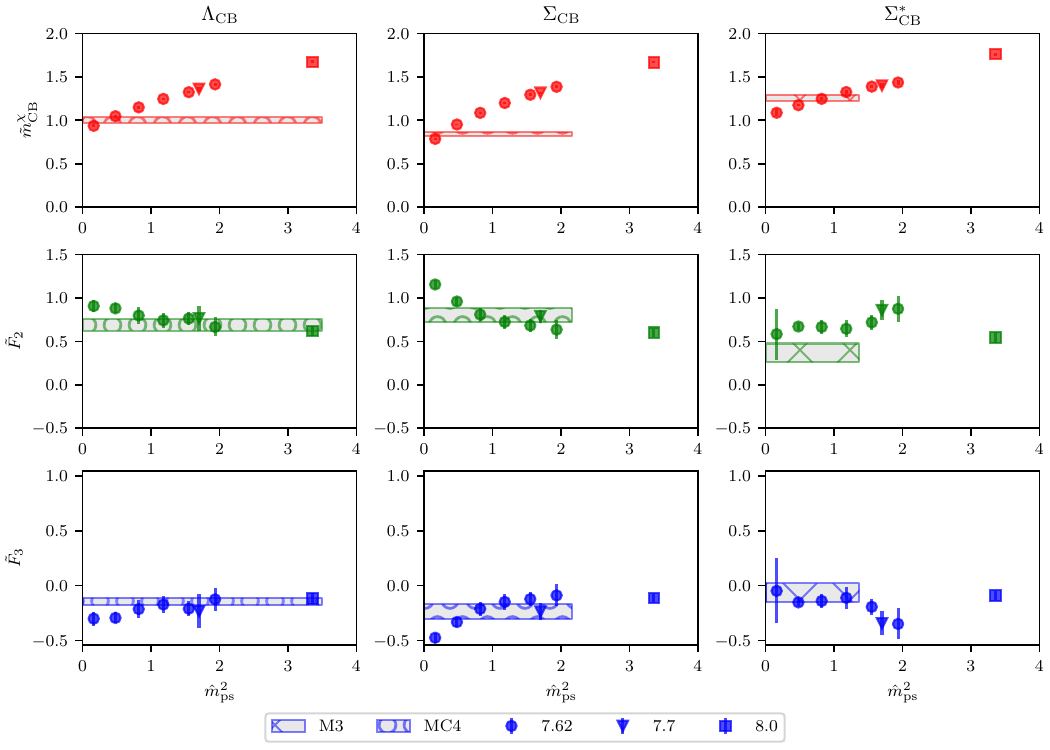}
	\caption{
Result of the cross checks based upon Eq.~(\ref{eq:sense_AS}).  The horizontal axis denotes the value of $\hat{m}_{\rm ps}$.  As indicated in the legend below the plots, different markers stand for measurements performed on different ensembles, and the bands represent results of $\hat{m}^{\chi}_{\rm CB}$, $F_{2}$ and $F_{3}$ from the best global-fit analysis procedures (see Tab.~\ref{tab:best_fit}).
 The height and the width of each band correspond to the size of the statistical error and the value of $\hat{m}_{\rm ps, cut}$, respectively.
 }
	\label{fig:fix_AS}
\end{figure}
\begin{figure}[t]
	\centering
	\includegraphics[width=1\textwidth]{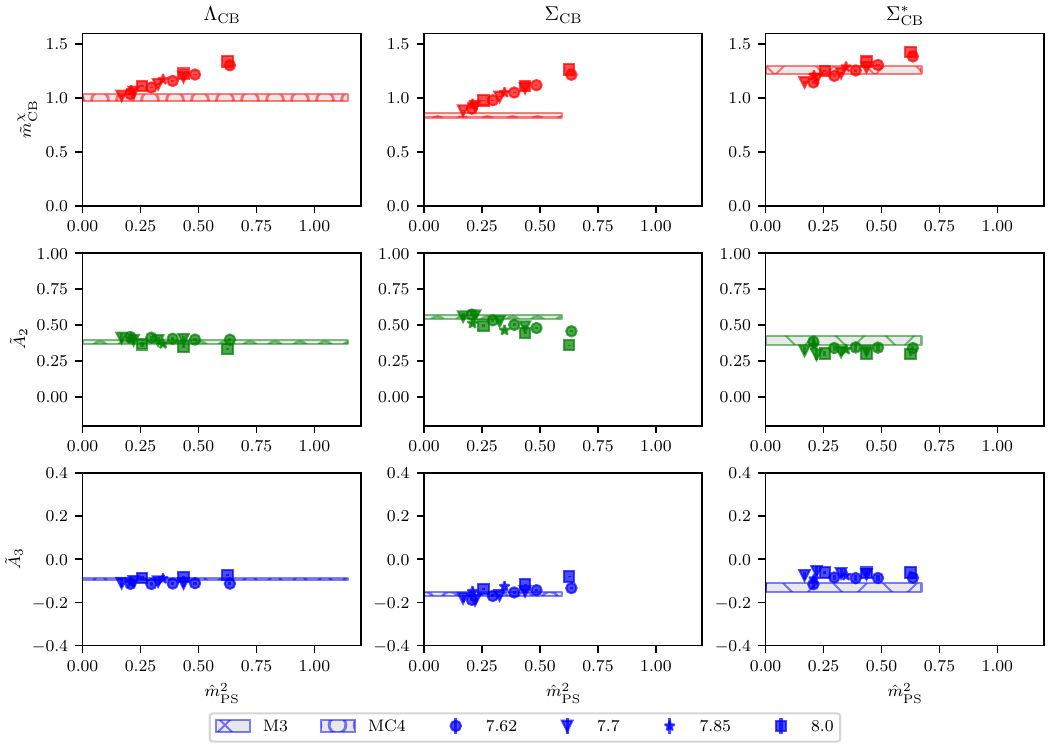}
	\caption{
     Result of the cross checks based upon Eq.~(\ref{eq:sense_f}).  The horizontal axis denotes the value of $\hat{m}_{\rm ps}$.  As indicated in the legend below the plots, different markers stand for measurements performed on different ensembles, and the bands represent results of $\hat{m}^{\chi}_{\rm CB}$, $A_{2}$ and $A_{3}$ from the best global-fit analysis procedures (see Tab.~\ref{tab:best_fit}).
     The height and the width of each band correspond to the size of the statistical error and the value of $\hat{m}_{\rm PS, cut}$, respectively.
     }
	\label{fig:fix_f}
\end{figure}

In Fig.~\ref{fig:fix_AS}, we present results of the fitted $\tilde{m}^{\chi}_{\mathrm{CB}}$, $\tilde{F}_{2}$ and $\tilde{F}_{3}$ in Eq.~(\ref{eq:sense_AS}) for three values of lattice spacing, corresponding to $\beta=7.62$, 7.7 and 8.0 listed in Tab.~\ref{tab:ENS}.  
As discussed above, these three parameters should depend upon $\hat{m}_{\mathrm{ps}}$ and the lattice spacing.
The plots in Fig.~\ref{fig:fix_AS} indicate that lattice artifacts are small.
Yet, notice that most results presented in this figure are from the coarsest lattice ($\beta=7.62$).
This is because the number of data points in the other two ensembles, when fixing $\hat{m}_{\mathrm{ps}}$, is small and in many cases does not allow us to carry out this exercise.  This is also the reason why we cannot perform the cross check on the other two ensembles ($\beta=7.85$ and $8.2$) listed in Tab.~\ref{tab:ENS}. 

The plots in Fig.~\ref{fig:fix_AS} demonstrate that $\tilde{m}^{\chi}_{\mathrm{CB}}$ and $\tilde{F}_{2}$ have non-negligible $\hat{m}_{\mathrm{ps}}$-dependence. 
The bands in each plot represent the global fit results, namely $m_{\rm CB}^\chi$, $F_2$, and $F_3$, respectively, obtained from the best analysis procedures in Tab.~\ref{tab:best_fit}.  The height and width of each band correspond to the size of the statistical error and the value of $\hat{m}_{\rm ps, cut}$, respectively. 
It can be seen that $\tilde{m}^{\chi}_{\mathrm{CB}}$ is compatible with the value of $m^{\chi}_{\mathrm{CB}}$ obtained from the best global-fit analysis procedure for $\hat{m}_{\Lambda_{\rm CB}}$ and $\hat{m}_{\Sigma_{\rm CB}}$ in the small-$\hat{m}_{\rm ps}$ regime.
This is not the case for  $\hat{m}_{\Sigma^{\ast}_{\rm CB}}$, for which we find a larger value of  $L_{1}$ 
 (see Tab.~\ref{tab:best_fit}), indicating the presence of more significant lattice artifacts.
We further observe that results of $\tilde{F}_{2}$ for $\hat{m}_{\Lambda_{\rm CB}}$ and $\hat{m}_{\Sigma_{\rm CB}}$ show non-negligible dependence upon the chosen value of $\hat{m}_{\rm ps}$, indicating the need to include the cross term, $C_{4} \hat{m}^{2}_{\rm PS}\hat{m}^{2}_{\rm ps}$, in the global-fit analysis.

 Analogously, we conduct a similar cross check by fixing the value of $\hat{m}_{\rm PS}$, as described by Eq.~(\ref{eq:sense_f}).  Results of this process are displayed in Fig.~\ref{fig:fix_f}, which show similar features to those we discussed in commenting on 
 Fig.~\ref{fig:fix_AS}. 
Since this work is the first exploratory study of the spectrum of the chimera baryons in the $Sp(4)$ gauge theory, we do not attempt to estimate systematic errors affecting our results.
It has to be emphasized that the current calculation is performed in the quenched approximation, and we are interested in the qualitative feature of the spectrum at this stage.  More precise, dynamical computations are deferred to future work.  

\begin{figure}
	\centering
	\includegraphics[width=1\textwidth]{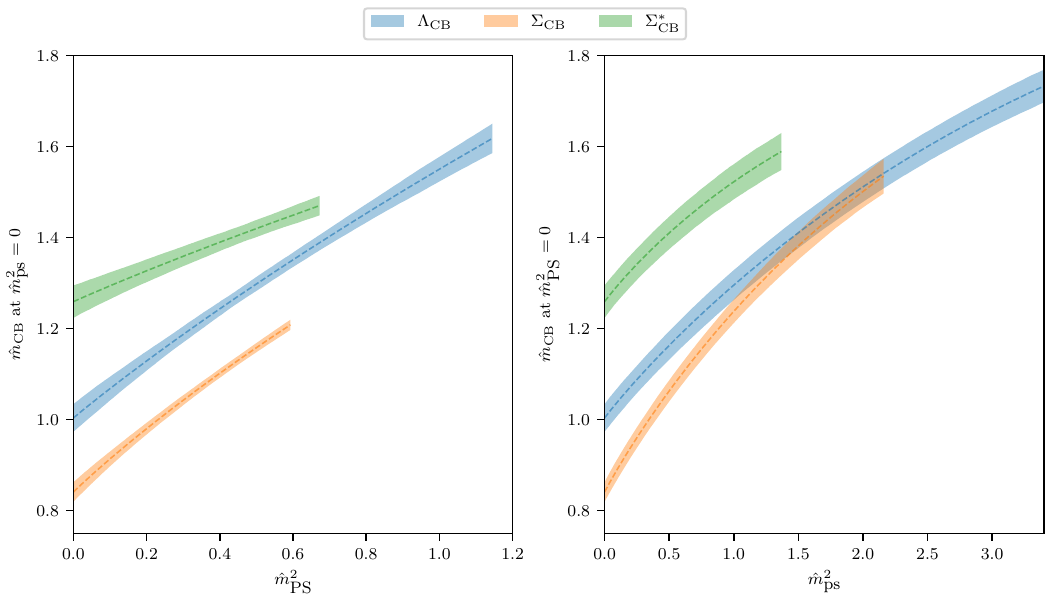}
	\caption{
	Dependence on $\hat{m}^{2}_{\rm PS}$ (left) and  $\hat{m}^{2}_{\rm ps}$ (right) of the mass of three chimera baryons, 
	${\Lambda_{\rm CB}}$, ${\Sigma_{\rm CB}} $, and ${\Sigma^{\ast}_{\rm CB}}$,     in the limit where the lattice spacing vanishes, 
	while $\hat{m}^{2}_{\rm ps} = 0$ (left) and  $\hat{m}^{2}_{\rm PS}=0$ (right).  These plots are generated using the best-fit LECs in Tab.~\ref{tab:best_fit}, with the bands representing the statistical errors.
	These bands straddle in the horizontal direction between zero and the optimal choices of $\hat{m}^{2}_{\rm PS, cut}$ (left) and $\hat{m}^{2}_{\rm ps, cut}$ (right).}
	\label{fig:m_massless}
\end{figure}
\begin{figure}
	\centering
	\includegraphics[width=1\textwidth]{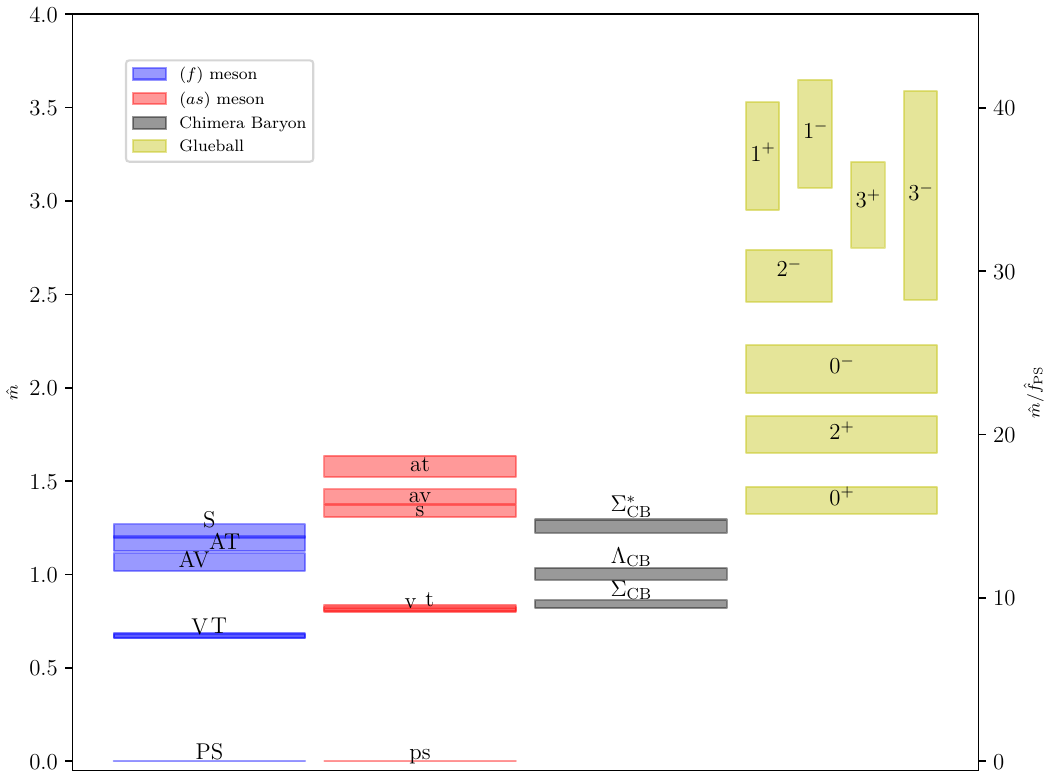}
	\caption{
	Quenched spectrum of the $Sp(4)$ gauge theory in the continuum and massless-hyperquark limit. The glueball states are labelled using the $J^{P}$ notation, while PS (ps), V (v), T(t), AV (av), AT (at) and S (s) denote the pseudoscalar, vector, tensor, axial-vector, axial-tensor and scalar mesons composed of fundamental (antisymmetric)  hyperquarks.
 The results of mesons and glueballs are taken from our previous works in Ref.~\cite{Bennett:2019cxd, Bennett:2020qtj}.
 The results for the chimera baryons are original to this work.
    }
	\label{fig:quench_spec}
\end{figure}

Using the results summarized  in Tab.~\ref{tab:best_fit}, for  the  LECs denoted as $\hat{m}^{\chi}_{\rm CB}$, $F_{2,3}$, and $A_{2,3}$, we present the dependence on  $\hat{m}^{2}_{\rm PS}$ and $\hat{m}^{2}_{\rm ps}$ of the chimera-baryon masses in the continuum limit, in Fig.~\ref{fig:m_massless}.  That is, the plots in this figure are generated using Eq.~(\ref{eq:fitting_func}) with $\hat{a}=0$.  The left (right) panel of this figure shows the evolution of $\hat{m}_{\Lambda_{\rm CB}}$, $\hat{m}_{\Sigma_{\rm CB}}$, and $\hat{m}_{\Sigma_{\rm CB}}$ as a function of  $\hat{m}_{\rm PS}$ ($\hat{m}_{\rm ps}$) in the limit where $\hat{m}_{\rm ps} = 0$ ($\hat{m}_{\rm PS} = 0$).
The color bands represent the statistical errors, and they straddle in the horizontal direction from 0 to the values $\hat{m}_{\rm PS} = \hat{m}_{\rm PS, cut}$ (left) and $\hat{m}_{\rm ps} = \hat{m}_{\rm ps, cut}$ (right).  The  mass hierarchy,
\beq
\label{eq:m_cb_hierarchy}
 \hat{m}_{\Sigma_{\rm CB}} \lesssim  \hat{m}_{\Lambda_{\rm CB}} < \hat{m}_{\Sigma^{\ast}_{\rm CB}} \, ,
\eeq
emerges in the whole range of hyperquark masses investigated in this work.
The masses $\hat{m}_{\Lambda_{\rm CB}}$ and $\hat{m}_{\Sigma_{\rm CB}}$ become compatible with one another only in the regime of heavier $(as)$ hyperquarks.  The hierarchy in Eq.~(\ref{eq:m_cb_hierarchy}) can have non-trivial implications in constructing viable models for top partial compositeness~\cite{Banerjee:2022izw}.

It is interesting to compare the masses of the chimera baryons with those of other states in the theory, as we do in Fig.~\ref{fig:quench_spec}. Meson and glueball masses are taken from our previous measurements in the quenched approximation~\cite{Bennett:2019cxd, Bennett:2020qtj} (see also Refs.~\cite{Bennett:2020hqd,Bennett:2022gdz} for related studies).  In this figure, mesons denoted by capital letters are those composed of  $(f)$ hyperquarks, while those expressed by lowercase letters contain $(as)$ hyperquarks only.  All the masses presented in the plot have been extrapolated to the continuum and massless-hyperquark limit, and are shown in both gradient-flow units (vertical axis on the left-hand side), as well as in units of the fundamental pseudoscalar meson decay constant~\cite{Bennett:2019cxd} (vertical axis on the right-hand side). The height of the bands represents statistical errors.
As shown in the figure, we find that the masses of the top-partner candidates, $\Lambda_{\rm CB}$ and $\Sigma_{\rm CB}$, are comparable to those of the $(as)$ vector mesons.

The spectrum of CHMs with top partial compositeness has also been studied using other methods, such as Schwinger-Dyson equations, Nambu-Jona-Lasinio models, or in the framework of holography~\cite{Elander:2021bmt,Elander:2020nyd,Erdmenger:2023hkl,Erdmenger:2020flu,Erdmenger:2020lvq,Abt:2019tas}.
To facilitate comparison with these as well as other future studies, and in view of possible phenomenological applications,  we express our final results for the massless-hyperquark and continuum extrapolations for the chimera baryon states  in units of the mass, $m_{\rm v}$, of the lightest vector meson with $(as)$-type constituents. We find
\beq
 m_{\Lambda_{\rm CB}}/m_{\rm v}=1.234(32)\,,~
 m_{\Sigma_{\rm CB}}/m_{\rm v}=1.016(25)\,,
 ~\rm{and}~ 
 m_{\Sigma^*_{\rm CB}}/m_{\rm v}=1.576(47)\,,
\eeq
where the quoted error is statistical errors without including systematic effects, for example due to the quenched approximation.

\section{Summary and Outlook}
\label{Sec:summary}

The strongly interacting $Sp(4)$ gauge theory coupled to $N_{f}=2$ fundamental, $(f)$,  and $n_{f}=3$ two-index antisymmetric, $(as)$, Dirac fermions (hyperquarks) is the minimal model amenable to lattice investigations that can provide a UV completion of CHMs with top partial compositeness~\cite{Barnard:2013zea}.
Chimera baryons are composite states formed by two $(f)$ and one $(as)$ hyperquarks, and are sourced by the operators ${\mathcal{O}}^{5}$  in Eq.~(\ref{eq:chim_bar_src}) and ${\mathcal{O}}^{\mu}$ in Eq.~(\ref{eq:chim_bar_src_mu}).  The lightest state sourced by ${\mathcal{O}}^{5}$ is the spin-1/2 chimera baryon, $\Lambda_{\mathrm{CB}}$, while ${\mathcal{O}}^{\mu}$ can source spin-1/2 and -3/2 baryons, and we denote by $\Sigma_{\mathrm{CB}}$ and $\Sigma_{\mathrm{CB}}^{\ast}$, respectively, the two lightest states with definite spin.  Either  $\Lambda_{\mathrm{CB}}$ or $\Sigma_{\mathrm{CB}}$ are candidate top partners~\cite{Banerjee:2022izw}.

Because this is the first systematic lattice calculation of the chimera baryon spectrum in the $Sp(4)$ gauge theory, we perform  it in the quenched approximation, in which the hyperquark determinant in the path integral is set to a constant---see Ref.~\cite{Bennett:2022yfa} for pioneering work on the theory with $N_f=2$ $(f)$-type and $n_f=3$ $(as)$-type dynamical fermions.
Working in the quenched approximation not only makes the numerical computation significantly less demanding, but it also
allows us to scan a large region of parameter space and gather useful information for future dynamical calculations.
The interpolating operators, ${\mathcal{O}}^{5}$ and ${\mathcal{O}}^{\mu}$, source states with both  even and odd parity.
As discussed in Section~\ref{Sec:projection}, having established  that states with even parity are lighter, we implement projections to the parity-even states in our analysis.
Furthermore, spin projectors are introduced to distinguish between $\Sigma_{\mathrm{CB}}$ and $\Sigma_{\mathrm{CB}}^{\ast}$ states, that are sourced by the same operator, ${\mathcal{O}}^{\mu}$.

The main focus  of this  study is the hyperquark-mass dependence of the chimera baryon masses. 
As we use the Wilson-Dirac formulation for hyperquark fields, we find it convenient to express this dependence in terms of the
mass of the pseudoscalar mesons, which we denote as $\hat{m}_{\rm PS}$ and $\hat{m}_{\rm ps}$, respectively, for mesons
built of $(f)$-type and $(as)$-type hyperquarks.
As is expected, the three chimera baryon masses approach one another when increasing $\hat{m}_{\rm PS}$ and $\hat{m}_{\rm ps}$.
Working under the assumption  that the hyperquark masses are sufficiently light to make it viable, we use an effective description
 inspired by baryon chiral EFT~\cite{Jenkins:1990jv, Jenkins:1991ne}.
We include only polynomial terms in the continuum and massless-hyperquark extrapolations.
In the range of hyperquark masses probed in this work, we find the mass hierarchy $m^{\ast}_{\Sigma_{\mathrm{CB}}} > m_{\Lambda_{\mathrm{CB}}} \gtrsim m_{\Sigma_{\mathrm{CB}}}$. 
Our measurements show that the ratio $m_{\Lambda_{\mathrm{CB}}}/m_{\Sigma_{\mathrm{CB}}}$ decreases when $m_{\rm ps}$ increases, and that,
for the heaviest available values of $m_{\rm ps}$, this ratio is compatible with unity.
These findings suggest that the hierarchy may not hold true in other regimes of hyperquark masses, which warrants further, more extensive future investigations.

The use of EFT-inspired relations also enables the extrapolation to the continuum limit, by including effects of lattice-artifact that break the global symmetries of the system explicitly~\cite{Beane:2003xv}.
As explained in detail in Section~\ref{Sec:mexpt}, we implement various ansatze for this simultaneous continuum and massless-hyperquark extrapolation.
Leveraging the AIC~\cite{Akaike:1998zah,Jay:2020jkz} to assess the fit quality and performing cross checks by fixing the pseudoscalar-meson masses in each of the representations, we find the optimal choice of the fit procedure for the chimera baryons in our analysis.
The evolution of the chimera baryon masses as a function of the pseudoscalar masses in the continuum limit are displayed in Fig.~\ref{fig:m_massless}.
Furthermore, in Fig.~\ref{fig:quench_spec} we display the complete ground state spectrum of the theory in the limit of vanishing hyperquark masses, displaying together with chimera baryon results also meson and glueball masses
taken from our earlier studies~\cite{Bennett:2019cxd,Bennett:2020qtj}. 

This investigation of chimera baryon mass spectra sets the stage for lattice simulations with  dynamical matter fields. Understanding the intricate mass relations between chimera baryons and hyperquarks is pivotal for navigating the multi-faced lattice parameter space and constructing physically meaningful models. 
However, due to the quenching effects, especially the lack of fermion dynamics, it does not describe the validity of the fully dynamical two-representation $Sp(4)$ model as a viable composite Higgs model. This result should be taken cautiously when applied to particle phenomenology. In particular, to study the anomalous dimension, the fermion dynamics are essential to make the theory (near-)conformal. Nevertheless, it is encouraging that the lightest chimera baryon is as light as the vector meson in the antisymmetric representation.
The commitment to precise and accurate physics is central to our quest for a deeper understanding of composite Higgs and top partial compositeness models and their potential implications for particle physics.
We here established a robust analysis framework, that will be crucial in future high-precision studies, particularly when confronting computationally demanding calculations that require control over the numerous lattice inputs.

\begin{acknowledgments}

The work of EB has been supported by the UKRI Science and Technology Facilities Council (STFC)
Research Software Engineering Fellowship EP/V052489/1, and by the ExaTEPP project EP/X017168/1.

The work of DKH was supported by Basic Science Research Program through the National Research Foundation of Korea (NRF) funded by the Ministry of Education (NRF-2017R1D1A1B06033701).

The work of JWL was supported in part by the National Research Foundation of Korea (NRF) grant funded 
by the Korea government(MSIT) (NRF-2018R1C1B3001379) and by IBS under the project code, IBS-R018-D1. 

The work of DKH and JWL was further supported by the National Research Foundation of Korea (NRF) grant funded by the Korea government (MSIT) (2021R1A4A5031460).

The work of HH and CJDL is supported by the Taiwanese MoST grant 109-2112-M-009-006-MY3 and NSTC grant 112-2112-M-A49-021-MY3. 

The work of BL and MP has been supported in part by the STFC 
Consolidated Grants  No. ST/P00055X/1, ST/T000813/1, and ST/X000648/1.
 BL and MP received funding from
the European Research Council (ERC) under the European
Union’s Horizon 2020 research and innovation program
under Grant Agreement No.~813942. 
The work of BL is further supported in part 
by the Royal Society Wolfson Research Merit Award 
WM170010 and by the Leverhulme Trust Research Fellowship No. RF-2020-4619.

The work of DV is supported by STFC under Consolidated
Grant No.~ST/X000680/1.

Numerical simulations have been performed on the 
Swansea University SUNBIRD cluster (part of the Supercomputing Wales project) and AccelerateAI A100 GPU system,
on the local HPC
clusters in Pusan National University (PNU) and in National Yang Ming Chiao Tung University (NYCU),
and on the DiRAC Data Intensive service at Leicester.
The Swansea University SUNBIRD system and AccelerateAI are part funded by the European Regional Development Fund (ERDF) via Welsh Government.
The DiRAC Data Intensive service at Leicester is operated by 
the University of Leicester IT Services, which forms part of 
the STFC DiRAC HPC Facility (www.dirac.ac.uk). The DiRAC 
Data Intensive service equipment at Leicester was funded 
by BEIS capital funding via STFC capital grants ST/K000373/1 
and ST/R002363/1 and STFC DiRAC Operations grant ST/R001014/1. 
DiRAC is part of the National e-Infrastructure.

{\bf Open Access Statement}---For the purpose of open access, the authors have applied a Creative Commons 
Attribution (CC BY) license  to any Author Accepted Manuscript version arising.

{\bf Research Data Access Statement}---The data generated for this manuscript can be downloaded in machine-readable format from Ref.~\cite{DATA}.
\end{acknowledgments}

\begin{table}
\begin{center}
\input{tabs/table_8.tex}
\end{center}
   \caption{
   Numerical values  of the  bare masses, $am_0^{(f)}$ and $am_0^{(as)}$, used in the measurements on the ensemble QB1, that has gradient flow scale $w_0/a=1.448(3)$.
   For each set of bare masses, we present the pseudoscalar meson  mass, $am_{\rm PS}$ and $am_{\rm ps}$, the mass ratio between pseudoscalar and vector mesons  in both representations, $m_{\rm PS}/m_{\rm V}$ and $m_{\rm ps}/m_{\rm v}$, and the chimera-baryon masses $am_{\Lambda_{\rm CB}}$, $am_{\Sigma_{\rm CB}}$ and $am_{\Sigma^\ast_{\rm CB}}$.
   }
   \label{tab:QB1_bare_mass}
\end{table}

\begin{table}
\begin{center}
\input{tabs/table_9.tex}
\end{center}
   \caption{
   Numerical values  of the  bare masses, $am_0^{(f)}$ and $am_0^{(as)}$, used in the measurements on the ensemble QB2, that has gradient flow scale $w_0/a=1.6070(19)$.
   For each set of bare masses, we present the pseudoscalar meson  mass, $am_{\rm PS}$ and $am_{\rm ps}$, the mass ratio between pseudoscalar and vector mesons  in both representations, $m_{\rm PS}/m_{\rm V}$ and $m_{\rm ps}/m_{\rm v}$, and the chimera-baryon masses $am_{\Lambda_{\rm CB}}$, $am_{\Sigma_{\rm CB}}$ and $am_{\Sigma^\ast_{\rm CB}}$.
   }
   \label{tab:QB2_bare_mass}
\end{table}

\begin{table}
\begin{center}
\input{tabs/table_10.tex}
\end{center}
   \caption{
   Numerical values  of the  bare masses, $am_0^{(f)}$ and $am_0^{(as)}$, used in the measurements on the ensemble QB3, that has gradient flow scale $w_0/a=1.944(3)$.
   For each set of bare masses, we present the pseudoscalar meson mass, $am_{\rm PS}$ and $am_{\rm ps}$, the mass ratio between pseudoscalar and vector mesons  in both representations, $m_{\rm PS}/m_{\rm V}$ and $m_{\rm ps}/m_{\rm v}$, and the chimera-baryon masses $am_{\Lambda_{\rm CB}}$, $am_{\Sigma_{\rm CB}}$ and $am_{\Sigma^\ast_{\rm CB}}$.
   }
   \label{tab:QB3_bare_mass}
\end{table}

\begin{table}
\begin{center}
\input{tabs/table_11.tex}
\end{center}
   \caption{
   Numerical values  of the  bare masses, $am_0^{(f)}$ and $am_0^{(as)}$, used in the measurements on the ensemble QB4, that has gradient flow scale $w_0/a=2.3149(12)$.
   For each set of bare masses, we present the pseudoscalar meson  mass, $am_{\rm PS}$ and $am_{\rm ps}$, the mass ratio between pseudoscalar and vector mesons  in both representations, $m_{\rm PS}/m_{\rm V}$ and $m_{\rm ps}/m_{\rm v}$, and the chimera-baryon masses $am_{\Lambda_{\rm CB}}$, $am_{\Sigma_{\rm CB}}$ and $am_{\Sigma^\ast_{\rm CB}}$.
   }
   \label{tab:QB4_bare_mass}
\end{table}

\begin{table}
\begin{center}
\input{tabs/table_12.tex}
\end{center}
   \caption{
   Numerical values  of the  bare masses, $am_0^{(f)}$ and $am_0^{(as)}$, used in the measurements on the ensemble QB5, that has gradient flow scale $w_0/a=2.8812(21)$.
   For each set of bare masses, we present the pseudoscalar meson  mass, $am_{\rm PS}$ and $am_{\rm ps}$, the mass ratio between pseudoscalar and vector mesons  in both representations, $m_{\rm PS}/m_{\rm V}$ and $m_{\rm ps}/m_{\rm v}$, and the chimera-baryon masses $am_{\Lambda_{\rm CB}}$, $am_{\Sigma_{\rm CB}}$ and $am_{\Sigma^\ast_{\rm CB}}$.
   }
   \label{tab:QB5_bare_mass}
\end{table}

\appendix

\section{Extracted masses of mesons and chimera baryons}\label{app:data}

In this appendix, we provide a comprehensive summary of the choices of bare hyperquark masses we used in the measurements we made, along with the resulting masses of composite states, all expressed in lattice units.
Our dataset includes the masses of pseudoscalar mesons composed of $(f)$ and $(as)$ hyperquarks, which we denote as $m_{\rm PS}$ and $m_{\rm ps}$, respectively, as well as the masses of the chimera baryons $\Lambda_{\rm CB}$, $\Sigma_{\rm CB}$, and $\Sigma^\ast_{\rm CB}$.
We also report the mass ratios between pseudoscalar and vector mesons, which serves as an indicator of the amount of explicit
breaking of the global symmetry of the theory.
We have organized the data into tables based on the ensemble in which the measurements were performed, 
individual ensembles differing both by the lattice coupling $\beta$ and the
volume (see Tab.~\ref{tab:ENS}).
Specifically, Tab.~\ref{tab:QB1_bare_mass}--\ref{tab:QB5_bare_mass} correspond to measurements on ensembles QB1--QB5, respectively.
All numbers presented here are available in machine-readable format in the data release associated with this work~\cite{DATA}.
The software workflow used to analyze the data and prepare the plots and tables are made available in Ref.~\cite{CODE}.
Further technical details, such as smearing parameters, fitting ranges applied to the mass extraction, and values of the resulting $\chi^2/N_{\rm d.o.f.}$, are presented in Appendix~\ref{app:fitting}.

\section{Smearing techniques}\label{app:smearing}

Wuppertal smearing~\cite{Gusken:1989qx} and APE smearing~\cite{APE:1987ehd} are well-developed lattice techniques, which are normally applied simultaneously, to improve the overlap of a ground state.
The former amounts to a modification of the operators used to define the 2-point functions, in particular on
the position of the hyperquarks constituting mesons and chimera baryons. 
The latter consists of a smoothening of the gauge configurations, that removes short-distance fluctuations.

In calculations involving point sources, a hyperquark propagator, $S_R$, involving fermions transforming in the $R$ representation of the gauge group, is obtained by solving the Dirac equation
\beq
D^{R}_{a\alpha,b\beta}(x,y) S^{\,b\beta}_{R\,\,\,\,c\gamma} (y) = \delta_{x,0}\delta_{\alpha\gamma}\delta_{ac}\,,
\eeq
where $D^{R}_{a\alpha,b\beta}$ is the Wilson-Dirac operator in Eq.~(\ref{eq:WDO}).
Smearing of the source is obtained by replacing the spatial delta function, $\delta_{x,0}$, by a new 
source, $q^{(n)}(x)$, constructed with the Wuppertal smearing function, which is defined recursively by
the relation
\beq\label{eq:Wuppertal_smearing}
q^{(n+1)}(x) = \frac{1}{1+6\varepsilon} \left [ q^{(n)}(x) + \varepsilon \sum_{\mu= \pm1}^{\pm 3} U_\mu(x)q^{(n)}(x+\hat{\mu}) \right]\,,
\eeq
with $n=0\,,\cdots\,, N_{\text{W}}$.
The initial condition is $q^{(0)} (x) = \delta_{x,0}$ refers to a point source.
The tunable parameters, $\varepsilon$ and $N_{\text{W}}$, are referred to as step size and number of iterations, respectively.
The smearing of the sink  is obtained by replacing $q^{(0)}$ with a hyperquark propagator $S_R$.

We supplement  Wuppertal smearing by smoothening gauge links with APE smearing.
The smearing function is 
\beq\label{eq:APE_smearing}
    U^{(n+1)}_\mu(x) = P \left\{ (1-\alpha)U^{(n)}_\mu(x) + \frac{\alpha}{6} S^{(n)}_\mu(x) \right \} \,,
\eeq
where $S_{\mu}$ denotes the staple operator, defined as
\beq
    S_\mu(x) = \sum_{\pm \nu \neq \mu} U_\nu(x)U_\mu(x+\hat{\nu})U^\dagger_\nu(x+\hat{\mu}).
\eeq
The  iteration number, $n=0\,,\cdots\,,N_{\text{APE}}$, and step size, $\alpha$, are tunable parameters.
Because of the summation over neighboring gauge links in Eq.~(\ref{eq:APE_smearing}), the smeared gauge links should be projected back to the group.
A project $P$ is provided by the re-symplectisation algorithm in the HB calculations, which inherits the numerical stability of the Gram-Schmidt process.
It takes advantage of the symplectic structure of $Sp(2N)$: once the first $N$ columns, $col_{j}$, with $j=1,\,\cdots,\,N$, of the $Sp(2N)$ matrix are given, the remaining ones are given by 
\beq
col_{j+N} = -\Omega \,col^\ast_j\,
\eeq
Having normalized the first column, the $(N+1)$-th one is obtained algebraically.
The second column is computed by orthonormalizing the first and the $(N+1)$-th.
By repeating the process for every column, one arrives at a complete $Sp(2N)$ matrix.

In previous work~\cite{Bennett:2017kga,Bennett:2019jzz,Bennett:2019cxd}, we used stochastic wall sources~\cite{Boyle:2008rh} for the meson calculations.
To improve the signal of chimera baryons, in this study, we apply Wuppertal and APE smearing simultaneously to obtain $(f)$ and $(as)$ hyperquark propagators.
Wuppertal smearing step sizes are chosen differently for $(f)$ and $(as)$ hyperquark propagators, and we denote them as $\epsilon^{(f)}$ and $\epsilon^{(as)}$, respectively.
We fix the number of iterations at the source, and select individually the number of iterations at the sink that display the optimal plateau for each meson and each chimera baryon.
Our choices of the Wuppertal smearing parameters are presented in Appendix~\ref{app:fitting}.
The APE smearing parameters are fixed in all the calculations to be $\alpha=0.4$ and $N_{\rm APE}=50$.

The same techniques are also applied to our study on the spectrum of $Sp(4)$ gauge theory with $n_f = 3$ antisymmetric fermions~\cite{Lee:2022m}, particularly for the calculation of the first excited state of the vector and tensor mesons.
Additionally, these techniques have been utilized as a cross-verification in the excited state subtraction method applied to the computation of a singlet meson---see Appendix B of Ref.~\cite{Bennett:2023rsl}.

\section{Details about fitting procedures}\label{app:fitting}

In this appendix, we tabulate  numerical information relevant to the mass extractions for mesons and chimera baryons. 
In Tab.~\ref{tab:Meson_f_fit_pars}, we list values of the smearing parameters of Wuppertal smearing, the fitting interval for the mass extraction of mesons made of $(f)$ hyperquark constituents and the corresponding $\chi^2/N_{\rm d.o.f.}$ of the fits.
Similar information for mesons made of $(as)$ hyperquarks is presented in Tab.~\ref{tab:Meson_as_fit_pars}.
For  chimera baryons, we provide the relevant details separately for each ensemble in Tabs.~\ref{tab:CB_fit_pars_QB1}--\ref{tab:CB_fit_pars_QB5}, where the number of iterations of Wuppertal smearing at the source and sink, the fitting intervals imposed to fit the correlation functions, as well as the resulting $\chi^2/N_{\rm d.o.f.}$ are displayed. 
While we set the number of iterations to be the same for both $(f)$ and $(as)$ hyperquarks, we set the step size differently for each type of hyperquark.
All parameters presented are also  available in machine-readable format in the data release associated with this publication~\cite{DATA}.

\begin{table}
\begin{center}
\input{tabs/table_13.tex}
\end{center}
   \caption{
   Technical details about the computation of the masses of the pseudoscalar and vector mesons constituted by hyperquarks in $(f)$ representation. 
   For Wuppertal smearing, we denote the step size by $\epsilon^{(f)}$, the number of iterations at the source by $N_{\rm W}^{\rm source}$ and the number of iterations at the sink by $N_{\rm W}^{\rm sink}$.
   The APE smearing parameters $\alpha$ and $N_{\rm APE}$ are fixed to $0.4$ and $50$, respectively, in all the calculations.
   For each choice of bare mass, $am_0^{(f)}$, and each meson, we show the fitting intervals as Euclidean time $I=[t_i, t_f]$, between the initial time $t_i$ and the final time $t_f$.
   We perform a correlated fit with standard $\chi^2$-minimization to the function in Eq.~(\ref{eq:meson_corr}). We report the values of $\chi^2$ normalized by the number degrees of freedom, $\chi^2/N_{\rm d.o.f.}$.
   }
   \label{tab:Meson_f_fit_pars}
\end{table}

\begin{table}
\begin{center}
   \input{tabs/table_14.tex}
 \end{center}
   \caption{
   Technical details about the computation of the masses of the pseudoscalar and vector mesons constituted by hyperquarks in $(as)$ representation. 
   For Wuppertal smearing, we denote the step size by $\epsilon^{(as)}$, the number of iterations at the source by $N_{\rm W}^{\rm source}$ and the number of iterations at the sink by $N_{\rm W}^{\rm sink}$.
   The APE smearing parameters $\alpha$ and $N_{\rm APE}$ are fixed to $0.4$ and $50$, respectively, in all the calculations.
   For each choice of bare mass, $am_0^{(as)}$, and each meson, we show the fitting intervals as Euclidean time $I=[t_i, t_f]$, between the initial time $t_i$ and the final time $t_f$.
   We perform a correlated fit with standard $\chi^2$-minimization to the function in Eq.~(\ref{eq:meson_corr}). We report the values of $\chi^2$ normalized by the number degrees of freedom, $\chi^2/N_{\rm d.o.f.}$.
   }
   \label{tab:Meson_as_fit_pars}
\end{table}

\begin{table}
\begin{center}
\input{tabs/table_15.tex}
\end{center}
   \caption{
   Technical details about the computation of the masses of $\Lambda_{\rm CB}$, $\Sigma_{\rm CB}$, and $\Sigma_{\rm CB}^\ast$, with bare masses $am_0^{(f)}$ and $am_0^{(as)}$, on ensemble QB1.
   For Wuppertal smearing parameters, we represent the number of iterations at the source as $N_{\rm W}^{\rm source}$ and at the sink as $N_{\rm W}^{\rm sink}$.
   For each chimera baryon, we select the number of sink iterations to present the optimal plateau, considering both its length and error size.
   APE smearing parameters $\alpha$ and $N_{\rm APE}$ are $0.4$ and $50$, respectively, for all the calculations.
   For each set of bare masses and each chimera baryon, we report the fitting intervals as the Euclidean time $I=[t_i, t_f]$, between the initial time $t_i$ and the final time $t_f$.
   We perform a correlated fit with the standard $\chi^2$-minimization to the function Eq.~(\ref{eq:cCB}). We report the values of $\chi^2$ normalized by the degrees of freedom, $\chi^2/N_{\rm d.o.f.}$.
   }
   \label{tab:CB_fit_pars_QB1}
\end{table}

\begin{table}
\begin{center}
    \input{tabs/table_16.tex}
\end{center}
   \caption{
   Technical details about the computation of the masses of $\Lambda_{\rm CB}$, $\Sigma_{\rm CB}$, and $\Sigma_{\rm CB}^\ast$, with bare masses $am_0^{(f)}$ and $am_0^{(as)}$, on ensemble QB2.
   For Wuppertal smearing parameters, we represent the number of iterations at the source as $N_{\rm W}^{\rm source}$ and at the sink as $N_{\rm W}^{\rm sink}$.
   For each chimera baryon, we select the number of sink iterations to present the optimal plateau, considering both its length and error size.
   APE smearing parameters $\alpha$ and $N_{\rm APE}$ are $0.4$ and $50$, respectively, for all the calculations.
   For each set of bare masses and each chimera baryon, we report the fitting intervals as the Euclidean time $I=[t_i, t_f]$, between the initial time $t_i$ and the final time $t_f$.
   We perform a correlated fit with the standard $\chi^2$-minimization to the function Eq.~(\ref{eq:cCB}). We report the values of $\chi^2$ normalized by the degrees of freedom, $\chi^2/N_{\rm d.o.f.}$.
   }
   \label{tab:CB_fit_pars_QB2}
\end{table}

\begin{table}
\begin{center}
  \input{tabs/table_17.tex}
\end{center}
   \caption{
   Technical details about the computation of the masses of $\Lambda_{\rm CB}$, $\Sigma_{\rm CB}$, and $\Sigma_{\rm CB}^\ast$, with bare masses $am_0^{(f)}$ and $am_0^{(as)}$, on ensemble QB3.
   For Wuppertal smearing parameters, we represent the number of iterations at the source as $N_{\rm W}^{\rm source}$ and at the sink as $N_{\rm W}^{\rm sink}$.
   For each chimera baryon, we select the number of sink iterations to present the optimal plateau, considering both its length and error size.
   APE smearing parameters $\alpha$ and $N_{\rm APE}$ are $0.4$ and $50$, respectively, for all the calculations.
   For each set of bare masses and each chimera baryon, we report the fitting intervals as the Euclidean time $I=[t_i, t_f]$, between the initial time $t_i$ and the final time $t_f$.
   We perform a correlated fit with the standard $\chi^2$-minimization to the function Eq.~(\ref{eq:cCB}). We report the values of $\chi^2$ normalized by the degrees of freedom, $\chi^2/N_{\rm d.o.f.}$.
   }
   \label{tab:CB_fit_pars_QB3}
\end{table}

\begin{table}
\begin{center}
   \input{tabs/table_18.tex}
\end{center}
   \caption{
   Technical details about the computation of the masses of $\Lambda_{\rm CB}$, $\Sigma_{\rm CB}$, and $\Sigma_{\rm CB}^\ast$, with bare masses $am_0^{(f)}$ and $am_0^{(as)}$, on ensemble QB4.
   For Wuppertal smearing parameters, we represent the number of iterations at the source as $N_{\rm W}^{\rm source}$ and at the sink as $N_{\rm W}^{\rm sink}$.
   For each chimera baryon, we select the number of sink iterations to present the optimal plateau, considering both its length and error size.
   APE smearing parameters $\alpha$ and $N_{\rm APE}$ are $0.4$ and $50$, respectively, for all the calculations.
   For each set of bare masses and each chimera baryon, we report the fitting intervals as the Euclidean time $I=[t_i, t_f]$, between the initial time $t_i$ and the final time $t_f$.
   We perform a correlated fit with the standard $\chi^2$-minimization to the function Eq.~(\ref{eq:cCB}). We report the values of $\chi^2$ normalized by the degrees of freedom, $\chi^2/N_{\rm d.o.f.}$.
   }
   \label{tab:CB_fit_pars_QB4}
\end{table}

\begin{table}
\begin{center}
    \input{tabs/table_19.tex}
 \end{center}
   \caption{
   Technical details about the computation of the masses of $\Lambda_{\rm CB}$, $\Sigma_{\rm CB}$, and $\Sigma_{\rm CB}^\ast$, with bare masses $am_0^{(f)}$ and $am_0^{(as)}$, on ensemble QB5.
   For Wuppertal smearing parameters, we represent the number of iterations at the source as $N_{\rm W}^{\rm source}$ and at the sink as $N_{\rm W}^{\rm sink}$.
   For each chimera baryon, we select the number of sink iterations to present the optimal plateau, considering both its length and error size.
   APE smearing parameters $\alpha$ and $N_{\rm APE}$ are $0.4$ and $50$, respectively, for all the calculations.
   For each set of bare masses and each chimera baryon, we report the fitting intervals as the Euclidean time $I=[t_i, t_f]$, between the initial time $t_i$ and the final time $t_f$.
   We perform a correlated fit with the standard $\chi^2$-minimization to the function Eq.~(\ref{eq:cCB}). We report the values of $\chi^2$ normalized by the degrees of freedom, $\chi^2/N_{\rm d.o.f.}$.
   }
   \label{tab:CB_fit_pars_QB5}
\end{table}

\bibliography{refs}

\end{document}

%% file: tabs/table_2.tex
\begin{tabular}{| c | c | c | c | c | c|}\hline\hline 
Ensemble & $\beta$   & $N_t\times N^3_s$ & $\left < P \right >$  & $w_0/a$  \\ \hline 
QB1	    & $7.62$    & $48\times24^3$   & 0.6018898(94)	& 1.448(3)      \\ 
QB2	    & $7.7$     & $60\times48^3$    & 0.6088000(35)	& 1.6070(19)    \\ 
QB3	    & $7.85$    & $60\times48^3$   & 0.6203809(28)	& 1.944(3)      \\ 
QB4	    & $8.0$     & $60\times48^3$    & 0.6307425(27)	& 2.3149(12)    \\ 
QB5	    & $8.2$     & $60\times48^3$    & 0.6432302(25)	& 2.8812(21)    \\ \hline 
\end{tabular}

%% file: tabs/table_4.tex
 \begin{tabular}{|c|C{1.4cm}C{1.4cm}C{1.4cm}C{1.4cm}C{1.4cm}|} 
 \hline\hline& \multicolumn{5}{c|}{$\Lambda_{\rm CB}$}   \\ 
Ansatz&  $\hat{m}_{\rm PS,cut}$ &  $\hat{m}_{\rm ps,cut}$ & $\chi^2/N_{\rm d.o.f.}$ &  AIC & $W$ \\ \hline 
M2 & 0.77 & 0.97 & 0.56 & 199.76 & $\sim 10^{-38}$  \\ 
M3 & 1.07 & 1.87 & 0.85 & 136.16 & $\sim 10^{-10}$  \\ 
MF4 & 1.07 & 1.87 & 0.85 & 138.16 & $\sim 10^{-11}$  \\ 
MA4 & 1.07 & 1.87 & 0.83 & 134.97 & $\sim 10^{-9}$  \\ 
MC4 & 1.07 & 1.87 & 0.67 & 114.91 & 0.63  \\ 
\hline 
 \end{tabular}

%% file: tabs/table_5.tex
 \begin{tabular}{|c|C{1.4cm}C{1.4cm}C{1.4cm}C{1.4cm}C{1.4cm}|} 
 \hline\hline & \multicolumn{5}{c|}{$\Sigma_{\rm CB}$}  \\ 
Ansatz & $\hat{m}_{\rm PS,cut}$ &  $\hat{m}_{\rm ps,cut}$  & $\chi^2/N_{\rm d.o.f.}$ &  AIC & $W$ \\ 
\hline 
M2 & 0.62 & 0.72 & 0.67 & 235.41 & $\sim 10^{-29}$  \\ 
M3 & 0.62 & 1.37 & 0.77 & 218.29 & $\sim 10^{-22}$  \\ 
MF4 & 0.57 & 1.82 & 0.77 & 219.74 & $\sim 10^{-22}$  \\ 
MA4 & 0.57 & 1.87 & 0.65 & 209.31 & $\sim 10^{-18}$  \\ 
MC4 & 0.77 & 1.47 & 0.78 & 169.92 & 0.97  \\ 
\hline 
 \end{tabular}

%% file: tabs/table_6.tex
 \begin{tabular}{|c|C{1.4cm}C{1.4cm}C{1.4cm}C{1.4cm}C{1.4cm}|} 
 \hline\hline & \multicolumn{5}{c|}{$\Sigma^\ast_{\rm CB}$}  \\ 
Ansatz & $\hat{m}_{\rm PS,cut}$ &  $\hat{m}_{\rm ps,cut}$  & $\chi^2/N_{\rm d.o.f.}$ &  AIC & $W$ \\ 
\hline 
M2 & 0.82 & 0.87 & 1.05 & 233.08 & $\sim 10^{-9}$  \\ 
M3 & 0.82 & 1.17 & 1.01 & 214.42 & 0.64  \\ 
MF4 & 0.82 & 1.17 & 1.05 & 218.5 & 0.01  \\ 
MA4 & 0.82 & 1.67 & 1.4 & 217.59 & 0.03  \\ 
MC4 & 0.82 & 1.17 & 1.01 & 215.53 & 0.21  \\ 
\hline 
 \end{tabular}

%% file: tabs/table_7.tex
\begin{tabular}{| C{1cm}  | C{1cm}   ||c  |c  |c  |c  |c  |c  |c |c  |c |} 
 \hline\hline 
CB & Ansatz &$\hat{m}_{\rm CB}^\chi $ & $ F_2  $ 	&  $A_2 $   &  $L_1 $  & $F_3 $  &   $A_3 $  & $L_{2F} $  &   $L_{2A}$ & $C_4$ \\ 
 \hline 
$\Lambda_{\rm CB}$ & MC4 & 1.003(30) & 0.691(66) & 0.383(13) & -0.14(47) & -0.14(35) & -0.091(48) & 0.091(76) & 0.002(14) & -0.024(63)\\ \hline 
 $\Sigma_{\rm CB}$ & MC4 & 0.840(21) & 0.801(83) & 0.558(13) & -0.13(33) & -0.23(67) & -0.161(83) & 0.190(66) & -0.02(17) & -0.079(66)\\ \hline 
 $\Sigma^\ast_{\rm CB}$ & M3 & 1.259(35) & 0.36(10) & 0.394(32) & -0.33(55) & -0.06(85) & -0.13(19) & 0.334(84) & 0.007(31) & -\\ \hline 
 \end{tabular} 

%% file: tabs/table_8.tex
\begin{tabular}{| c  c || c  | c | c | c | c| c | c| } 
\hline\hline 
$am_0^{(f)}$ &  $am_0^{(as)}$ &$am_{\rm PS}$	& $m_{\rm PS} /  m_{\rm V}$ & 
$am_{\rm ps}$	& $m_{\rm ps} / m_{\rm v}$ & 
$am_{\Lambda_{\rm CB}}$ & $am_{\Sigma_{\rm CB}}$ & $am_{\Sigma^\star_{\rm CB}}$ \\ \hline 
-0.7 & -0.8 & 0.55051(63) & 0.8814(14)& \multirow{5}{*}{1.14239(68)} & \multirow{5}{*}{0.96157(30)} & 1.3165(30) & 1.3033(27) & 1.3418(33)\\
-0.73 & -0.8 & 0.48083(59) & 0.8431(18) & & & 1.2643(34) & 1.2505(31) & 1.2914(38)\\
-0.75 & -0.8 & 0.43054(60) & 0.8061(23) & & & 1.2294(40) & 1.2143(36) & 1.2575(45)\\
-0.77 & -0.8 & 0.37585(69) & 0.742(11) & & & 1.1937(57) & 1.1786(47) & 1.2213(63)\\
-0.79 & -0.8 & 0.31357(79) & 0.6802(35) & & & 1.1613(70) & 1.1427(61) & 1.1889(70)\\
\hline
-0.7 & -0.9 & 0.55051(63) & 0.8814(14)& \multirow{5}{*}{0.96083(76)} & \multirow{5}{*}{0.93732(66)} & 1.2251(28) & 1.2047(26) & 1.2539(33)\\
-0.73 & -0.9 & 0.48083(59) & 0.8431(18) & & & 1.1705(31) & 1.1498(29) & 1.2023(38)\\
-0.75 & -0.9 & 0.43054(60) & 0.8061(23) & & & 1.1343(36) & 1.1129(32) & 1.1680(45)\\
-0.77 & -0.9 & 0.37585(69) & 0.742(11) & & & 1.0984(46) & 1.0757(41) & 1.1331(51)\\
-0.79 & -0.9 & 0.31357(79) & 0.6802(35) & & & 1.0633(49) & 1.0427(47) & 1.0918(76)\\
\hline
-0.7 & -0.95 & 0.55051(63) & 0.8814(14)& \multirow{5}{*}{0.86018(70)} & \multirow{5}{*}{0.91745(72)} & 1.1759(27) & 1.1510(26) & 1.2069(32)\\
-0.73 & -0.95 & 0.48083(59) & 0.8431(18) & & & 1.1208(29) & 1.0947(27) & 1.1541(39)\\
-0.75 & -0.95 & 0.43054(60) & 0.8061(23) & & & 1.0841(33) & 1.0570(29) & 1.1197(45)\\
-0.77 & -0.95 & 0.37585(69) & 0.742(11) & & & 1.0470(38) & 1.0195(37) & 1.0845(52)\\
-0.79 & -0.95 & 0.31357(79) & 0.6802(35) & & & 1.0093(51) & 0.9838(46) & 1.0483(66)\\
\hline
-0.7 & -1.0 & 0.55051(63) & 0.8814(14)& \multirow{5}{*}{0.74972(70)} & \multirow{5}{*}{0.88675(95)} & 1.1264(30) & 1.0940(26) & 1.1599(34)\\
-0.73 & -1.0 & 0.48083(59) & 0.8431(18) & & & 1.0689(28) & 1.0363(26) & 1.1049(38)\\
-0.75 & -1.0 & 0.43054(60) & 0.8061(23) & & & 1.0314(31) & 0.9977(29) & 1.0704(41)\\
-0.77 & -1.0 & 0.37585(69) & 0.742(11) & & & 0.9935(36) & 0.9591(36) & 1.0352(47)\\
-0.79 & -1.0 & 0.31357(79) & 0.6802(35) & & & 0.9551(47) & 0.9220(44) & 0.9996(61)\\
\hline
-0.7 & -1.05 & 0.55051(63) & 0.8814(14)& \multirow{5}{*}{0.62516(68)} & \multirow{5}{*}{0.8347(19)} & 1.0689(28) & 1.0311(24) & 1.1048(34)\\
-0.73 & -1.05 & 0.48083(59) & 0.8431(18) & & & 1.0109(31) & 0.9718(25) & 1.0518(38)\\
-0.75 & -1.05 & 0.43054(60) & 0.8061(23) & & & 0.9720(35) & 0.9318(30) & 1.0168(42)\\
-0.77 & -1.05 & 0.37585(69) & 0.742(11) & & & 0.9325(45) & 0.8914(34) & 0.9817(47)\\
-0.79 & -1.05 & 0.31357(79) & 0.6802(35) & & & 0.8935(60) & 0.8517(43) & 0.9472(61)\\
\hline
-0.7 & -1.1 & 0.55051(63) & 0.8814(14)& \multirow{5}{*}{0.47811(61)} & \multirow{5}{*}{0.7401(26)} & 1.0070(30) & 0.9624(24) & 1.0521(31)\\
-0.73 & -1.1 & 0.48083(59) & 0.8431(18) & & & 0.9495(32) & 0.9011(25) & 1.0002(34)\\
-0.75 & -1.1 & 0.43054(60) & 0.8061(23) & & & 0.9105(33) & 0.8593(27) & 0.9655(34)\\
-0.77 & -1.1 & 0.37585(69) & 0.742(11) & & & 0.8708(37) & 0.8165(29) & 0.9309(37)\\
-0.79 & -1.1 & 0.31357(79) & 0.6802(35) & & & 0.8295(48) & 0.7724(35) & 0.8966(43)\\
\hline
-0.77 & -1.12 & 0.37585(69) & 0.742(11)& \multirow{3}{*}{0.40811(66)} & \multirow{3}{*}{0.6845(26)} & 0.8420(37) & 0.7821(30) & 0.9004(50)\\
-0.78 & -1.12 & 0.34593(72) & 0.7200(43) & & & 0.8214(41) & 0.7594(33) & 0.8829(53)\\
-0.79 & -1.12 & 0.31357(79) & 0.6802(35) & & & 0.8003(47) & 0.7362(35) & 0.8655(60)\\
\hline
-0.77 & -1.14 & 0.37585(69) & 0.742(11)& \multirow{3}{*}{0.32632(76)} & \multirow{3}{*}{0.5944(38)} & 0.8133(41) & 0.7461(31) & 0.8763(50)\\
-0.78 & -1.14 & 0.34593(72) & 0.7200(43) & & & 0.7926(46) & 0.7223(33) & 0.8582(52)\\
-0.79 & -1.14 & 0.31357(79) & 0.6802(35) & & & 0.7713(52) & 0.6972(38) & 0.8401(57)\\
\hline
-0.7 & -1.15 & 0.55051(63) & 0.8814(14)& \multirow{5}{*}{0.27757(81)} & \multirow{5}{*}{0.5289(31)} & 0.9399(34) & 0.8842(29) & 0.9908(44)\\
-0.73 & -1.15 & 0.48083(59) & 0.8431(18) & & & 0.8804(37) & 0.8190(30) & 0.9314(71)\\
-0.75 & -1.15 & 0.43054(60) & 0.8061(23) & & & 0.8400(39) & 0.7737(31) & 0.8953(80)\\
-0.77 & -1.15 & 0.37585(69) & 0.742(11) & & & 0.7989(46) & 0.7266(34) & 0.8689(46)\\
-0.79 & -1.15 & 0.31357(79) & 0.6802(35) & & & 0.7564(56) & 0.6765(40) & 0.8275(62)\\
\hline
\end{tabular} 

%% file: tabs/table_9.tex
\begin{tabular}{| c  c || c  | c | c | c | c| c | c| } 
\hline\hline 
$am_0^{(f)}$ &  $am_0^{(as)}$ &$am_{\rm PS}$	& $m_{\rm PS} /  m_{\rm V}$ & 
$am_{\rm ps}$	& $m_{\rm ps} / m_{\rm v}$ & 
$am_{\Lambda_{\rm CB}}$ & $am_{\Sigma_{\rm CB}}$ & $am_{\Sigma^\star_{\rm CB}}$ \\ \hline 
-0.7 & -0.89 & 0.46136(51) & 0.8541(12)& \multirow{4}{*}{0.89541(64)} & \multirow{4}{*}{0.93561(91)} & 1.0912(38) & 1.0745(27) & 1.1190(37)\\
-0.72 & -0.89 & 0.41055(54) & 0.8151(16) & & & 1.0522(46) & 1.0363(29) & 1.0833(42)\\
-0.74 & -0.89 & 0.35509(61) & 0.7649(27) & & & 1.0166(42) & 0.9981(37) & 1.0468(50)\\
-0.77 & -0.89 & 0.25517(63) & 0.6240(41) & & & 0.9622(74) & 0.9407(61) & 0.9867(82)\\
\hline
-0.72 & -0.91 & 0.41055(54) & 0.8151(16)& \multirow{3}{*}{0.85430(49)} & \multirow{3}{*}{0.92741(77)} & 1.0350(29) & 1.0154(25) & 1.0656(42)\\
-0.74 & -0.91 & 0.35509(61) & 0.7649(27) & & & 0.9964(34) & 0.9761(28) & 1.0299(50)\\
-0.77 & -0.91 & 0.25517(63) & 0.6240(41) & & & 0.9400(58) & 0.9161(43) & 0.9726(76)\\
\hline
-0.72 & -0.92 & 0.41055(54) & 0.8151(16)& \multirow{3}{*}{0.83312(49)} & \multirow{3}{*}{0.92381(87)} & 1.0243(29) & 1.0034(25) & 1.0555(41)\\
-0.74 & -0.92 & 0.35509(61) & 0.7649(27) & & & 0.9855(34) & 0.9644(28) & 1.0197(49)\\
-0.77 & -0.92 & 0.25517(63) & 0.6240(41) & & & 0.9290(57) & 0.9042(43) & 0.9625(74)\\
\hline
-0.7 & -0.93 & 0.46136(51) & 0.8541(12)& \multirow{5}{*}{0.81140(63)} & \multirow{5}{*}{0.9182(11)} & 1.0494(36) & 1.0302(26) & 1.0788(37)\\
-0.72 & -0.93 & 0.41055(54) & 0.8151(16) & & & 1.0135(32) & 0.9916(24) & 1.0429(41)\\
-0.74 & -0.93 & 0.35509(61) & 0.7649(27) & & & 0.9748(36) & 0.9526(28) & 1.0063(49)\\
-0.76 & -0.93 & 0.29175(62) & 0.6873(44) & & & 0.9362(45) & 0.9129(36) & 0.9677(63)\\
-0.77 & -0.93 & 0.25517(63) & 0.6240(41) & & & 0.9185(56) & 0.8923(41) & 0.9462(78)\\
\hline
-0.76 & -0.97 & 0.29175(62) & 0.6873(44)& \multirow{1}{*}{0.72046(63)} & \multirow{1}{*}{0.8930(14)} & 0.8906(43) & 0.8626(32) & 0.9254(62)\\
\hline
-0.76 & -1.01 & 0.29175(62) & 0.6873(44)& \multirow{1}{*}{0.62049(64)} & \multirow{1}{*}{0.8545(17)} & 0.8397(41) & 0.8077(30) & 0.8810(61)\\
\hline
-0.72 & -1.09 & 0.41055(54) & 0.8151(16)& \multirow{3}{*}{0.36948(65)} & \multirow{3}{*}{0.6720(49)} & 0.8171(33) & 0.7688(22) & 0.8615(44)\\
-0.74 & -1.09 & 0.35509(61) & 0.7649(27) & & & 0.7750(35) & 0.7237(24) & 0.8250(53)\\
-0.76 & -1.09 & 0.29175(62) & 0.6873(44) & & & 0.7322(45) & 0.6768(29) & 0.7898(69)\\
\hline
-0.72 & -1.1 & 0.41055(54) & 0.8151(16)& \multirow{4}{*}{0.32806(68)} & \multirow{4}{*}{0.6275(56)} & 0.8026(34) & 0.7518(28) & 0.8495(48)\\
-0.74 & -1.1 & 0.35509(61) & 0.7649(27) & & & 0.7603(38) & 0.7061(30) & 0.8130(57)\\
-0.76 & -1.1 & 0.29175(62) & 0.6873(44) & & & 0.7175(47) & 0.6573(27) & 0.7780(78)\\
-0.77 & -1.1 & 0.25517(63) & 0.6240(41) & & & 0.6965(39) & 0.6326(31) & 0.7618(58)\\
\hline
-0.72 & -1.11 & 0.41055(54) & 0.8151(16)& \multirow{4}{*}{0.28138(69)} & \multirow{4}{*}{0.5674(75)} & 0.7883(38) & 0.7351(35) & 0.8352(55)\\
-0.74 & -1.11 & 0.35509(61) & 0.7649(27) & & & 0.7458(42) & 0.6858(26) & 0.8040(52)\\
-0.76 & -1.11 & 0.29175(62) & 0.6873(44) & & & 0.7027(50) & 0.6369(28) & 0.7707(63)\\
-0.77 & -1.11 & 0.25517(63) & 0.6240(41) & & & 0.6810(43) & 0.6114(31) & 0.7493(61)\\
\hline
-0.72 & -1.12 & 0.41055(54) & 0.8151(16)& \multirow{4}{*}{0.22629(62)} & \multirow{4}{*}{0.4959(61)} & 0.7737(43) & 0.7140(26) & 0.8269(50)\\
-0.74 & -1.12 & 0.35509(61) & 0.7649(27) & & & 0.7309(49) & 0.6655(27) & 0.7904(56)\\
-0.76 & -1.12 & 0.29175(62) & 0.6873(44) & & & 0.6875(57) & 0.6152(30) & 0.7549(64)\\
-0.77 & -1.12 & 0.25517(63) & 0.6240(41) & & & 0.6650(48) & 0.5888(33) & 0.7371(68)\\
\hline
\end{tabular} 

%% file: tabs/table_10.tex
\begin{tabular}{| c  c || c  | c | c | c | c| c | c| } 
\hline\hline 
$am_0^{(f)}$ &  $am_0^{(as)}$ &$am_{\rm PS}$	& $m_{\rm PS} /  m_{\rm V}$ & 
$am_{\rm ps}$	& $m_{\rm ps} / m_{\rm v}$ & 
$am_{\Lambda_{\rm CB}}$ & $am_{\Sigma_{\rm CB}}$ & $am_{\Sigma^\star_{\rm CB}}$ \\ \hline 
-0.66 & -0.85 & 0.41447(68) & 0.8644(16)& \multirow{4}{*}{0.83372(63)} & \multirow{4}{*}{0.94174(74)} & 0.9889(30) & 0.9742(22) & 1.0151(21)\\
-0.68 & -0.85 & 0.36183(67) & 0.8229(21) & & & 0.9465(30) & 0.9335(19) & 0.9789(24)\\
-0.7 & -0.85 & 0.30299(59) & 0.7545(24) & & & 0.9071(41) & 0.8938(22) & 0.9449(35)\\
-0.72 & -0.85 & 0.23505(56) & 0.6485(35) & & & 0.8641(58) & 0.8538(31) & 0.9103(55)\\
\hline
-0.66 & -0.87 & 0.41447(68) & 0.8644(16)& \multirow{4}{*}{0.79115(62)} & \multirow{4}{*}{0.93379(83)} & 0.9682(28) & 0.9509(16) & 0.9959(24)\\
-0.68 & -0.87 & 0.36183(67) & 0.8229(21) & & & 0.9275(33) & 0.9107(18) & 0.9599(27)\\
-0.7 & -0.87 & 0.30299(59) & 0.7545(24) & & & 0.8854(40) & 0.8701(21) & 0.9243(34)\\
-0.72 & -0.87 & 0.23505(56) & 0.6485(35) & & & 0.8421(57) & 0.8295(31) & 0.8897(54)\\
\hline
-0.66 & -0.9 & 0.41447(68) & 0.8644(16)& \multirow{4}{*}{0.72422(62)} & \multirow{4}{*}{0.9187(10)} & 0.9335(25) & 0.9146(16) & 0.9626(20)\\
-0.68 & -0.9 & 0.36183(67) & 0.8229(21) & & & 0.8923(29) & 0.8738(18) & 0.9259(23)\\
-0.7 & -0.9 & 0.30299(59) & 0.7545(24) & & & 0.8497(37) & 0.8326(21) & 0.8895(27)\\
-0.72 & -0.9 & 0.23505(56) & 0.6485(35) & & & 0.8079(55) & 0.7917(29) & 0.8528(39)\\
\hline
-0.66 & -0.93 & 0.41447(68) & 0.8644(16)& \multirow{4}{*}{0.65283(65)} & \multirow{4}{*}{0.8979(12)} & 0.9008(27) & 0.8772(16) & 0.9312(25)\\
-0.68 & -0.93 & 0.36183(67) & 0.8229(21) & & & 0.8594(31) & 0.8356(18) & 0.8949(27)\\
-0.7 & -0.93 & 0.30299(59) & 0.7545(24) & & & 0.8165(38) & 0.7934(20) & 0.8590(34)\\
-0.72 & -0.93 & 0.23505(56) & 0.6485(35) & & & 0.7719(52) & 0.7510(30) & 0.8244(50)\\
\hline
-0.72 & -0.97 & 0.23505(56) & 0.6485(35)& \multirow{2}{*}{0.54933(59)} & \multirow{2}{*}{0.8575(14)} & 0.7203(54) & 0.6921(27) & 0.7770(52)\\
-0.73 & -0.97 & 0.19352(66) & 0.5635(45) & & & 0.6953(68) & 0.6705(37) & 0.7610(72)\\
\hline
-0.72 & -0.99 & 0.23505(56) & 0.6485(35)& \multirow{2}{*}{0.49152(59)} & \multirow{2}{*}{0.8256(17)} & 0.6926(53) & 0.6599(26) & 0.7516(53)\\
-0.73 & -0.99 & 0.19352(66) & 0.5635(45) & & & 0.6667(69) & 0.6376(35) & 0.7350(72)\\
\hline
-0.72 & -1.01 & 0.23505(56) & 0.6485(35)& \multirow{2}{*}{0.42835(59)} & \multirow{2}{*}{0.7811(23)} & 0.6634(53) & 0.6253(26) & 0.7247(57)\\
-0.73 & -1.01 & 0.19352(66) & 0.5635(45) & & & 0.6366(72) & 0.6021(32) & 0.7073(79)\\
\hline
-0.72 & -1.03 & 0.23505(56) & 0.6485(35)& \multirow{2}{*}{0.35724(59)} & \multirow{2}{*}{0.7152(29)} & 0.6326(57) & 0.5879(26) & 0.6962(61)\\
-0.73 & -1.03 & 0.19352(66) & 0.5635(45) & & & 0.6051(75) & 0.5634(30) & 0.6774(89)\\
\hline
-0.7 & -1.04 & 0.30299(59) & 0.7545(24)& \multirow{3}{*}{0.31740(54)} & \multirow{3}{*}{0.6741(25)} & 0.6652(37) & 0.6183(21) & 0.7192(34)\\
-0.71 & -1.04 & 0.27064(52) & 0.7087(28) & & & 0.6409(44) & 0.5938(23) & 0.6999(36)\\
-0.72 & -1.04 & 0.23505(56) & 0.6485(35) & & & 0.6155(54) & 0.5689(25) & 0.6808(42)\\
\hline
-0.7 & -1.05 & 0.30299(59) & 0.7545(24)& \multirow{3}{*}{0.27275(51)} & \multirow{3}{*}{0.6150(30)} & 0.6498(39) & 0.5986(23) & 0.7042(41)\\
-0.71 & -1.05 & 0.27064(52) & 0.7087(28) & & & 0.6254(47) & 0.5735(25) & 0.6847(44)\\
-0.72 & -1.05 & 0.23505(56) & 0.6485(35) & & & 0.6000(54) & 0.5481(27) & 0.6653(53)\\
\hline
-0.7 & -1.06 & 0.30299(59) & 0.7545(24)& \multirow{3}{*}{0.22070(52)} & \multirow{3}{*}{0.5309(37)} & 0.6341(43) & 0.5776(23) & 0.6900(47)\\
-0.71 & -1.06 & 0.27064(52) & 0.7087(28) & & & 0.6089(49) & 0.5516(25) & 0.6704(52)\\
-0.72 & -1.06 & 0.23505(56) & 0.6485(35) & & & 0.5821(60) & 0.5248(27) & 0.6508(63)\\
\hline
\end{tabular} 

%% file: tabs/table_11.tex
\begin{tabular}{| c  c || c  | c | c | c | c| c | c| } 
\hline\hline 
$am_0^{(f)}$ &  $am_0^{(as)}$ &$am_{\rm PS}$	& $m_{\rm PS} /  m_{\rm V}$ & 
$am_{\rm ps}$	& $m_{\rm ps} / m_{\rm v}$ & 
$am_{\Lambda_{\rm CB}}$ & $am_{\Sigma_{\rm CB}}$ & $am_{\Sigma^\star_{\rm CB}}$ \\ \hline 
-0.6 & -0.81 & 0.44096(43) & 0.9122(14)& \multirow{7}{*}{0.79148(42)} & \multirow{7}{*}{0.95387(47)} & 0.9475(18) & 0.9379(16) & 0.9643(18)\\
-0.62 & -0.81 & 0.39299(44) & 0.8899(19) & & & 0.9071(21) & 0.8981(17) & 0.9269(20)\\
-0.64 & -0.81 & 0.34160(44) & 0.8518(24) & & & 0.8655(25) & 0.8580(19) & 0.8890(22)\\
-0.66 & -0.81 & 0.28512(43) & 0.7896(23) & & & 0.8245(28) & 0.8182(21) & 0.8508(26)\\
-0.68 & -0.81 & 0.21852(47) & 0.6825(38) & & & 0.7817(43) & 0.7788(28) & 0.8153(36)\\
-0.69 & -0.81 & 0.17780(57) & 0.5922(49) & & & 0.7651(45) & 0.7609(35) & 0.7988(48)\\
-0.7 & -0.81 & 0.12461(75) & 0.4505(96) & & & 0.7486(68) & 0.7409(54) & 0.7818(90)\\
\hline
-0.62 & -0.82 & 0.39299(44) & 0.8899(19)& \multirow{4}{*}{0.77014(45)} & \multirow{4}{*}{0.95136(50)} & 0.8963(21) & 0.8867(17) & 0.9164(20)\\
-0.64 & -0.82 & 0.34160(44) & 0.8518(24) & & & 0.8547(25) & 0.8464(19) & 0.8784(21)\\
-0.68 & -0.82 & 0.21852(47) & 0.6825(38) & & & 0.7700(42) & 0.7680(29) & 0.8051(35)\\
-0.7 & -0.82 & 0.12461(75) & 0.4505(96) & & & 0.7370(67) & 0.7287(53) & 0.7713(91)\\
\hline
-0.62 & -0.84 & 0.39299(44) & 0.8899(19)& \multirow{4}{*}{0.72632(46)} & \multirow{4}{*}{0.94418(57)} & 0.8743(21) & 0.8633(17) & 0.8950(19)\\
-0.64 & -0.84 & 0.34160(44) & 0.8518(24) & & & 0.8326(24) & 0.8227(18) & 0.8567(21)\\
-0.68 & -0.84 & 0.21852(47) & 0.6825(38) & & & 0.7475(41) & 0.7433(28) & 0.7830(34)\\
-0.7 & -0.84 & 0.12461(75) & 0.4505(96) & & & 0.7132(63) & 0.7036(51) & 0.7498(89)\\
\hline
-0.64 & -0.91 & 0.34160(44) & 0.8518(24)& \multirow{2}{*}{0.55722(50)} & \multirow{2}{*}{0.9003(11)} & 0.7493(22) & 0.7324(17) & 0.7750(26)\\
-0.66 & -0.91 & 0.28512(43) & 0.7896(23) & & & 0.7064(25) & 0.6903(20) & 0.7371(25)\\
\hline
-0.64 & -0.93 & 0.34160(44) & 0.8518(24)& \multirow{2}{*}{0.50263(48)} & \multirow{2}{*}{0.8779(13)} & 0.7234(21) & 0.7039(17) & 0.7504(27)\\
-0.66 & -0.93 & 0.28512(43) & 0.7896(23) & & & 0.6804(25) & 0.6610(20) & 0.7122(24)\\
\hline
-0.64 & -0.95 & 0.34160(44) & 0.8518(24)& \multirow{2}{*}{0.44385(51)} & \multirow{2}{*}{0.8466(17)} & 0.6949(22) & 0.6738(17) & 0.7237(28)\\
-0.66 & -0.95 & 0.28512(43) & 0.7896(23) & & & 0.6525(26) & 0.6298(19) & 0.6865(26)\\
\hline
-0.66 & -0.99 & 0.28512(43) & 0.7896(23)& \multirow{2}{*}{0.30571(47)} & \multirow{2}{*}{0.7232(25)} & 0.5948(23) & 0.5603(18) & 0.6347(28)\\
-0.68 & -0.99 & 0.21852(47) & 0.6825(38) & & & 0.5478(31) & 0.5117(24) & 0.5965(34)\\
\hline
-0.66 & -1.01 & 0.28512(43) & 0.7896(23)& \multirow{2}{*}{0.21414(43)} & \multirow{2}{*}{0.5735(32)} & 0.5627(26) & 0.5198(20) & 0.6063(31)\\
-0.68 & -1.01 & 0.21852(47) & 0.6825(38) & & & 0.5148(35) & 0.4677(24) & 0.5688(40)\\
\hline
-0.66 & -1.015 & 0.28512(43) & 0.7896(23)& \multirow{4}{*}{0.18561(39)} & \multirow{4}{*}{0.5161(38)} & 0.5536(30) & 0.5099(21) & 0.5993(34)\\
-0.67 & -1.015 & 0.25347(60) & 0.7468(39) & & & 0.5295(34) & 0.4836(22) & 0.5814(45)\\
-0.68 & -1.015 & 0.21852(47) & 0.6825(38) & & & 0.5056(41) & 0.4567(25) & 0.5623(40)\\
-0.69 & -1.015 & 0.17780(57) & 0.5922(49) & & & 0.4826(56) & 0.4290(30) & 0.5475(60)\\
\hline
\end{tabular} 

%% file: tabs/table_12.tex
\begin{tabular}{| c  c || c  | c | c | c | c| c | c| } 
\hline\hline 
$am_0^{(f)}$ &  $am_0^{(as)}$ &$am_{\rm PS}$	& $m_{\rm PS} /  m_{\rm V}$ & 
$am_{\rm ps}$	& $m_{\rm ps} / m_{\rm v}$ & 
$am_{\Lambda_{\rm CB}}$ & $am_{\Sigma_{\rm CB}}$ & $am_{\Sigma^\star_{\rm CB}}$ \\ \hline 
-0.62 & -0.95 & 0.25078(68) & 0.8158(29)& \multirow{3}{*}{0.22264(47)} & \multirow{3}{*}{0.6751(34)} & 0.4901(28) & 0.4600(17) & 0.5203(31)\\
-0.64 & -0.95 & 0.18239(82) & 0.6997(57) & & & 0.4371(37) & 0.4061(21) & 0.4779(40)\\
-0.646 & -0.95 & 0.15759(89) & 0.6397(76) & & & 0.4203(45) & 0.3897(24) & 0.4660(45)\\
\hline
-0.62 & -0.956 & 0.25078(68) & 0.8158(29)& \multirow{3}{*}{0.19303(45)} & \multirow{3}{*}{0.6200(42)} & 0.4792(29) & 0.4469(18) & 0.5111(33)\\
-0.64 & -0.956 & 0.18239(82) & 0.6997(57) & & & 0.4251(40) & 0.3918(22) & 0.4682(42)\\
-0.646 & -0.956 & 0.15759(89) & 0.6397(76) & & & 0.4078(47) & 0.3749(25) & 0.4562(49)\\
\hline
-0.62 & -0.961 & 0.25078(68) & 0.8158(29)& \multirow{3}{*}{0.16488(48)} & \multirow{3}{*}{0.5568(52)} & 0.4693(37) & 0.4350(20) & 0.5033(35)\\
-0.64 & -0.961 & 0.18239(82) & 0.6997(57) & & & 0.4145(52) & 0.3781(25) & 0.4601(46)\\
-0.646 & -0.961 & 0.15759(89) & 0.6397(76) & & & 0.3971(61) & 0.3605(29) & 0.4481(52)\\
\hline
\end{tabular} 

%% file: tabs/table_13.tex
\begin{tabular}{ | c| c | c  | c c c c| c c c  c|} 
 \hline\hline 
& 	& &     \multicolumn{4}{c|}{PS}   &   \multicolumn{4}{c|}{V}  \\ 
 Ensemble & $am_0^{(f)}$ & 	$\epsilon^{(f)}$     &  $N_{\rm W}^{\rm source}$  	 &  $N_{\rm W}^{\rm sink}$ 	&  I 	   &  $\chi^2/N_{\rm d.o.f.}$ &   $N_{\rm W}^{\rm source}$ 	 &  $N_{\rm W}^{\rm sink}$   &   I 	   &  $\chi^2/N_{\rm d.o.f.}$  \\ 
 \hline 
\multirow{6}{*}{QB1} 
 & -0.7 & 0.05 & 100 & 0 & [13 24] & 0.97 & 100 & 0 & [15 23] & 0.99 \\ 
 & -0.73 & 0.2 & 100 & 0 & [13 24] & 1.13 & 100 & 0 & [15 23] & 0.94 \\ 
 & -0.75 & 0.2 & 100 & 0 & [13 24] & 1.22 & 100 & 0 & [15 23] & 0.82 \\ 
 & -0.77 & 0.18 & 60 & 0 & [14 24] & 1.32 & 60 & 20 & [15 23] & 0.37 \\ 
 & -0.78 & 0.18 & 60 & 0 & [14 24] & 1.37 & 60 & 40 & [10 22] & 0.52 \\ 
 & -0.79 & 0.18 & 60 & 0 & [14 24] & 1.4 & 60 & 40 & [8 22] & 0.92 \\ 
\hline 
\multirow{5}{*}{QB2} 
 & -0.7 & 0.18 & 50 & 0 & [13 29] & 0.74 & 50 & 0 & [12 29] & 1.71 \\ 
 & -0.72 & 0.18 & 50 & 0 & [13 29] & 0.67 & 50 & 0 & [12 25] & 1.67 \\ 
 & -0.74 & 0.18 & 50 & 0 & [19 29] & 0.59 & 50 & 0 & [12 20] & 1.16 \\ 
 & -0.76 & 0.18 & 50 & 0 & [19 29] & 0.5 & 50 & 50 & [12 20] & 0.98 \\ 
 & -0.77 & 0.18 & 50 & 0 & [19 29] & 0.42 & 50 & 0 & [12 20] & 1.87 \\ 
\hline 
\multirow{6}{*}{QB3} 
 & -0.66 & 0.18 & 50 & 0 & [19 26] & 0.28 & 50 & 0 & [21 30] & 0.78 \\ 
 & -0.68 & 0.18 & 50 & 0 & [19 26] & 0.43 & 50 & 0 & [21 30] & 0.88 \\ 
 & -0.7 & 0.18 & 50 & 0 & [19 30] & 0.8 & 50 & 0 & [14 26] & 0.48 \\ 
 & -0.71 & 0.18 & 50 & 0 & [18 30] & 0.61 & 50 & 0 & [14 26] & 0.54 \\ 
 & -0.72 & 0.18 & 50 & 0 & [20 30] & 0.6 & 50 & 0 & [14 26] & 0.68 \\ 
 & -0.73 & 0.18 & 50 & 0 & [22 30] & 0.52 & 50 & 0 & [14 29] & 1.13 \\ 
\hline 
\multirow{8}{*}{QB4} 
 & -0.6 & 0.18 & 100 & 40 & [19 30] & 1.27 & 100 & 40 & [23 30] & 1.16 \\ 
 & -0.62 & 0.18 & 60 & 40 & [18 30] & 1.21 & 60 & 20 & [18 28] & 0.95 \\ 
 & -0.64 & 0.18 & 60 & 40 & [18 30] & 1.21 & 60 & 10 & [18 28] & 0.76 \\ 
 & -0.66 & 0.18 & 50 & 0 & [15 28] & 2.1 & 50 & 0 & [18 29] & 1.1 \\ 
 & -0.67 & 0.18 & 50 & 20 & [23 30] & 1.29 & 50 & 0 & [22 30] & 0.94 \\ 
 & -0.68 & 0.18 & 50 & 0 & [15 28] & 1.6 & 50 & 0 & [18 29] & 0.45 \\ 
 & -0.69 & 0.18 & 50 & 0 & [15 28] & 1.13 & 50 & 0 & [18 29] & 0.47 \\ 
 & -0.7 & 0.18 & 50 & 0 & [15 28] & 1.04 & 50 & 0 & [18 27] & 1.04 \\ 
\hline 
\multirow{3}{*}{QB5} 
 & -0.62 & 0.18 & 50 & 0 & [25 30] & 0.26 & 50 & 0 & [25 30] & 0.15 \\ 
 & -0.64 & 0.18 & 50 & 0 & [25 30] & 0.41 & 50 & 0 & [25 30] & 0.77 \\ 
 & -0.646 & 0.18 & 50 & 0 & [25 30] & 0.42 & 50 & 20 & [23 30] & 0.9 \\ 
\hline 
\end{tabular}

%% file: tabs/table_14.tex
\begin{tabular}{ | c| c | c  | c c c c| c c c  c|} 
 \hline\hline 
& 	& &     \multicolumn{4}{c|}{ps}   &   \multicolumn{4}{c|}{v}  \\ 
 Ensemble & $am_0^{(as)}$ & 	$\epsilon^{(as)}$     &  $N_{\rm W}^{\rm source}$  	 &  $N_{\rm W}^{\rm sink}$ 	&  I 	   &  $\chi^2/N_{\rm d.o.f.}$ &   $N_{\rm W}^{\rm source}$ 	 &  $N_{\rm W}^{\rm sink}$   &   I 	   &  $\chi^2/N_{\rm d.o.f.}$  \\ 
 \hline 
\multirow{9}{*}{QB1} & -0.8 & 0.01 & 100 & 0 & [8 24] & 0.57 & 100 & 0 & [8 24] & 0.65 \\ 
 & -0.9 & 0.01 & 100 & 20 & [15 23] & 0.32 & 100 & 20 & [8 24] & 0.44 \\ 
 & -0.95 & 0.01 & 100 & 0 & [14 23] & 0.48 & 100 & 0 & [10 24] & 0.61 \\ 
 & -1.0 & 0.01 & 100 & 0 & [14 23] & 0.63 & 100 & 0 & [10 24] & 0.82 \\ 
 & -1.05 & 0.01 & 100 & 0 & [10 23] & 0.44 & 100 & 0 & [15 24] & 0.6 \\ 
 & -1.1 & 0.08 & 100 & 0 & [12 24] & 0.46 & 100 & 0 & [12 24] & 1.52 \\ 
 & -1.12 & 0.18 & 60 & 0 & [12 24] & 0.64 & 60 & 0 & [9 20] & 1.33 \\ 
 & -1.14 & 0.18 & 60 & 0 & [12 24] & 0.83 & 60 & 0 & [9 20] & 0.68 \\ 
 & -1.15 & 0.1 & 100 & 0 & [10 24] & 0.82 & 100 & 0 & [7 18] & 0.53 \\ 
\hline 
\multirow{10}{*}{QB2} & -0.89 & 0.02 & 50 & 0 & [24 30] & 0.84 & 50 & 20 & [16 28] & 1.27 \\ 
 & -0.91 & 0.02 & 50 & 0 & [17 29] & 1.42 & 50 & 0 & [14 25] & 1.5 \\ 
 & -0.92 & 0.02 & 50 & 0 & [17 29] & 1.41 & 50 & 0 & [17 27] & 1.58 \\ 
 & -0.93 & 0.02 & 50 & 0 & [23 29] & 0.79 & 50 & 0 & [16 25] & 1.33 \\ 
 & -0.97 & 0.02 & 50 & 0 & [24 30] & 0.64 & 50 & 0 & [16 25] & 0.85 \\ 
 & -1.01 & 0.02 & 50 & 0 & [24 30] & 0.63 & 50 & 0 & [16 25] & 0.54 \\ 
 & -1.09 & 0.12 & 50 & 0 & [24 30] & 1.11 & 50 & 0 & [16 25] & 0.45 \\ 
 & -1.1 & 0.12 & 50 & 0 & [23 30] & 0.78 & 50 & 0 & [15 23] & 0.51 \\ 
 & -1.11 & 0.14 & 50 & 0 & [23 30] & 0.61 & 50 & 0 & [15 23] & 0.49 \\ 
 & -1.12 & 0.2 & 50 & 0 & [18 29] & 1.33 & 50 & 0 & [12 21] & 0.58 \\ 
\hline 
\multirow{11}{*}{QB3} & -0.85 & 0.01 & 50 & 0 & [21 29] & 0.33 & 50 & 0 & [19 29] & 0.47 \\ 
 & -0.87 & 0.01 & 50 & 0 & [21 29] & 0.38 & 50 & 0 & [19 29] & 0.48 \\ 
 & -0.9 & 0.01 & 50 & 0 & [21 29] & 0.49 & 50 & 0 & [19 28] & 0.57 \\ 
 & -0.93 & 0.01 & 50 & 0 & [21 29] & 0.65 & 50 & 0 & [19 29] & 0.61 \\ 
 & -0.97 & 0.01 & 50 & 0 & [16 25] & 0.32 & 50 & 0 & [15 26] & 0.75 \\ 
 & -0.99 & 0.01 & 50 & 0 & [16 25] & 0.28 & 50 & 0 & [15 26] & 0.92 \\ 
 & -1.01 & 0.01 & 50 & 0 & [16 25] & 0.27 & 50 & 0 & [15 26] & 1.0 \\ 
 & -1.03 & 0.01 & 50 & 0 & [15 25] & 0.29 & 50 & 0 & [14 26] & 0.83 \\ 
 & -1.04 & 0.16 & 50 & 0 & [15 25] & 0.39 & 50 & 0 & [12 21] & 0.73 \\ 
 & -1.05 & 0.18 & 50 & 0 & [15 25] & 0.42 & 50 & 0 & [12 21] & 0.78 \\ 
 & -1.06 & 0.2 & 50 & 0 & [15 29] & 0.93 & 50 & 0 & [12 23] & 0.95 \\ 
\hline 
\multirow{9}{*}{QB4} & -0.81 & 0.1 & 50 & 0 & [18 30] & 1.0 & 50 & 0 & [19 30] & 1.61 \\ 
 & -0.82 & 0.01 & 60 & 0 & [19 30] & 0.88 & 60 & 0 & [19 30] & 1.32 \\ 
 & -0.84 & 0.01 & 60 & 0 & [18 30] & 0.76 & 60 & 0 & [19 30] & 1.33 \\ 
 & -0.91 & 0.01 & 60 & 0 & [18 30] & 0.69 & 60 & 0 & [19 30] & 1.06 \\ 
 & -0.93 & 0.01 & 60 & 0 & [18 30] & 0.77 & 60 & 0 & [19 30] & 1.08 \\ 
 & -0.95 & 0.01 & 60 & 0 & [18 30] & 0.92 & 60 & 0 & [19 30] & 1.18 \\ 
 & -0.99 & 0.16 & 50 & 0 & [17 30] & 1.55 & 50 & 0 & [16 30] & 0.84 \\ 
 & -1.01 & 0.16 & 50 & 0 & [14 30] & 1.52 & 50 & 0 & [12 28] & 2.11 \\ 
 & -1.015 & 0.2 & 50 & 20 & [11 30] & 0.9 & 50 & 40 & [11 25] & 2.01 \\ 
\hline 
\multirow{3}{*}{QB5} & -0.95 & 0.2 & 50 & 0 & [20 30] & 0.76 & 50 & 0 & [17 29] & 1.07 \\ 
 & -0.956 & 0.2 & 50 & 0 & [18 30] & 1.12 & 50 & 0 & [17 29] & 1.16 \\ 
 & -0.961 & 0.2 & 50 & 0 & [18 30] & 1.14 & 50 & 0 & [17 29] & 1.16 \\ 
\hline 
\end{tabular}

%% file: tabs/table_15.tex
\begin{tabular}{ | c c | c c | c c c c | c c c c| c c c c|} 
  \hline\hline 
& 	&  &  & \multicolumn{4}{c|}{$\Lambda_{\rm CB}$}  &   \multicolumn{4}{c|}{$\Sigma_{\rm CB}$} &   \multicolumn{4}{c|}{$\Sigma^\ast_{\rm CB}$}  \\ 
 $am_0^{(f)}$ & $am_0^{(as)}$ & $\epsilon^{(f)}$  & $\epsilon^{(as)}$   &  $N_{\rm W}^{\rm source}$ 	 &  $N_{\rm W}^{\rm sink}$ 	&  I 	   &  $\chi^2/N_{\rm d.o.f.}$ & $N_{\rm W}^{\rm source}$ 	 &  $N_{\rm W}^{\rm sink}$   &   I 	   &  $\chi^2/N_{\rm d.o.f.}$& $N_{\rm W}^{\rm source}$ 	 &  $N_{\rm W}^{\rm sink}$   &   I 	   &  $\chi^2/N_{\rm d.o.f.}$ \\ 
 \hline-0.7 & -0.8 & 0.05 & 0.01 & 100 & 10 &  [12 24]  & 0.45 & 100 & 0 & [12 24] & 0.37 & 100 & 0 & [12 24] & 0.68 \\
-0.73 & -0.8 & 0.05 & 0.01 & 100 & 0 &  [12 24]  & 0.53 & 100 & 0 & [12 24] & 0.37 & 100 & 0 & [12 24] & 0.64 \\
-0.75 & -0.8 & 0.05 & 0.01 & 100 & 0 &  [12 24]  & 0.59 & 100 & 0 & [14 24] & 0.26 & 100 & 0 & [12 24] & 0.54 \\
-0.77 & -0.8 & 0.05 & 0.01 & 100 & 0 &  [12 20]  & 1.03 & 100 & 0 & [14 22] & 0.42 & 100 & 0 & [14 20] & 0.26 \\
-0.79 & -0.8 & 0.05 & 0.01 & 100 & 0 &  [12 24]  & 0.91 & 100 & 0 & [14 24] & 0.57 & 100 & 0 & [12 24] & 0.38 \\
\hline
-0.7 & -0.9 & 0.05 & 0.01 & 100 & 0 &  [12 24]  & 0.35 & 100 & 0 & [14 24] & 0.23 & 100 & 0 & [12 24] & 0.71 \\
-0.73 & -0.9 & 0.08 & 0.01 & 100 & 0 &  [12 24]  & 0.34 & 100 & 0 & [14 24] & 0.25 & 100 & 0 & [12 24] & 0.66 \\
-0.75 & -0.9 & 0.08 & 0.01 & 100 & 0 &  [12 24]  & 0.4 & 100 & 0 & [14 24] & 0.28 & 100 & 0 & [12 24] & 0.57 \\
-0.77 & -0.9 & 0.08 & 0.01 & 100 & 0 &  [12 24]  & 0.54 & 100 & 0 & [14 24] & 0.42 & 100 & 0 & [12 24] & 0.46 \\
-0.79 & -0.9 & 0.08 & 0.01 & 100 & 0 &  [10 24]  & 0.73 & 100 & 0 & [12 24] & 0.83 & 100 & 0 & [14 24] & 0.4 \\
\hline
-0.7 & -0.95 & 0.05 & 0.01 & 100 & 0 &  [12 24]  & 0.33 & 100 & 0 & [14 24] & 0.24 & 100 & 0 & [12 24] & 0.69 \\
-0.73 & -0.95 & 0.2 & 0.01 & 100 & 0 &  [10 24]  & 0.5 & 100 & 0 & [12 24] & 0.47 & 100 & 0 & [12 24] & 0.64 \\
-0.75 & -0.95 & 0.2 & 0.01 & 100 & 0 &  [10 24]  & 0.62 & 100 & 0 & [12 24] & 0.5 & 100 & 0 & [12 24] & 0.59 \\
-0.77 & -0.95 & 0.2 & 0.01 & 100 & 0 &  [10 24]  & 0.76 & 100 & 0 & [12 24] & 0.65 & 100 & 0 & [12 24] & 0.53 \\
-0.79 & -0.95 & 0.2 & 0.01 & 100 & 0 &  [10 24]  & 0.86 & 100 & 0 & [12 24] & 0.97 & 100 & 0 & [12 24] & 0.5 \\
\hline
-0.7 & -1.0 & 0.05 & 0.01 & 100 & 0 &  [10 20]  & 0.68 & 100 & 0 & [12 22] & 0.53 & 100 & 0 & [11 20] & 1.03 \\
-0.73 & -1.0 & 0.1 & 0.01 & 100 & 0 &  [10 24]  & 0.43 & 100 & 0 & [12 24] & 0.46 & 100 & 0 & [12 24] & 0.58 \\
-0.75 & -1.0 & 0.1 & 0.01 & 100 & 0 &  [10 24]  & 0.49 & 100 & 0 & [12 24] & 0.54 & 100 & 0 & [12 24] & 0.5 \\
-0.77 & -1.0 & 0.1 & 0.01 & 100 & 0 &  [10 24]  & 0.64 & 100 & 0 & [12 24] & 0.74 & 100 & 0 & [12 24] & 0.4 \\
-0.79 & -1.0 & 0.1 & 0.01 & 100 & 0 &  [10 24]  & 0.83 & 100 & 0 & [12 24] & 1.07 & 100 & 0 & [12 24] & 0.41 \\
\hline
-0.7 & -1.05 & 0.05 & 0.01 & 100 & 0 &  [12 24]  & 0.65 & 100 & 0 & [12 24] & 0.45 & 100 & 0 & [12 24] & 0.55 \\
-0.73 & -1.05 & 0.08 & 0.01 & 100 & 0 &  [12 24]  & 0.5 & 100 & 0 & [12 24] & 0.49 & 100 & 0 & [12 24] & 0.48 \\
-0.75 & -1.05 & 0.08 & 0.01 & 100 & 0 &  [12 24]  & 0.53 & 100 & 0 & [12 24] & 0.59 & 100 & 0 & [12 24] & 0.4 \\
-0.77 & -1.05 & 0.08 & 0.01 & 100 & 0 &  [12 24]  & 0.71 & 100 & 0 & [12 24] & 0.78 & 100 & 0 & [12 24] & 0.33 \\
-0.79 & -1.05 & 0.08 & 0.01 & 100 & 0 &  [12 24]  & 1.06 & 100 & 0 & [12 24] & 1.02 & 100 & 0 & [12 24] & 0.41 \\
\hline
-0.7 & -1.1 & 0.05 & 0.08 & 100 & 0 &  [10 20]  & 1.23 & 100 & 0 & [12 22] & 0.51 & 100 & 0 & [10 20] & 0.87 \\
-0.73 & -1.1 & 0.05 & 0.08 & 100 & 0 &  [10 20]  & 1.16 & 100 & 0 & [12 22] & 0.5 & 100 & 0 & [10 20] & 0.82 \\
-0.75 & -1.1 & 0.05 & 0.08 & 100 & 0 &  [10 20]  & 1.08 & 100 & 0 & [12 22] & 0.51 & 100 & 0 & [10 20] & 0.82 \\
-0.77 & -1.1 & 0.05 & 0.08 & 100 & 0 &  [10 20]  & 1.01 & 100 & 0 & [12 22] & 0.52 & 100 & 0 & [10 20] & 0.85 \\
-0.79 & -1.1 & 0.05 & 0.08 & 100 & 0 &  [10 20]  & 0.96 & 100 & 0 & [12 22] & 0.51 & 100 & 0 & [10 20] & 0.78 \\
\hline
-0.77 & -1.12 & 0.18 & 0.18 & 60 & 0 &  [10 22]  & 1.36 & 60 & 0 & [12 24] & 0.51 & 60 & 0 & [12 20] & 0.57 \\
-0.78 & -1.12 & 0.18 & 0.18 & 60 & 0 &  [10 22]  & 1.28 & 60 & 0 & [12 24] & 0.52 & 60 & 0 & [12 20] & 0.56 \\
-0.79 & -1.12 & 0.18 & 0.18 & 60 & 0 &  [10 22]  & 1.21 & 60 & 0 & [12 24] & 0.54 & 60 & 0 & [12 20] & 0.52 \\
\hline
-0.77 & -1.14 & 0.18 & 0.18 & 60 & 0 &  [10 22]  & 1.2 & 60 & 0 & [12 24] & 0.6 & 60 & 0 & [11 20] & 0.59 \\
-0.78 & -1.14 & 0.18 & 0.18 & 60 & 0 &  [10 22]  & 1.12 & 60 & 0 & [12 24] & 0.59 & 60 & 0 & [11 20] & 0.53 \\
-0.79 & -1.14 & 0.18 & 0.18 & 60 & 0 &  [10 20]  & 1.33 & 60 & 0 & [12 22] & 0.7 & 60 & 0 & [11 20] & 0.48 \\
\hline
-0.7 & -1.15 & 0.05 & 0.1 & 100 & 0 &  [10 20]  & 1.49 & 100 & 0 & [12 22] & 0.53 & 100 & 0 & [11 20] & 0.98 \\
-0.73 & -1.15 & 0.05 & 0.1 & 100 & 0 &  [10 20]  & 1.26 & 100 & 0 & [12 22] & 0.59 & 100 & 0 & [14 20] & 0.19 \\
-0.75 & -1.15 & 0.05 & 0.1 & 100 & 0 &  [10 20]  & 1.08 & 100 & 0 & [12 22] & 0.65 & 100 & 0 & [14 20] & 0.21 \\
-0.77 & -1.15 & 0.05 & 0.1 & 100 & 0 &  [10 20]  & 0.96 & 100 & 0 & [12 22] & 0.73 & 100 & 0 & [10 20] & 0.85 \\
-0.79 & -1.15 & 0.05 & 0.1 & 100 & 0 &  [10 20]  & 0.87 & 100 & 0 & [12 22] & 0.82 & 100 & 0 & [11 20] & 0.54 \\
\hline
\end{tabular} 

%% file: tabs/table_16.tex
\begin{tabular}{ | c c | c c | c c c c | c c c c| c c c c|} 
  \hline\hline 
& 	&  &  & \multicolumn{4}{c|}{$\Lambda_{\rm CB}$}  &   \multicolumn{4}{c|}{$\Sigma_{\rm CB}$} &   \multicolumn{4}{c|}{$\Sigma^\ast_{\rm CB}$}  \\ 
 $am_0^{(f)}$ & $am_0^{(as)}$ & $\epsilon^{(f)}$  & $\epsilon^{(as)}$   &  $N_{\rm W}^{\rm source}$ 	 &  $N_{\rm W}^{\rm sink}$ 	&  I 	   &  $\chi^2/N_{\rm d.o.f.}$ & $N_{\rm W}^{\rm source}$ 	 &  $N_{\rm W}^{\rm sink}$   &   I 	   &  $\chi^2/N_{\rm d.o.f.}$& $N_{\rm W}^{\rm source}$ 	 &  $N_{\rm W}^{\rm sink}$   &   I 	   &  $\chi^2/N_{\rm d.o.f.}$ \\ 
 \hline-0.7 & -0.89 & 0.18 & 0.02 & 50 & 0 &  [17 29]  & 0.57 & 50 & 0 & [17 28] & 0.52 & 50 & 0 & [17 25] & 0.25 \\
-0.72 & -0.89 & 0.18 & 0.02 & 50 & 0 &  [17 29]  & 0.75 & 50 & 0 & [17 28] & 0.33 & 50 & 0 & [17 25] & 0.29 \\
-0.74 & -0.89 & 0.18 & 0.02 & 50 & 0 &  [15 29]  & 0.94 & 50 & 0 & [17 28] & 0.24 & 50 & 0 & [17 25] & 0.34 \\
-0.77 & -0.89 & 0.18 & 0.02 & 50 & 0 &  [15 29]  & 0.9 & 50 & 0 & [17 28] & 1.01 & 50 & 0 & [17 25] & 0.29 \\
\hline
-0.72 & -0.91 & 0.18 & 0.02 & 50 & 0 &  [14 27]  & 0.64 & 50 & 0 & [14 30] & 0.38 & 50 & 0 & [16 25] & 0.5 \\
-0.74 & -0.91 & 0.18 & 0.02 & 50 & 0 &  [14 27]  & 0.75 & 50 & 0 & [14 28] & 0.28 & 50 & 0 & [16 25] & 0.61 \\
-0.77 & -0.91 & 0.18 & 0.02 & 50 & 0 &  [14 27]  & 0.86 & 50 & 0 & [14 28] & 0.88 & 50 & 0 & [16 25] & 0.53 \\
\hline
-0.72 & -0.92 & 0.18 & 0.02 & 50 & 0 &  [14 27]  & 0.64 & 50 & 0 & [14 28] & 0.36 & 50 & 0 & [16 25] & 0.51 \\
-0.74 & -0.92 & 0.18 & 0.02 & 50 & 0 &  [14 27]  & 0.75 & 50 & 0 & [14 28] & 0.28 & 50 & 0 & [16 25] & 0.62 \\
-0.77 & -0.92 & 0.18 & 0.02 & 50 & 0 &  [14 27]  & 0.85 & 50 & 0 & [14 28] & 0.88 & 50 & 0 & [16 25] & 0.54 \\
\hline
-0.7 & -0.93 & 0.18 & 0.02 & 50 & 0 &  [17 28]  & 0.5 & 50 & 0 & [17 29] & 0.6 & 50 & 0 & [17 26] & 0.24 \\
-0.72 & -0.93 & 0.18 & 0.02 & 50 & 0 &  [14 25]  & 0.66 & 50 & 0 & [14 27] & 0.36 & 50 & 0 & [17 26] & 0.28 \\
-0.74 & -0.93 & 0.18 & 0.02 & 50 & 0 &  [14 25]  & 0.75 & 50 & 0 & [14 27] & 0.3 & 50 & 0 & [17 26] & 0.35 \\
-0.76 & -0.93 & 0.18 & 0.02 & 50 & 0 &  [14 26]  & 0.81 & 50 & 0 & [14 26] & 0.46 & 50 & 0 & [17 24] & 0.44 \\
-0.77 & -0.93 & 0.18 & 0.02 & 50 & 0 &  [14 25]  & 1.0 & 50 & 0 & [14 27] & 0.64 & 50 & 0 & [17 24] & 0.35 \\
\hline
-0.76 & -0.97 & 0.18 & 0.02 & 50 & 0 &  [14 26]  & 0.75 & 50 & 0 & [14 26] & 0.47 & 50 & 0 & [17 24] & 0.47 \\
\hline
-0.76 & -1.01 & 0.18 & 0.02 & 50 & 0 &  [13 26]  & 0.73 & 50 & 0 & [14 26] & 0.48 & 50 & 0 & [17 25] & 0.51 \\
\hline
-0.72 & -1.09 & 0.18 & 0.12 & 50 & 0 &  [12 22]  & 0.8 & 50 & 0 & [14 24] & 0.9 & 50 & 0 & [16 25] & 0.37 \\
-0.74 & -1.09 & 0.18 & 0.12 & 50 & 0 &  [12 22]  & 0.61 & 50 & 0 & [14 24] & 0.89 & 50 & 0 & [16 25] & 0.39 \\
-0.76 & -1.09 & 0.18 & 0.12 & 50 & 0 &  [12 22]  & 0.45 & 50 & 0 & [14 24] & 0.85 & 50 & 0 & [16 25] & 0.47 \\
\hline
-0.72 & -1.1 & 0.18 & 0.12 & 50 & 0 &  [12 23]  & 0.76 & 50 & 0 & [17 24] & 1.05 & 50 & 0 & [16 26] & 0.48 \\
-0.74 & -1.1 & 0.18 & 0.12 & 50 & 0 &  [12 23]  & 0.56 & 50 & 0 & [17 24] & 0.92 & 50 & 0 & [16 26] & 0.5 \\
-0.76 & -1.1 & 0.18 & 0.12 & 50 & 0 &  [12 23]  & 0.41 & 50 & 0 & [14 26] & 0.79 & 50 & 0 & [16 25] & 0.46 \\
-0.77 & -1.1 & 0.2 & 0.18 & 50 & 0 &  [9 22]  & 0.38 & 50 & 0 & [14 28] & 0.8 & 50 & 0 & [12 23] & 0.72 \\
\hline
-0.72 & -1.11 & 0.18 & 0.14 & 50 & 0 &  [12 23]  & 0.76 & 50 & 0 & [18 25] & 1.02 & 50 & 0 & [16 24] & 0.37 \\
-0.74 & -1.11 & 0.18 & 0.14 & 50 & 0 &  [12 23]  & 0.55 & 50 & 0 & [14 26] & 0.92 & 50 & 0 & [14 23] & 0.72 \\
-0.76 & -1.11 & 0.18 & 0.14 & 50 & 0 &  [12 23]  & 0.42 & 50 & 0 & [14 26] & 0.88 & 50 & 0 & [14 23] & 0.69 \\
-0.77 & -1.11 & 0.2 & 0.18 & 50 & 0 &  [9 22]  & 0.42 & 50 & 0 & [14 28] & 0.85 & 50 & 0 & [12 23] & 0.77 \\
\hline
-0.72 & -1.12 & 0.18 & 0.2 & 50 & 0 &  [12 23]  & 0.66 & 50 & 0 & [14 26] & 1.01 & 50 & 0 & [12 23] & 0.62 \\
-0.74 & -1.12 & 0.18 & 0.2 & 50 & 0 &  [12 23]  & 0.48 & 50 & 0 & [14 26] & 0.97 & 50 & 0 & [12 23] & 0.61 \\
-0.76 & -1.12 & 0.18 & 0.2 & 50 & 0 &  [12 23]  & 0.4 & 50 & 0 & [14 26] & 0.97 & 50 & 0 & [12 23] & 0.71 \\
-0.77 & -1.12 & 0.2 & 0.18 & 50 & 0 &  [9 22]  & 0.46 & 50 & 0 & [14 28] & 0.97 & 50 & 0 & [12 23] & 0.81 \\
\hline
\end{tabular} 

%% file: tabs/table_17.tex
\begin{tabular}{ | c c | c c | c c c c | c c c c| c c c c|} 
  \hline\hline 
& 	&  &  & \multicolumn{4}{c|}{$\Lambda_{\rm CB}$}  &   \multicolumn{4}{c|}{$\Sigma_{\rm CB}$} &   \multicolumn{4}{c|}{$\Sigma^\ast_{\rm CB}$}  \\ 
 $am_0^{(f)}$ & $am_0^{(as)}$ & $\epsilon^{(f)}$  & $\epsilon^{(as)}$   &  $N_{\rm W}^{\rm source}$ 	 &  $N_{\rm W}^{\rm sink}$ 	&  I 	   &  $\chi^2/N_{\rm d.o.f.}$ & $N_{\rm W}^{\rm source}$ 	 &  $N_{\rm W}^{\rm sink}$   &   I 	   &  $\chi^2/N_{\rm d.o.f.}$& $N_{\rm W}^{\rm source}$ 	 &  $N_{\rm W}^{\rm sink}$   &   I 	   &  $\chi^2/N_{\rm d.o.f.}$ \\ 
 \hline-0.66 & -0.85 & 0.18 & 0.01 & 50 & 0 &  [19 27]  & 1.17 & 50 & 0 & [19 29] & 1.69 & 50 & 0 & [16 30] & 0.78 \\
-0.68 & -0.85 & 0.18 & 0.01 & 50 & 0 &  [17 30]  & 1.0 & 50 & 0 & [16 30] & 1.13 & 50 & 0 & [16 28] & 0.78 \\
-0.7 & -0.85 & 0.18 & 0.01 & 50 & 0 &  [17 25]  & 0.37 & 50 & 0 & [16 27] & 1.03 & 50 & 0 & [18 28] & 0.62 \\
-0.72 & -0.85 & 0.18 & 0.01 & 50 & 0 &  [17 25]  & 0.35 & 50 & 0 & [16 27] & 0.74 & 50 & 0 & [18 28] & 0.6 \\
\hline
-0.66 & -0.87 & 0.18 & 0.01 & 50 & 0 &  [17 25]  & 0.88 & 50 & 0 & [16 27] & 1.55 & 50 & 0 & [18 28] & 0.62 \\
-0.68 & -0.87 & 0.18 & 0.01 & 50 & 0 &  [17 25]  & 0.66 & 50 & 0 & [16 27] & 1.4 & 50 & 0 & [18 28] & 0.6 \\
-0.7 & -0.87 & 0.18 & 0.01 & 50 & 0 &  [17 25]  & 0.39 & 50 & 0 & [16 27] & 1.08 & 50 & 0 & [18 28] & 0.59 \\
-0.72 & -0.87 & 0.18 & 0.01 & 50 & 0 &  [17 25]  & 0.37 & 50 & 0 & [16 27] & 0.76 & 50 & 0 & [18 28] & 0.56 \\
\hline
-0.66 & -0.9 & 0.18 & 0.01 & 50 & 0 &  [17 27]  & 1.13 & 50 & 0 & [16 29] & 1.43 & 50 & 0 & [16 30] & 0.85 \\
-0.68 & -0.9 & 0.18 & 0.01 & 50 & 0 &  [17 27]  & 1.01 & 50 & 0 & [16 29] & 1.3 & 50 & 0 & [16 30] & 1.04 \\
-0.7 & -0.9 & 0.18 & 0.01 & 50 & 0 &  [17 27]  & 1.12 & 50 & 0 & [16 29] & 0.99 & 50 & 0 & [16 30] & 1.19 \\
-0.72 & -0.9 & 0.18 & 0.01 & 50 & 0 &  [17 25]  & 0.4 & 50 & 0 & [16 25] & 0.52 & 50 & 0 & [16 26] & 0.68 \\
\hline
-0.66 & -0.93 & 0.18 & 0.01 & 50 & 0 &  [17 25]  & 1.02 & 50 & 0 & [16 27] & 1.67 & 50 & 0 & [18 28] & 0.57 \\
-0.68 & -0.93 & 0.18 & 0.01 & 50 & 0 &  [17 25]  & 0.79 & 50 & 0 & [16 27] & 1.52 & 50 & 0 & [18 28] & 0.5 \\
-0.7 & -0.93 & 0.18 & 0.01 & 50 & 0 &  [17 25]  & 0.5 & 50 & 0 & [16 27] & 1.19 & 50 & 0 & [18 28] & 0.46 \\
-0.72 & -0.93 & 0.18 & 0.01 & 50 & 0 &  [17 25]  & 0.45 & 50 & 0 & [16 27] & 0.78 & 50 & 0 & [18 28] & 0.41 \\
\hline
-0.72 & -0.97 & 0.18 & 0.01 & 50 & 0 &  [17 25]  & 0.56 & 50 & 0 & [16 27] & 0.75 & 50 & 0 & [18 28] & 0.3 \\
-0.73 & -0.97 & 0.18 & 0.01 & 50 & 0 &  [17 25]  & 0.69 & 50 & 0 & [16 27] & 0.67 & 50 & 0 & [18 28] & 0.4 \\
\hline
-0.72 & -0.99 & 0.18 & 0.01 & 50 & 0 &  [17 25]  & 0.61 & 50 & 0 & [16 27] & 0.71 & 50 & 0 & [18 28] & 0.28 \\
-0.73 & -0.99 & 0.18 & 0.01 & 50 & 0 &  [17 25]  & 0.76 & 50 & 0 & [16 27] & 0.62 & 50 & 0 & [18 28] & 0.35 \\
\hline
-0.72 & -1.01 & 0.18 & 0.01 & 50 & 0 &  [17 25]  & 0.66 & 50 & 0 & [16 27] & 0.67 & 50 & 0 & [18 28] & 0.3 \\
-0.73 & -1.01 & 0.18 & 0.01 & 50 & 0 &  [17 25]  & 0.79 & 50 & 0 & [16 27] & 0.58 & 50 & 0 & [18 28] & 0.34 \\
\hline
-0.72 & -1.03 & 0.18 & 0.01 & 50 & 0 &  [17 25]  & 0.71 & 50 & 0 & [16 27] & 0.66 & 50 & 0 & [18 28] & 0.38 \\
-0.73 & -1.03 & 0.18 & 0.01 & 50 & 0 &  [17 25]  & 0.79 & 50 & 0 & [16 27] & 0.61 & 50 & 0 & [18 28] & 0.36 \\
\hline
-0.7 & -1.04 & 0.18 & 0.16 & 50 & 0 &  [17 25]  & 0.92 & 50 & 0 & [17 27] & 1.08 & 50 & 0 & [15 27] & 0.58 \\
-0.71 & -1.04 & 0.18 & 0.16 & 50 & 0 &  [17 25]  & 0.82 & 50 & 0 & [17 27] & 0.91 & 50 & 0 & [15 27] & 0.44 \\
-0.72 & -1.04 & 0.18 & 0.16 & 50 & 0 &  [17 25]  & 0.92 & 50 & 0 & [17 27] & 0.76 & 50 & 0 & [15 27] & 0.36 \\
\hline
-0.7 & -1.05 & 0.18 & 0.18 & 50 & 0 &  [16 25]  & 0.87 & 50 & 0 & [17 27] & 1.13 & 50 & 0 & [16 27] & 0.6 \\
-0.71 & -1.05 & 0.18 & 0.18 & 50 & 0 &  [16 25]  & 0.76 & 50 & 0 & [17 27] & 0.99 & 50 & 0 & [16 27] & 0.47 \\
-0.72 & -1.05 & 0.18 & 0.18 & 50 & 0 &  [16 25]  & 0.77 & 50 & 0 & [17 27] & 0.88 & 50 & 0 & [16 27] & 0.41 \\
\hline
-0.7 & -1.06 & 0.18 & 0.2 & 50 & 0 &  [16 30]  & 0.8 & 50 & 0 & [16 30] & 1.13 & 50 & 0 & [16 28] & 0.5 \\
-0.71 & -1.06 & 0.18 & 0.2 & 50 & 0 &  [16 30]  & 0.8 & 50 & 0 & [16 30] & 1.19 & 50 & 0 & [16 28] & 0.41 \\
-0.72 & -1.06 & 0.18 & 0.2 & 50 & 0 &  [16 30]  & 0.86 & 50 & 0 & [16 30] & 1.28 & 50 & 0 & [16 28] & 0.38 \\
\hline
\end{tabular} 

%% file: tabs/table_18.tex
\begin{tabular}{ | c c | c c | c c c c | c c c c| c c c c|} 
  \hline\hline 
& 	&  &  & \multicolumn{4}{c|}{$\Lambda_{\rm CB}$}  &   \multicolumn{4}{c|}{$\Sigma_{\rm CB}$} &   \multicolumn{4}{c|}{$\Sigma^\ast_{\rm CB}$}  \\ 
 $am_0^{(f)}$ & $am_0^{(as)}$ & $\epsilon^{(f)}$  & $\epsilon^{(as)}$   &  $N_{\rm W}^{\rm source}$ 	 &  $N_{\rm W}^{\rm sink}$ 	&  I 	   &  $\chi^2/N_{\rm d.o.f.}$ & $N_{\rm W}^{\rm source}$ 	 &  $N_{\rm W}^{\rm sink}$   &   I 	   &  $\chi^2/N_{\rm d.o.f.}$& $N_{\rm W}^{\rm source}$ 	 &  $N_{\rm W}^{\rm sink}$   &   I 	   &  $\chi^2/N_{\rm d.o.f.}$ \\ 
 \hline-0.6 & -0.81 & 0.18 & 0.01 & 100 & 0 &  [20 29]  & 1.08 & 100 & 0 & [20 30] & 0.62 & 100 & 0 & [19 30] & 0.43 \\
-0.62 & -0.81 & 0.18 & 0.01 & 60 & 0 &  [20 30]  & 1.3 & 60 & 0 & [20 30] & 0.56 & 60 & 0 & [19 30] & 0.39 \\
-0.64 & -0.81 & 0.18 & 0.01 & 60 & 0 &  [20 30]  & 1.55 & 60 & 0 & [20 30] & 0.55 & 60 & 0 & [19 30] & 0.31 \\
-0.66 & -0.81 & 0.18 & 0.1 & 50 & 0 &  [19 30]  & 1.38 & 50 & 0 & [20 30] & 0.72 & 50 & 0 & [20 30] & 0.26 \\
-0.68 & -0.81 & 0.18 & 0.1 & 50 & 0 &  [19 27]  & 1.44 & 50 & 0 & [20 27] & 1.0 & 50 & 0 & [20 30] & 0.48 \\
-0.69 & -0.81 & 0.18 & 0.1 & 50 & 0 &  [17 27]  & 1.0 & 50 & 0 & [19 27] & 1.24 & 50 & 0 & [20 30] & 0.85 \\
-0.7 & -0.81 & 0.18 & 0.01 & 50 & 0 &  [16 27]  & 0.57 & 50 & 0 & [17 25] & 1.33 & 50 & 0 & [19 27] & 0.46 \\
\hline
-0.62 & -0.82 & 0.18 & 0.01 & 60 & 0 &  [20 30]  & 1.3 & 60 & 0 & [20 30] & 0.57 & 60 & 0 & [19 30] & 0.4 \\
-0.64 & -0.82 & 0.18 & 0.01 & 60 & 0 &  [20 30]  & 1.56 & 60 & 0 & [20 30] & 0.56 & 60 & 0 & [19 30] & 0.32 \\
-0.68 & -0.82 & 0.18 & 0.01 & 60 & 0 &  [19 27]  & 1.32 & 60 & 0 & [19 27] & 0.67 & 60 & 0 & [19 30] & 0.33 \\
-0.7 & -0.82 & 0.18 & 0.01 & 50 & 0 &  [16 27]  & 0.58 & 50 & 0 & [17 25] & 1.31 & 50 & 0 & [19 27] & 0.46 \\
\hline
-0.62 & -0.84 & 0.18 & 0.01 & 60 & 0 &  [20 30]  & 1.32 & 60 & 0 & [20 30] & 0.58 & 60 & 0 & [19 30] & 0.41 \\
-0.64 & -0.84 & 0.18 & 0.01 & 60 & 0 &  [20 30]  & 1.59 & 60 & 0 & [20 30] & 0.58 & 60 & 0 & [19 30] & 0.32 \\
-0.68 & -0.84 & 0.18 & 0.01 & 60 & 0 &  [19 27]  & 1.36 & 60 & 0 & [19 27] & 0.71 & 60 & 0 & [19 30] & 0.34 \\
-0.7 & -0.84 & 0.18 & 0.01 & 50 & 0 &  [16 27]  & 0.6 & 50 & 0 & [17 25] & 1.26 & 50 & 0 & [19 27] & 0.45 \\
\hline
-0.64 & -0.91 & 0.18 & 0.01 & 60 & 0 &  [20 30]  & 1.78 & 60 & 0 & [20 29] & 0.83 & 60 & 0 & [22 30] & 0.43 \\
-0.66 & -0.91 & 0.18 & 0.01 & 100 & 0 &  [19 29]  & 1.44 & 100 & 0 & [20 29] & 0.84 & 100 & 0 & [19 30] & 0.29 \\
\hline
-0.64 & -0.93 & 0.18 & 0.01 & 60 & 0 &  [20 30]  & 1.88 & 60 & 0 & [20 29] & 0.91 & 60 & 0 & [22 30] & 0.46 \\
-0.66 & -0.93 & 0.18 & 0.01 & 100 & 0 &  [19 29]  & 1.51 & 100 & 0 & [20 29] & 0.91 & 100 & 0 & [19 30] & 0.33 \\
\hline
-0.64 & -0.95 & 0.18 & 0.01 & 60 & 0 &  [20 26]  & 1.55 & 60 & 0 & [20 29] & 1.0 & 60 & 0 & [22 26] & 0.7 \\
-0.66 & -0.95 & 0.18 & 0.01 & 100 & 0 &  [20 28]  & 2.08 & 100 & 0 & [20 29] & 0.99 & 100 & 0 & [19 30] & 0.38 \\
\hline
-0.66 & -0.99 & 0.18 & 0.16 & 50 & 0 &  [18 29]  & 2.19 & 50 & 0 & [20 30] & 1.31 & 50 & 0 & [18 27] & 0.48 \\
-0.68 & -0.99 & 0.18 & 0.16 & 50 & 0 &  [18 29]  & 1.4 & 50 & 0 & [20 30] & 1.11 & 50 & 0 & [18 27] & 0.41 \\
\hline
-0.66 & -1.01 & 0.18 & 0.16 & 50 & 0 &  [17 30]  & 1.36 & 50 & 0 & [20 30] & 1.46 & 50 & 0 & [18 26] & 0.09 \\
-0.68 & -1.01 & 0.18 & 0.16 & 50 & 0 &  [17 30]  & 0.62 & 50 & 0 & [20 30] & 1.22 & 50 & 0 & [18 26] & 0.31 \\
\hline
-0.66 & -1.015 & 0.18 & 0.2 & 50 & 0 &  [18 29]  & 1.31 & 50 & 0 & [21 30] & 1.36 & 50 & 0 & [18 26] & 0.12 \\
-0.67 & -1.015 & 0.18 & 0.2 & 50 & 0 &  [18 29]  & 0.96 & 50 & 0 & [21 30] & 1.23 & 50 & 0 & [19 26] & 0.21 \\
-0.68 & -1.015 & 0.18 & 0.2 & 50 & 0 &  [18 29]  & 0.62 & 50 & 0 & [21 30] & 1.13 & 50 & 0 & [17 26] & 0.34 \\
-0.69 & -1.015 & 0.18 & 0.2 & 50 & 0 &  [18 29]  & 0.57 & 50 & 0 & [21 30] & 1.04 & 50 & 0 & [18 25] & 0.57 \\
\hline
\end{tabular} 

%% file: tabs/table_19.tex
\begin{tabular}{ | c c | c c | c c c c | c c c c| c c c c|} 
  \hline\hline 
& 	&  &  & \multicolumn{4}{c|}{$\Lambda_{\rm CB}$}  &   \multicolumn{4}{c|}{$\Sigma_{\rm CB}$} &   \multicolumn{4}{c|}{$\Sigma^\ast_{\rm CB}$}  \\ 
 $am_0^{(f)}$ & $am_0^{(as)}$ & $\epsilon^{(f)}$  & $\epsilon^{(as)}$   &  $N_{\rm W}^{\rm source}$ 	 &  $N_{\rm W}^{\rm sink}$ 	&  I 	   &  $\chi^2/N_{\rm d.o.f.}$ & $N_{\rm W}^{\rm source}$ 	 &  $N_{\rm W}^{\rm sink}$   &   I 	   &  $\chi^2/N_{\rm d.o.f.}$& $N_{\rm W}^{\rm source}$ 	 &  $N_{\rm W}^{\rm sink}$   &   I 	   &  $\chi^2/N_{\rm d.o.f.}$ \\ 
 \hline-0.62 & -0.95 & 0.18 & 0.2 & 50 & 20 &  [20 30]  & 0.36 & 50 & 20 & [20 30] & 0.36 & 50 & 0 & [21 30] & 0.89 \\
-0.64 & -0.95 & 0.18 & 0.2 & 50 & 20 &  [20 30]  & 0.49 & 50 & 20 & [20 29] & 0.38 & 50 & 0 & [21 30] & 0.57 \\
-0.646 & -0.95 & 0.18 & 0.2 & 50 & 20 &  [20 30]  & 0.49 & 50 & 20 & [20 29] & 0.4 & 50 & 0 & [21 30] & 0.52 \\
\hline
-0.62 & -0.956 & 0.18 & 0.2 & 50 & 20 &  [20 30]  & 0.39 & 50 & 20 & [20 29] & 0.29 & 50 & 0 & [21 30] & 0.82 \\
-0.64 & -0.956 & 0.18 & 0.2 & 50 & 20 &  [20 30]  & 0.53 & 50 & 20 & [20 29] & 0.43 & 50 & 0 & [21 30] & 0.5 \\
-0.646 & -0.956 & 0.18 & 0.2 & 50 & 20 &  [20 30]  & 0.54 & 50 & 20 & [20 29] & 0.43 & 50 & 0 & [21 30] & 0.48 \\
\hline
-0.62 & -0.961 & 0.18 & 0.2 & 50 & 0 &  [22 30]  & 0.54 & 50 & 0 & [23 29] & 0.1 & 50 & 0 & [21 30] & 0.75 \\
-0.64 & -0.961 & 0.18 & 0.2 & 50 & 0 &  [22 30]  & 0.77 & 50 & 0 & [23 29] & 0.16 & 50 & 0 & [21 30] & 0.48 \\
-0.646 & -0.961 & 0.18 & 0.2 & 50 & 0 &  [22 30]  & 0.82 & 50 & 0 & [23 29] & 0.14 & 50 & 0 & [21 30] & 0.5 \\
\hline
\end{tabular} 